\documentclass[11pt]{article}
\usepackage[pdftex,margin=1in]{geometry}
\usepackage[T1]{fontenc}
\usepackage[utf8x]{inputenc}
\usepackage{amsmath,amsthm,amssymb,array,graphicx,enumerate,booktabs,bigstrut,rotating,multirow,etoolbox,lmodern,comment,listings,moresize,bbm,float,subcaption,pdfescape,mathtools,enumitem,multicol,soul,import,tikz, placeins,ragged2e,lscape,dsfont,setspace,flafter,lineno,stackengine}
\usepackage[justification=centering,font=small]{caption}
\usepackage[flushleft]{threeparttable}
\usepackage[colorinlistoftodos]{todonotes}
\usepackage[comma]{natbib}
\usepackage[english]{babel}
\usepackage[small]{titlesec}
\DeclareMathOperator{\E}{\mathbb{E}}
\usepackage{lipsum}
\usepackage{hyperref}
\hypersetup{colorlinks,linkcolor={blue},citecolor={blue},urlcolor={blue}}  

\setcitestyle{citesep={;}}

\makeatletter

\renewcommand\subsubsection{\@startsection{subsubsection}{3}{\z@}%
	{-3.25ex\@plus -1ex \@minus -.2ex}%
	{-1.5ex \@plus -.2ex}%
	{\normalfont\normalsize\bfseries}
}
\def\@biblabel#1{\hspace*{-\labelsep}}
\makeatother
\def\sym#1{\ifmmode^{#1}\else\(^{#1}\)\fi}

\newcolumntype{L}[1]{>{\raggedright\let\newline\\\arraybackslash\hspace{0pt}}m{#1}}
\newcolumntype{C}[1]{>{\centering\let\newline\\\arraybackslash\hspace{0pt}}m{#1}}
\newcolumntype{R}[1]{>{\raggedleft\let\newline\\\arraybackslash\hspace{0pt}}m{#1}}

\begin{document}

\title{Inequality of Opportunity and Income Redistribution\thanks{Preuss: Cornell University, email, \href{mp2222@cornell.edu}{mp2222@cornell.edu}; Reyes: Middlebury College and IZA, email, \href{greyes@middlebury.edu}{greyes@middlebury.edu}; Somerville: Federal Reserve Bank of New York, email, \href{jason.somerville@ny.frb.org}{jason.somerville@ny.frb.org}; Wu: University of British Columbia, email, \href{joy.wu@sauder.ubc.ca}{joy.wu@sauder.ubc.ca}. For helpful comments and suggestions, we thank Peter Andre, Maxim Bakhtin, Hans-Peter Gr\"uner, Ori Heffetz, Margaret Jodlowski, Muriel Niederle, Alex Rees-Jones, Ted O'Donoghue, Ryan Oprea, Matthew Rabin, and seminar participants at the briq Institute on Behavior and Inequality, the Economic and Social Research Institute, the VfS Annual Conference 2022, the 2022 Big Data in Economics at ZBW-Leibniz Information Centre for Economics, the 2022 North-American Economic Science Association Conference, the 2023 CESifo Area Conference on Public Economics, the ORG seminar at LMU Munich, and the 2024 Spring Behavioral and Experimental Economics Research Conference. The views expressed in this paper are those of the authors and do not necessarily reflect the views of the Federal Reserve Bank of New York or the Federal Reserve System. The experiment reported in this paper was pre-registered in the AEA RCT Registry in October 2021, under the ID \href{https://www.socialscienceregistry.org/trials/8147}{AEARCTR-0008147}. Details on our experimental and survey items are available in our \href{https://joyzwu.github.io/papers/prsw_supplementary.pdf}{Supplementary Materials}. The experiment was reviewed and granted an exemption by the Institutional Review Board at Cornell University. This work has been supported in part by a grant from the Cornell Center for Social Sciences.}}

\author{
	         Marcel Preuss
		\and Germán Reyes
		\and Jason Somerville
		\and Joy Wu
	}

\renewcommand{\today}{\ifcase \month \or January\or February\or March\or %
		April\or May \or June\or July\or August\or September\or October\or November\or %
		December\fi \ \number \year} 
		
\date{\today}

\maketitle
\thispagestyle{empty}

\begin{abstract} 

\noindent Opportunities, such as access to education or family background, shape income inequality by influencing the chances of economic success. Unequal opportunities create uncertainty about whether success is merit- or luck-based. We examine how uncertainty impacts preferences for redistribution, comparing it to more transparent forms of luck. Using a U.S.-representative sample, we elicit redistribution decisions in two environments: when workers' earnings are influenced by luck directly (``lucky outcomes'') or indirectly (``lucky opportunities''). Spectators redistribute less and are less sensitive to differences in luck under opportunities. Inferential challenges drive this gap, which implies that individuals are less meritocratic than previous findings suggest.

    \vspace{10pt}
    \vspace{10pt}

 \end{abstract}

\clearpage

\section{Introduction}

The design of redistribution policies---from progressive taxation to social welfare programs---depends on what citizens consider to be a fair distribution of income. Experimental work has found that most people hold meritocratic fairness attitudes: they tolerate income disparities that are merit-based but choose to redistribute when income differences result from circumstances beyond individuals' control, such as luck \citep[e.g.,][]{cappelen2007pluralism, cappelen2013just, almaas2020cutthroat}. However, the widespread acceptance of meritocratic principles contrasts sharply with recent trends in income inequality in the United States, where support for redistribution has remained stagnant despite a rising impact of circumstances beyond people's control \citep{chetty2014united,ashok2015support, Bozio2024}.

We propose that this apparent disconnect between experimental findings and observed redistribution trends is partly due to limitations in how prior studies implemented luck in lab settings. Previous work has primarily focused on redistribution behavior in situations where luck arises in a simple and transparent way, making it easy to assess whether inequality is merit- or luck-based. By contrast, in a survey we conducted of U.S. households, we found that Americans identify unequal opportunities that merely affect someone's \textit{chances of success} as the most critical source of luck in life.\footnote{Approximately 1,000 panelists from the New York Fed's \textit{Survey of Consumer Expectations} participated in this survey. We provide a detailed description and analysis of the survey in Appendix \ref{sec:luck-survey}.} For example, the main sources of luck identified by these U.S. households include the family one is born into and the education one receives. The impact of these opportunities on an individual's success is hard to assess without also observing that individual's effort. Thus, a fundamental feature of unequal opportunities is that they create uncertainty about the reasons for someone's success. This uncertainty creates inferential challenges that prior studies have not examined.

To illustrate the difficulty of implementing meritocratic fairness ideals when there are unequal opportunities, consider whether the success of Microsoft founder Bill Gates was due to merit or luck. On one hand, he was fortunate to attend one of the few high schools that offered unlimited access to a computer programming terminal. This is just one of many lucky opportunities in Gates' lifetime that were instrumental to his eventual success \citep{frank2016success}. However, he is also known for a fierce work ethic, famously stating that he ``didn't believe in weekends; didn't believe in vacations.'' Both luck and effort were instrumental in his professional success, and assessing the exact impact of each factor poses a challenging inference problem. 

We recruited 2,400 workers on an online labor market to perform a real-effort task and randomly paired them to compete for a fixed prize in a winner-takes-all environment. Then, we asked 1,170 individuals (``spectators'') from the New York Fed's \textit{Survey of Consumer Expectations}---a nationally representative survey of U.S. households---to choose the final earnings allocation for pairs of workers. We randomly assigned spectators to one of two luck environments. In the \textit{lucky outcomes} environment, we selected the winner of each worker pair either by the number of tasks completed (merit) or by a coin flip (luck). In the \textit{lucky opportunities} environment, the winner of each pair was the worker with the higher number of tasks completed (merit) multiplied by a randomly assigned productivity multiplier (luck). These productivity multipliers capture, in reduced form, the type of advantages that create uncertainty about the source of success: circumstances beyond individual control such as family background, educational quality, and early-life conditions. We generate exogenous variation in the probability that inequality is merit-based---that is, the probability that the winner of a pair is the worker who completed more tasks---by varying the coin-flip probability (in lucky outcomes) or the relative productivity multipliers (in lucky opportunities). We use this probability to compare redistribution across the different luck environments while holding luck (the likelihood that the winner of a pair is the worker who performed better) constant.

In our baseline treatment, spectators are not informed about the probability that inequality is merit-based. Instead, spectators observe information about the environment that conveys the importance of merit and luck in determining the winner. The inferential challenge they face is using this information to infer how likely it is that the inequality between workers is merit-based. Under \textit{lucky outcomes}, spectators observe the probability that the winner was determined by a coin flip. This feature of the lucky outcomes environment is linearly related to the probability that inequality is merit-based. Under \textit{lucky opportunities}, they observe the productivity multiplier of each worker. Mapping this feature of the lucky opportunities environment to the probability that inequality is merit-based is more challenging since a small change in the relative advantage of the winner can imply large changes in the probability that the winner is the worker who completed more tasks.\footnote{The inferential challenges spectators face are not unique to our experimental design, but an inherent feature of many environments in which workers have unequal opportunities. Specifically, the probability that the outcome is merit-based is decreasing and convex in the winning worker's relative advantage. In Online Appendix Sections \ref{sec:convexity} and \ref{sec:headstarts}, we show that the same relationship prevails even if lucky opportunities are not multiplicative.}

Our main result is that spectators redistribute less when inequality arises from lucky opportunities compared to lucky outcomes. On average, spectators redistributed $27.6$ percent of earnings from winner to loser under lucky outcomes but only $23.3$ percent under lucky opportunities. This redistribution gap emerges only when there is uncertainty about whether inequality is merit-based. Specifically, when performance alone determines the winner (i.e., under equal productivity multipliers or zero probability of a coin flip), spectators redistribute similarly across environments. As the likelihood of luck-based inequality increases, redistribution responds strongly under lucky outcomes but remains more muted under lucky opportunities, creating a redistribution gap between environments. We estimate that a ten percentage point increase in the probability of luck-based inequality leads to a 3.7 percentage point increase in redistribution under lucky outcomes but only a 2.0 percentage point increase under lucky opportunities.

We introduce two additional treatments to investigate the mechanisms underlying the differences in redistribution across luck environments. First, to further examine the role of inferential challenges, we provide direct information about the likelihood that inequality is merit-based. Consistent with spectators facing inferential challenges, we find that this information significantly reduces the redistribution gap between luck environments. We show that the information acts by changing the way in which spectators incorporate unequal opportunities into their redistribution decisions. Second, to isolate the role of potential effort responses in driving our results, we implement an ``ex-post'' lucky opportunities condition in which workers learn about their multiplier only \textit{after} completing the task. We find no significant difference in redistribution between the baseline and ex-post conditions, indicating that effort responses do not explain the redistribution gap. 

We primarily contribute to a broad literature that studies how the source of inequality affects redistribution. Numerous experimental studies have demonstrated that people tolerate more inequality when earnings differences stem from differences in effort or individual choices, rather than luck \citep{cappelen2010responsibility,cappelen2013just, durante_preferences_2014, mollerstrom2015luck,cappelen2020merit,almaas2020cutthroat,cappelen2022meritocratic}.\footnote{Other research on the factors influencing redistribution has examined the role of other-regarding preferences \citep[e.g.,][]{charness_understanding_2002}, fairness ideals \citep[e.g.,][]{konow_fair_2000, cappelen2007pluralism} and perceptions regarding mobility and income inequality \citep[e.g.,][]{cruces2013biased,kuziemko_how_2015,alesina2018intergenerational,fehr2024social}.} We show that whether luck arises through unequal opportunities or unequal outcomes significantly affects how much inequality individuals tolerate. We also advance the methodology of redistribution experiments by introducing a common scale for quantifying the importance of luck under uncertainty. This scale provides a portable definition of luck that can be widely applied to different tournament environments.\footnote{Our approach builds on \cite{cappelen2022meritocratic}, who study redistribution under lucky outcomes using variation in the probability that inequality is merit-based. We extend their framework by deriving and implementing an analogous measure for opportunity-based luck, enabling us to systematically compare how redistribution responds to different forms of luck while holding the magnitude of luck constant.}

Our work also contributes to the emerging literature examining how individuals make redistribution decisions when luck stems from differences in workers' circumstances \citep{bhattacharyaandmollerstrom:2022, dongetal:2024, andre2024shallow}. \cite{andre2024shallow} investigates whether spectators hold workers responsible for the unequal incentives they face. He finds that disparities in randomly assigned piece-rate wages produce large differences in worker performance. However, spectators reward workers according to their effort and irrespective of how differences in piece rates impact workers' performance. \cite{dongetal:2024} examine redistribution when luck arises through different employment offers or training quality. They find that spectators only partially account for these differences in advantages when making redistribution decisions. Similarly, \cite{bhattacharyaandmollerstrom:2022} examine redistribution decisions when luck determines whether individuals can work at all. They find that spectators tolerate more inequality when luck determines who is allowed to work than when luck directly determines outcomes. 

Our paper advances this literature in two fundamental ways. First, we study an environment where the impact of unequal opportunities is uncertain, mirroring real-world settings where the relationship between advantages and success is ambiguous. This design reveals that inferential challenges---the difficulty in assessing how much luck versus merit drives inequality---are key to understanding redistribution decisions. Second, while prior work focuses on how spectators reward performance differences that arise from unequal opportunities, we examine situations where equally hard-working individuals can achieve vastly different outcomes purely due to their circumstances. This shows that spectators have different attitudes towards redistribution even when there are no performance differences across workers. 

Finally, we contribute to the literature that examines heuristics and biases in the inference process by documenting their consequences in an important economic context. We document that in the lucky opportunities environment, redistribution is close to linear on the workers' multiplier difference. This ``linearization heuristic'' leads spectators to underestimate the impact of unequal opportunities and consequently to redistribute less than they would under lucky outcomes. This finding is consistent with previous work showing that individuals often struggle to estimate nonlinear relationships \citep{larrick2008mpg, levy2016exponential, rees2020measuring}, and fail to solve even simple Bayesian updating problems \citep{benjamin2019errors}, particularly in environments with uncertainty that require contingent reasoning \citep{martinez2019failures}. The results from our information provision treatment are consistent with work showing that drawing attention to an attribute increases individuals' sensitivity to that attribute \citep{conlon2025} and that people focus on readily available information \citep{enke2020you, graeber2023inattentive}.

\section{Theoretical Framework}\label{sec:theory}

This section presents a stylized model of spectators' redistribution decisions when spectators care about the source of inequality but worker performance is unobservable. Our aim is not to test a specific model of spectator behavior, but rather to use the model to clarify our experimental setup and provide a common framework for quantifying the impact of luck on outcomes, regardless of its source. The model setup closely follows that of \cite{cappelen2022meritocratic} and applies the framework to different luck environments. 

Consider an impartial spectator who observes initial earnings in a winner-takes-all environment in which two randomly paired workers compete for a fixed prize. Spectator $i$'s task is to choose $r_i$, the fraction of income to redistribute from the winner to the loser. We model $r_i$ as a function of fairness views and beliefs about how likely it is that outcomes are based on merit. Formally, let $f_i$ denote the share of total income for the worker with the lower performance that spectator $i$ deems to be fair, and let $1-f_i$ denote the fair share for the higher performance worker. We assume that spectators choose $r_i$ to minimize the quadratic difference between the fair allocation $(f_i,1-f_i)$ and the actual allocation $(r_i,1-r_i)$:
\begin{align}\label{spec_prob}
   U(r_i,f_i) =  -(r_i - f_i)^2.
\end{align}
 
If spectators know with certainty that the winner is the worker who performed better, then they implement the fair allocation, $r_i^* = f_i$. However, in our experiment, as is often the case in reality, spectators do not observe workers' performance. This creates uncertainty about whether the best performer won or lost. As a result, spectators maximize the expected utility
\begin{align} \label{eq_exp_ut}
    \E(U(r_i,f_i)) &= - \pi\Big(r_i  - f_i\Big)^2  - (1 -\pi)\Big(r_i - (1-f_i)\Big)^2,
\end{align}
where $\pi$ denotes the probability that the best performer won, and $1-\pi$ is the probability that the best performer lost. In other words, $\pi$ represents the probability that the outcome is merit-based, while $1-\pi$ represents the probability that the outcome is luck-based. Thus, we often refer to $1-\pi$ as our measure of the degree of luck in the environment, which we use to compare redistribution choices across different luck conditions in our experiment.

In an interior solution to (\ref{eq_exp_ut}), the optimal level of redistribution is 
\begin{align} \label{eq_opt_y}
     r^*_i = {\pi}f_i + (1- {\pi})(1 - f_i).
\end{align}
 Equation \eqref{eq_opt_y} highlights that redistribution depends on both the spectator preferences about the fair share of earnings for the lower- and higher-effort worker ($f_i$) and how likely it is that the outcome was due to luck ($1-\pi$). Provided the fair share for the lower performance worker satisfies $f_i<1/2$, the optimal level of redistribution increases in $1-\pi$. In other words, the more likely it is that the better-performing worker lost, the more spectators redistribute. When performance solely determines the winner ($1-\pi=0$), spectators redistribute the fair share to the loser, $r_i^* = f_i$. When outcomes are due to pure luck ($1-\pi = 1/2$), spectators equalize earnings and choose $r_i^* = 1/2$. 

\subsection{Outcome Luck and Opportunity Luck}\label{sec:opportunities_outcomes}

We consider two work environments with different processes for generating worker earnings. In the \textit{lucky outcomes} environment, the winner of each worker pair is determined by either luck \textit{or} performance. Specifically, we assume that with some probability $q \in [0,1]$, a coin flip determines the winner of the match. Conversely, with probability $1 - q$ the winner is the worker who exerted more effort.

In the \textit{lucky opportunities} environment, each worker is assigned an individual-specific productivity multiplier, $m_i$, which influences their earnings. We assume that the distribution of productivity multipliers is exogenous to the distribution of effort, meaning that a worker's productivity is determined by factors outside of their control, such as innate abilities or acquired skills \citep[e.g.,][]{mirrlees_exploration_1971}.\footnote{This approach captures the idea that some workers may be more fortunate than others due to their inherent talents or the opportunities they have had to develop their skills, leading to higher earnings even when putting in the same level of effort as their peers.} The winner of each worker pair is determined by comparing their final output, which is calculated by multiplying each worker's productivity multiplier by their performance level.

We assume that spectators observe the features of the environment that determine whether the outcome is merit-based $\pi$, but do not observe $\pi$ directly. In the lucky outcomes environment, spectators observe the coin-flip probability $q$. To infer $\pi$ from $q$, spectators must use Bayesian updating, which implies $\pi=1-q/2$. In the lucky opportunities environment, spectators observe the productivity multipliers of each worker. Spectators can use this information to estimate $\pi$. Without loss of generality, assume that worker $1$ wins the tournament, which means $m_1 e_1 > m_2 e_2$. The probability that the worker who won is the worker who exerted more effort given the information observed by the spectator is
\vspace{-2mm}
\begin{align}\label{eq_pi}
    \pi &=  \text{Pr} \Big(e_{1}\geq e_2 \Big| m_1 e_1 > m_2 e_2,m_1,m_2 \Big).
\end{align}

When inferring $\pi$ from the features of the environment, spectators can encounter cases with or without uncertainty about whether the best performer lost. First, if $m_1 \leq m_2$, then $\pi = 1$ and there is no uncertainty. Intuitively, if worker $1$ wins despite being less or equally productive as worker $2$, then they must have exerted more effort than worker $2$. Second, if $m_1 > m_2$, there is uncertainty and equation \eqref{eq_pi} becomes
\vspace{-2mm}
\begin{align} \label{pr_high_e}
    \pi &= \frac{\Pr\Big(e_1 \geq e_2,m_1,m_2 \Big)}{\Pr\Big( m_1 e_1 > m_2 e_2,m_1,m_2 \Big)}=\frac{1/2}{\Pr\left(e_2/e_1<m_1/m_2\right) }\geq \frac{1}{2},
\end{align}
where the equality follows under the assumption that the distribution of effort is independent of the multiplier.\footnote{This assumption is not essential for our analysis but simplifies the exposition of the theory. We also show in Section \ref{sec:effort} that this assumption is consistent with our experimental data.} Expression \eqref{pr_high_e} shows that $\pi$ depends on the relative multiplier $m_1/m_2$ or, equivalently, on the relative advantage conferred to worker $1$. We show in Appendix \ref{sec:convexity} that $\pi$ is decreasing and convex in this relative advantage for any log-normal distribution of effort. Thus, the probability that the best worker lost $(1-\pi)$ or, equivalently, the probability that the winner was luck-based, is increasing and concave in the winning worker's relative advantage. Intuitively, this concavity arises because the density of $e_2/e_1$, which measures the relative effort of the disadvantaged worker compared to the advantaged worker, is centered around one (equal effort) and decreases for larger values of $e_2/e_1$. Therefore, even small relative advantages can greatly impact who wins, whereas the incremental effect of increasing this relative advantage matters less.\footnote{The convexity of $\pi$ (concavity of $1-\pi$) in the relative advantage is not unique to the way we chose to model unequal opportunities. We show in Appendix \ref{sec:headstarts} that implementing opportunity luck as an additive headstart also leads to a convex mapping from lucky headstarts to the likelihood that the winning worker exerted more effort ($\pi$) for the empirical distribution of worker effort that we observe in the experiment. This is because $e_2-e_1$, which is the relevant measure of relative effort in this case, has the most probability mass near zero (equal effort), and therefore, the convexity intuition above still applies (see the last paragraph in Section \ref{sec:convexity} for the complete theoretical argument). We also show in Appendix \ref{sec:convexity} that the same intuition readily extends to other common effort distributions.} 

This framework illustrates how differences in spectators' beliefs about $\pi$ can lead to differences in redistribution when luck affects workers indirectly through unequal opportunities versus when luck can affect workers' earnings directly. A reason for potential variation in beliefs about $\pi$ across our luck environments is that estimating $\pi$ may be more challenging under lucky opportunities, particularly if spectators fail to appreciate that even a small multiplier advantage can correspond to a considerable probability that the winning worker is not the one with the higher performance.

\section{Experimental Design} \label{sec:design}

\subsection{Production, Earnings, and Redistribution Stages}\label{sub:spectators}

The experiment follows the impartial-spectator paradigm in \cite{cappelen2013just}. It consists of three stages: a production stage, an earnings stage, and a redistribution stage. In the production stage, workers complete a real-effort task to encrypt three-letter ``words'' into numerical code \citep{erkal2011relative}. Workers have five minutes to correctly encrypt as many words as possible using a dynamic and randomly generated codebook for each word \citep{benndorf2019minimizing}. In the earnings stage, we randomly pair workers and determine the winner based on some combination of performance and luck. We initially allocate earnings of \$5 to the winner and \$0 to the losers. The exact interaction between luck and performance forms our main experimental treatments, which we describe in Section \ref{spectator-groups}.\footnote{We designed all experimental programs in oTree \citep{chen2016otree}. Screenshots of our experimental design and procedure are available in the \href{https://joyzwu.github.io/papers/prsw_supplementary.pdf}{Supplementary Materials}.}  Workers know that a third party may influence their final earnings, but the spectator's identity is entirely anonymous to them.\footnote{To combat the influence of anchoring effects in the spectators' redistribution choices, we inform spectators that we did not tell workers whether they won or lost nor the exact amount they will earn in each case. Spectators know that we only informed workers that they could earn up to \$5 and that winning against their randomly assigned opponent increases their chances of earning those \$5. This design removes any confounding issues relating to spectators' unwillingness to take earnings away from what workers might expect.} 

In the redistribution stage, spectators choose how much income to redistribute from the winner to the loser. Spectators make 12 redistribution decisions involving different real pairs of workers. To incentivize spectators to respond truthfully, we randomly select and implement one of their 12 decisions. Spectators may redistribute any amount between \$0 and \$5 in \$0.50 increments at no cost.\footnote{We implemented costless redistribution to preserve comparability with prior experimental work on redistributive behavior \citep[e.g.,][]{cappelen2013just, almaas2020cutthroat}. This approach also allows us to isolate how spectators respond to varying sources of luck.} We present each decision as an adjustment schedule (see Appendix Figure~\ref{fig:CF_IOp} for an example of a redistribution choice). For example, an adjustment of \$1.50 implies \$3.50 for the winner and \$1.50 for the loser. The first option is always a \$0.00 adjustment, which we label a ``no''-adjustment choice. The remaining $\{\$0.50, \ldots, \$5.00\}$ redistribution choices are labeled as a ``yes''-adjustment choice and denote the final earnings for both the winner and the loser: that is, \{(winner gets, loser gets)\} $=$ \{$(\$4.50, \$0.50)$, $(\$4.00, \$1.00)$, \ldots, $(\$0.50, \$4.50)$, $(\$0.00, \$5.00)$\}.

\subsection{Spectator Treatments}\label{spectator-groups}

We implement within-subject variation in how likely it was that the winner was merit-based (i.e., $\pi$). Spectators always observe features of the environment that are informative about the probability that the inequality between workers is merit-based. Moreover, the experiment embeds between-subject variation in the source of luck (lucky opportunities vs. lucky outcomes), the timing of when luck is revealed to the workers (ex-post vs. ex-ante), the information available to spectators about the importance of luck (full vs. partial), and the timing of workers' awareness about the source of luck (after vs. before). Appendix Table \ref{tab:treatments} provides an overview of these between-subject treatments. 

\subsubsection{Lucky Outcomes vs. Lucky Opportunities.}

We randomly assign one-third of the spectators to redistribute earnings under lucky outcomes and two-thirds to redistribute earnings under lucky opportunities. In the lucky outcomes condition, we select the winner based on a coin flip with probability $q$ and based on the higher number of correct encryptions with probability $1 - q$. In the lucky opportunities condition, we generate inequality of opportunity by randomly assigning productivity multipliers to workers. For example, a worker with a multiplier of 1.2 who solved 20 encryptions would have a score of 24, while a worker with a multiplier of 3.0 who solved ten encryptions would have a score of 30. The winner in each pair is the worker who has the higher score. We draw the multiplier for each worker $i$ from the distribution: $m_i=1$ with probability $0.05$, $m_i=4$ with probability $0.05$, and $m_i \sim U(1,4)$ with probability $0.9$.\footnote{We chose a relatively high upper bound of $m = 4$ to ensure that it is possible to generate $\pi = 1/2$ in our lucky opportunities treatment ($\pi = 1/2$ when the advantaged worker wins always; as if pure chance determines the winner).} We round all multipliers to the nearest tenth.

We do not inform spectators about the workers' actual performance, though we do provide some information about how likely it is that luck determined the winner. In the lucky outcomes condition, spectators know the probability $q$ that we determine the winner by a coin flip, but not whether a coin flip actually determined the winner. Spectators also know that we do not reveal this probability to workers, though workers know that there is some unstated chance that a coin flip determines their outcomes. In the lucky opportunities condition, spectators know each worker's multiplier. We inform spectators that workers only know of their own multiplier and the distribution from which it is drawn, but do not know anything about the worker they compete against. Appendix Figure \ref{fig:CF_IOp} provides an example of a redistribution decision in our lucky outcomes condition and an example of a redistribution decision in our lucky opportunities condition. 

\subsubsection{Varying the Importance of Luck.}\label{sec:design-pi-importance}

We implement within-subject variation in how likely it is that the better-performing worker won across worker pairs. We implement this variation in the lucky outcomes environment by changing $q$ across matches. We implement this variation in the lucky opportunities environment by varying the relative workers' multipliers across worker pairs. We control for the importance of luck by introducing a portable common metric across environments: the probability that the winner in a given pair completed more encryptions ($\pi$). When $\pi = 0.50$, there was a 50 percent chance that the winner of the match was the one who performed better; for example, when a coin flip determined the outcome ($q = 1$) in the lucky outcomes environment or when the ratio between worker multipliers is sufficiently large (so that the worker with the high multiplier always won the match) in the lucky opportunities environment. When $\pi = 1$, there was a 100 percent chance that the match's winner performed better; for example, when $q = 0$ or when the winning worker had a weakly lower productivity multiplier than the losing worker.

Spectators make redistribution decisions for a total of 12 worker pairs. Each worker pair corresponds to a unique value of $\pi$ drawn from one of the following 12 bins:
\begin{align}\label{eq:pi_bins}
	\pi \in \Big\{ \underbrace{\{0.50\}}_{\text{Bin 1}}, \underbrace{\{0.51, ..., 0.54\}}_{\text{Bin 2}}, \underbrace{\{0.55, 0.56, ..., 0.59\}}_{\text{Bin 3}}, ...,  \underbrace{\{0.95, 0.96, ..., 0.99\}}_{\text{Bin 11}}, \underbrace{\{1\}}_{\text{Bin 12}} \Big\}.
\end{align}
For each spectator, we randomly draw one value of $\pi$ from each of the 12 bins. This ensures that every spectator makes a decision with $\pi=0.5$, $\pi=1$, and that the remaining values are evenly distributed throughout the support of $\pi$. We present the 12 decisions in random order. 

The key information we present on each decision is the multiplier of each worker pair, $(m_i,m_j)$, or the ex-ante probability that a coin flip $q$ determined the winner. Every possible pair $(m_i,m_j)$ and coin-flip probability $q$ correspond to a unique $\pi$ value. The mapping from $\pi$ to $q$ is given by the formula $q = 2(1 - \pi)$. To map $\pi$ to a multiplier pair, $(m_i,m_j)$, it is sufficient to consider the relative multiplier $m \equiv \max\{m_i,m_j\}/\min\{m_i,m_j\}$. Given any relative multiplier $m$, we examine all possible worker pairs and compute the fraction of times that the winner was the worker who solved more encryptions, which is the empirical counterpart to the probability that the winner is the worker who performed better, $\pi$.\footnote{With 800 workers per condition, there are ${800 \choose 2} = 319,600$ possible pairings. Since we can assign the higher multiplier to either worker, that creates 639,200 observations that we can use to calculate $\pi$ for each relative multiplier, $m$. Using this method, we compute, for each $m$, the fraction of all possible pairings in which the winner completed more encryptions than the loser.} This method yields a mapping from $m$ to $\pi$ as depicted in Panel B in Figure \ref{fig:worker-eff}, which shows that $\pi$ converges to $1/2$ for all $m>3$. For a given $\pi$, we then select a random worker pairing with a corresponding relative multiplier. 

\subsubsection{Timing of Opportunity Luck.}\label{sec:design_timing}
We also randomly vary the timing of when luck occurs. In our baseline lucky opportunities condition, we inform workers of their multipliers before they begin working on the encryption task. In the ex-post lucky opportunities condition, workers learn their multipliers only after completing the task. We assigned half of the spectators in the lucky opportunities conditions to the baseline treatment and the other half to the ex-post treatment. 

\subsubsection{Information Intervention.}
We randomly assign half of the spectators in each treatment to receive full information about $\pi$. In the lucky opportunities condition, we present the following additional text on the redistribution decision screen: ``Based on historical data for these multipliers, there is a [$\pi\times100$]\% chance that the winner above completed more transcriptions than the loser.'' In the lucky outcomes condition, the equivalent text is: ``There is a [$\pi\times100$]\% chance that the winner above completed more encryptions than the loser.'' As noted above, the value of $\pi$ varies from decision to decision. Appendix Figure~\ref{fig:CF_IOp} shows an example of the decision screens for the information treatments for both luck environments.  

\subsubsection{Workers' Awareness about Rules.}\label{sec:design_rules}
Finally, we vary the timing of when workers learn how luck plays a role in determining outcomes. In the rules-before condition, we inform workers that effort multipliers or a coin flip can influence the outcome \textit{before} they start the task. In the rules-after condition, we inform workers that multipliers or a coin flip can influence the outcome \textit{after} they complete the task. We randomly assign half of the spectators in the ex-post lucky opportunities and lucky outcomes conditions to the rules-before treatment and half to the rules-after treatment. By construction, we assign all participants in the baseline lucky opportunities condition to the rules-before treatment. Spectators have complete information about when workers learned how we determine the winner.

\subsection{Comprehension Checks and Belief Elicitation}\label{sec:design-comprehension}

To ensure that spectators understand the design details, we implement several comprehension questions after they see the initial instructions about the worker task. These questions test spectators' understanding of how luck can affect outcomes and their awareness of when workers learn about the importance of luck. Spectators can only continue once they select the correct answer, and we briefly explain why the answer is correct once they submit it. Therefore, these questions serve as both a comprehension check and as reminders that reinforce the critical aspects of the worker task that are central to our design.

After the 12 redistribution decisions, spectators complete a brief exit survey. The first part of the exit survey consists of three questions. First, we randomly select one of the 12 decisions that the spectators made and present the same information to them. We then ask spectators in the lucky outcomes condition how many encryptions they think workers solved on average. For spectators in the lucky opportunities condition, we randomly draw a multiplier and ask how many encryptions they think workers with that multiplier solved on average. Finally, we asked them how much they would allocate to the winner if they knew they had solved more encryptions. 

The second part of the exit survey asks spectators to select their level of agreement with several belief statements in a five-point Likert scale grid. It probes their views on various topics relating to income redistribution and the role of the government. We also embed  an attention check in one of the rows that states: ``Select disagree if you are reading this.'' 

\subsection{Recruitment of Workers and Spectators} 

We recruited 2,416 participants on Amazon Mechanical Turk to participate in the worker task in September 2021. We restricted participation to workers that were U.S.-based, had a 95 percent minimum approval rate, and had at least 500 approved human intelligence tasks (HITs). We excluded 16 participants who completed less than one encryption per minute for a final sample of 2,400 workers. We paid all workers a fixed participation fee of \$2 upon task completion. Workers also received an additional payment of up to \$5 based on the decision of a randomly chosen spectator approximately six weeks after completing the task.

\begin{table}[H]\caption{Average spectator characteristics by treatment condition}
	\vspace{-15pt}
     \setlength{\tabcolsep}{0.18em} 
	\label{tab:summ-spectator}
	{\scriptsize
		\begin{centering} 
			\protect
			\begin{tabular}{lC{1.2cm}cC{1.3cm}C{1.85cm}C{1.95cm}cC{1.3cm}C{1.85cm}C{1.95cm}}
				\addlinespace \addlinespace \midrule			
				&     && \multicolumn{3}{c}{Baseline condition} && \multicolumn{3}{c}{Information Treatment} \\\cmidrule{4-6}\cmidrule{8-10} 
				& All && Lucky Outcomes & Lucky Opportunities & Ex-Post Lucky Opportunities && Lucky Outcomes & Lucky Opportunities & Ex-Post Lucky Opportunities \\	
				& (1) && (2)  & (3)   &  (4)     && (5)  & (6)  & (7)  \\
				\midrule 	
				\multicolumn{8}{l}{\textbf{Panel A. Demographic characteristics and race}} \\ 
				Age &49.09 & &49.50 &50.06 &47.28 & &48.68 &49.63 &49.37 \\
Male &0.48 & &0.50 &0.47 &0.48 & &0.48 &0.52 &0.45 \\
Married &0.62 & &0.61 &0.63 &0.68 & &0.57 &0.66 &0.60 \\
Nr. of children under 18 &0.60 & &0.51 &0.57 &0.61 & &0.53 &0.74 &0.62 \\
White &0.86 & &0.82 &0.88 &0.84 & &0.84 &0.88 &0.90 \\
Black &0.08 & &0.12 &0.06 &0.07 & &0.07 &0.07 &0.07 \\
Hispanic &0.08 & &0.09 &0.08 &0.09 & &0.08 &0.08 &0.06 \\
 \midrule
				
				\multicolumn{8}{l}{\textbf{Panel B. Education and employment}} \\ 
				Completed college &0.62 & &0.62 &0.62 &0.65 & &0.60 &0.67 &0.58 \\
Numeracy index &4.09 & &4.00 &4.12 &4.16 & &4.11 &4.15 &4.03 \\
Works full-time &0.58 & &0.62 &0.53 &0.64 & &0.55 &0.56 &0.58 \\
Works part-time &0.15 & &0.15 &0.14 &0.12 & &0.18 &0.14 &0.15 \\
Retired &0.20 & &0.16 &0.22 &0.19 & &0.19 &0.23 &0.22 \\
Houseowner &0.72 & &0.68 &0.77 &0.70 & &0.69 &0.74 &0.77 \\
 \midrule
				
				\multicolumn{8}{l}{\textbf{Panel C. Household Income}} \\ 
				Income below 40k &0.24 & &0.25 &0.22 &0.19 & &0.23 &0.24 &0.27 \\
Income btw. 40k and 75k &0.28 & &0.30 &0.27 &0.28 & &0.28 &0.25 &0.26 \\
Income btw. 75k and 100k &0.16 & &0.17 &0.14 &0.16 & &0.17 &0.15 &0.17 \\
Income over 100k &0.32 & &0.25 &0.36 &0.38 & &0.31 &0.36 &0.28 \\
 \midrule												
				
				\multicolumn{8}{l}{\textbf{Panel D. Region}} \\ 
				Lives in the Midwest &0.23 & &0.30 &0.20 &0.26 & &0.20 &0.20 &0.20 \\
Lives in the Northeast &0.21 & &0.20 &0.20 &0.19 & &0.23 &0.22 &0.23 \\
Lives in the South &0.35 & &0.32 &0.30 &0.36 & &0.31 &0.36 &0.43 \\
Lives in the West &0.22 & &0.18 &0.30 &0.20 & &0.25 &0.22 &0.15 \\
 \midrule
				
				Used a mobile device &0.37 & &0.38 &0.39 &0.35 & &0.42 &0.31 &0.38 \\
Minutes spent in experiment &14.72 & &14.67 &15.06 &13.62 & &14.83 &15.37 &15.09 \\
Passed attention check &0.89 & &0.86 &0.85 &0.92 & &0.93 &0.91 &0.88 \\\midrule
\(N\) (Spectators) &1,170 & &197 &194 &193 & &197 &196 &193 \\
\(N\) (Redistributive decisions) &14,040 & &2,364 &2,328 &2,316 & &2,364 &2,352 &2,316 \\
 \midrule				
			\end{tabular}
			\par\end{centering}
		\singlespacing\justify\footnotesize
        \vspace{-7pt}
	\textbf{Notes}: This table shows the demographic composition of our spectator sample, comparing spectators treated with and without information about the likelihood that the best performer is the winner, between lucky outcomes (columns 2 and 5), lucky opportunities (columns 3 and 6), and ex-post lucky opportunities (columns 4 and 7) conditions.
		
	}
\end{table}

We recruited 1,170 panelists from the Federal Reserve Bank of New York's \textit{Survey of Consumer Expectations} (SCE) to participate in the spectator task in October and November 2021. This survey targets a non-convenience, nationally representative panel of U.S. heads of households \citep{armantier2017overview}.\footnote{The SCE recruitment process works as follows. Prospective panelists receive an initial in-take survey that is distributed via mail to a random sample of US addresses, which contains a \$1 bill and offers an additional \$5 payment for completing and returning the paper-based survey that collects detailed demographic information and some initial economic expectations. Respondents are informed that they may be eligible to participate in additional surveys and asked to provide their email address if they would like to participate. Stratified random sampling is then used to selectively invite respondents to join the core SCE panel to achieve a nationally representative sample of US household heads in terms of income, gender, age, race/ethnicity, and Census division. Respondents participate in the core SCE survey for up to 12 months and receive a \$15 payment for each survey. They can also receive additional payments for completing special surveys, such as the one we fielded for this study. See Section 3 in \cite{armantier2017overview} for additional details about the SCE intake and recruitment process.} We conducted two waves of recruitment over two and a half weeks to reach our target sample size. Our recruitment sample includes all recent panelists who have taken part in SCE over the past three years but excludes those who are still actively participating in the SCE core module. We targeted a representative sample of US households based on the same criteria used for entry into the core SCE sample. We invited a total of 3,500 panelists to participate in the experiment over the two waves and had an average response rate of 33.4\%.

The experimental interface was mobile-friendly to encourage hard-to-reach demographic groups to participate in our experiment. Given the high-quality nature of the SCE sample, we pre-registered to not impose any exclusion criteria for our spectator sample. Our in-sample data-quality metrics support this decision: The median spectator spent 15 minutes on the survey, 89 percent passed the attention check, and 77 percent passed all four comprehension questions on their first attempt. No spectator failed to answer more than two comprehension questions. We paid all respondents a \$5 Amazon gift card for completing the survey. 

Table \ref{tab:summ-spectator} reports descriptive statistics for the spectator sample.\footnote{See Table \ref{tab:summ-wrk} in the Appendix for summary statistics on our sample of workers. Panel A of Appendix Figure \ref{fig:worker-eff} plots the distribution of worker performance.} The average spectator is 49 years old; 52 percent are female, 62 percent are married, and 14 percent are non-white. More than 62 percent of spectators attained a college degree, 58 percent work full-time, 15 percent work part-time, and 20 percent are retired. About a quarter (24 percent) of spectators have a household income below \$40,000 per year and about a third (32 percent) more than \$100,000 per year. Our sample includes individuals living in all 50 states plus Washington, DC. About 23 percent of spectators live in the Midwest, 21 percent in the Northeast, 35 percent in the South, and 22 percent in the West. Columns 2--7 show that spectator characteristics are similar across conditions. 

\section{Main Results} \label{sec:redistribution}

The primary outcome we examine is the fraction of earnings, $r_{ip}$, that spectator $i$ redistributes from the winner to the loser in worker pair $p$. We refer to the worker who initially receives the total earnings as the ``winner'' and the worker who initially receives no earnings as the ``loser.'' When $r_{ip}=0$, the loser gets none of the total earnings, and the winner retains all the earnings. Both workers receive half of the total earnings when $r_{ip}=0.5$. The primary outcomes, hypotheses, and empirical specifications described in this section were pre-registered in our pre-analysis plan. We note any deviations in the main text.

\subsection{Redistribution under Lucky Opportunities and Lucky Outcomes}\label{sub:results}

Spectators redistribute less income when luck is manifested through unequal opportunities. Panel A of Table \ref{tab:redist-pi} reports the average level of redistribution across our luck treatments. Under lucky outcomes, spectators redistributed 27.6 percent of earnings from the winner to the loser on average. In contrast, spectators redistributed only 23.4 percent of earnings on average under lucky opportunities. Thus, spectators redistributed 4.2 percentage points less of total income when luck was experienced through unequal opportunities ($p < 0.01$, column 3). This difference equates to a 15.3 percent decrease in the final earnings for the losing worker. If we restrict our analysis to observations in which spectators face uncertainty about whether the inequality is merit-based, we find that spectators redistributed 5 percentage points ($p < 0.01$) less of total income under lucky opportunities, implying 17.3 percent lower earnings for the losing worker.\footnote{We refer to situations in which $\pi=1$ as situations without uncertainty because the information available to spectators in those situations allows them to conclude without a doubt that the winner of a worker pair is the best performer of the pair. For example, this is the case when the coin flip probability is zero in the lucky outcomes condition or when both workers have the same multiplier in the lucky opportunities condition. Upon excluding these observations, average redistribution equals 28.9 percent in the lucky outcomes environment versus 23.9 percent in the lucky opportunities environment.}

\begin{figure}[H]\caption{Redistribution and the probability that the best performer lost ($1-\pi_b$)} \label{fig:redist-pi}
	\centering
	\vspace{-7pt}\includegraphics[width=.75\linewidth]{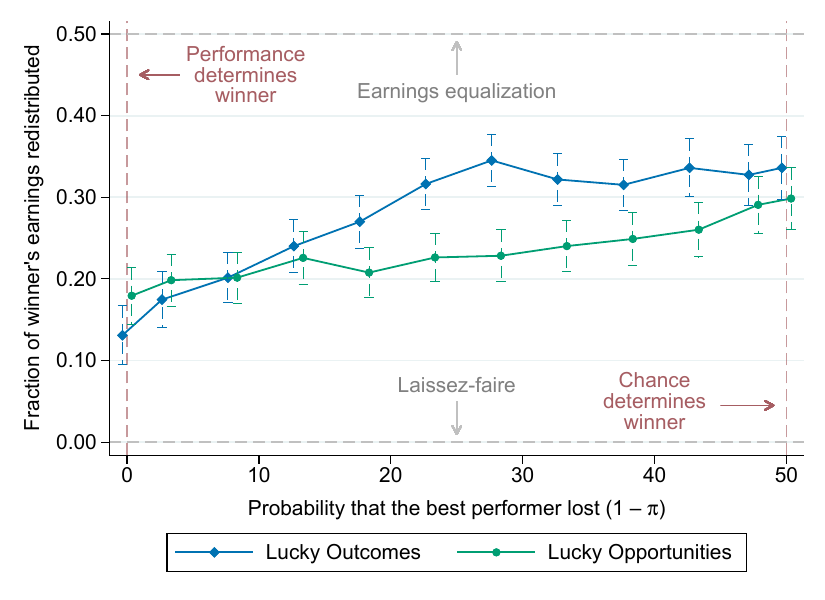}
	\vspace{-15pt}
	\footnotesize
	\singlespacing \justify \footnotesize
	\textbf{Notes:} This figure shows the average share of earnings redistributed between workers (from the higher-earning winner to the lower-earning loser) relative to the likelihood that the best performer lost ($1-\pi$). The two series represent our main experimental conditions: lucky outcomes (blue) and lucky opportunities (green). The vertical dashed lines denote 95\% confidence intervals.
\end{figure}

Next, we examine how redistribution responds to the importance of luck in determining outcomes. Figure~\ref{fig:redist-pi} plots mean redistribution in each luck treatment against the probability that the best performer lost: $1-\pi$. Intuitively, $1-\pi$ can be thought of as the probability that initial earnings inequality was due to luck or, equivalently, the probability that earnings inequality is not merit-based. We depict $1-\pi$ on the x-axis, rather than $\pi$, because this makes it easier to discuss the effect of increasing the impact of luck, which facilitates the exposition of our results. The intercept in Figure~\ref{fig:redist-pi} shows mean redistribution across worker matches when the best performer was always the winner--- because workers drew equal multipliers or because there was zero probability of earnings being determined by a coin flip. The slope shows how redistribution changes when there is an increase in the probability that initial earnings inequality was due to luck.

We first examine worker pairs where performance alone determined the winner, eliminating any uncertainty about whether initial earnings inequality stemmed from worker performance or luck. In these merit-based outcomes, spectators behave similarly across luck environments.\footnote{Notably, redistribution is slightly \textit{higher} under lucky opportunities in these merit-based outcomes, contrary to the overall redistribution gap documented in Panel A of Table \ref{tab:redist-pi}. This pattern may partly reflect spectator confusion in cases where the winner had the \textit{lower} multiplier. While winners had higher multipliers in 11 of 12 decisions, for the merit-based outcome case ($1-\pi=0$), the winning worker had either a lower or equal multiplier. When the winning worker had the lower multiplier, some spectators may not have noticed this difference from their other decisions. Indeed, when the multipliers were exactly equal, the average redistribution was 10.7 percent, compared to 19.2 percent when the winner had a lower multiplier. Although the confidence intervals are relatively large when we analyze these sub-samples separately, this suggests that some spectators may have erroneously attributed the lower multiplier to the loser for the pure merit decision.}

We next examine how redistribution responds to an increase in the probability that the best performer lost. Under lucky outcomes, spectators' redistribution decisions initially react strongly to this probability. As the probability that the best performer lost increases from 0 to about $25$--$30$ percent, the share of total income that spectators redistribute increases from 13.1 to 34.5 percent. Beyond that point, redistribution does not respond to further increases in the probability that the best performer lost. In contrast, spectators' reactions are more muted when there are lucky opportunities. As the probability that the best performer lost increases from 0 to the range $25$--$30$ percent, average redistribution increases from 17.9 to 22.8 percent. However, spectators continue to respond to the increased role that luck plays in determining outcomes beyond that point, increasing the amount they redistribute in an approximately linear manner. 

To summarize the relationship between average redistribution and the likelihood that the best performer in match $p$ lost in each luck environment, we estimate linear models that relate the share of earnings redistributed to $1 - \pi_{ip}$: 
\begin{align}\label{eq:pi-lin}
	r_{ip} = \alpha + \beta (1 - \pi_{ip}) + \varepsilon_{ip},
\end{align}
where $\varepsilon_{ip}$ is an error term. The main parameter of interest is $\beta = \partial\E(r_{ip})/\partial\E(1-\pi_{ip})$, which measures the elasticity of redistribution with respect to $1-\pi_{ip}$. The exogenous within-subject variation in $\pi_{ip}$ allows us to identify $\beta$. 

Spectators respond less to changes in the probability that the outcome was luck-based when luck stems from unequal opportunities instead of coin flips. Panel B of Table \ref{tab:redist-pi} presents estimates of $\beta$ across the different luck environments. A ten percentage point increase in $1-\pi_{ip}$ leads to a $\hat{\beta} = 3.7$ percentage point increase in the share of earnings redistributed in the lucky outcomes condition (column 1). In contrast, redistribution is less elastic to changes in $1-\pi_{ip}$ in the lucky opportunities condition: A ten percentage point increase in $1-\pi_{ip}$ leads to a $\hat{\beta} = 2.0$ percentage point increase in redistribution when luck emerges through unequal opportunities (column 2). The difference between estimated elasticities is statistically significant ($p < 0.01$, column 3).\footnote{These results remain robust when we exclude the two extreme cases $1-\pi_{ip}=0$ and $1-\pi_{ip}=0.50$. When excluding all observations that correspond to these extreme cases from the analysis, we find that a ten percentage point increase in $1-\pi_{ip}$ leads to a 3.5 percentage point increase in redistribution under lucky outcomes and a 1.8 percentage point increase under lucky opportunities, maintaining the 1.7 percentage point gap in responsiveness between the two environments.}

This differential sensitivity to changes in $1-\pi_{ip}$ leads to a substantial redistribution gap across our experimental treatments. Column (3) of panel C in Table \ref{tab:redist-pi} presents the average difference in redistribution for each $\pi$ bin, along with the estimated standard error.\footnote{Formally, in Panel C of Table \ref{tab:redist-pi}, we estimate regressions of the form: 
\begin{align} \label{eq:pi-bins}
	r_{ib} = \sum_{b=1}^{12} \gamma_{b} (1-\pi_{b})  + \varepsilon_{ib},
\end{align}
where $1-\pi_{b}$ is an indicator that equals one if $1-\pi_{ip}$ is in bin $b$. We estimate equation \eqref{eq:pi-bins} separately for each treatment and interact the bins with treatment dummies to assess formally whether mean redistribution is the same across luck treatments at a given $\pi$ bin. We cluster standard errors at the spectator level in all specifications.} We observe a significant difference in redistribution for six out of the ten $\pi$ bins in which spectators face inferential challenges when trying to determine whether the inequality was merit-based ($p$ < 0.05).\footnote{When making decisions for two of the twelve $\pi$ bins (i.e., in the two extreme cases: $1-\pi_{ip}=0$ and $1-\pi_{ip}=0.50$), spectators trying to assess whether merit could have determined the winner do not face the same inferential challenges as in the other ten decisions. This is because $1-\pi_{ip}=0$ corresponds to cases in which the probability of a coin flip is zero (in lucky outcomes) or the winner does not have any relative advantage (in lucky opportunities) so that the winner is for sure the better performer. Similarly, $1-\pi_{ip}=0.5$ corresponds to cases in which winning is independent of performance because the coin flip probability is $100$\% (in lucky outcomes) or the winning worker's relative advantage was too large to be overcome (in lucky opportunities).}

\begin{table}[h!]
\caption{Fraction redistributed as a function of the probability that the best performer lost ($1-\pi$)} \label{tab:redist-pi}
\vspace{-15pt}
	{\footnotesize 
		\begin{center}
			\newcommand\w{2}
			\begin{tabular}{l@{}lR{\w cm}@{}L{0.43cm}R{\w cm}@{}L{0.43cm}R{\w cm}@{}L{0.43cm}R{\w cm}@{}L{0.43cm}R{\w cm}@{}L{0.43cm}R{\w cm}@{}L{0.43cm}}
				\midrule
				&& \multicolumn{6}{c}{Outcome: Fraction of earnings redistributed}  \\
				\cmidrule{3-8} 
				& & Lucky & &  Lucky & & \multirow{2}{*}{Difference} & \\
				& & Outcomes & & Opportunities & &  & \\
				&& (1) && (2) && (3)  \\
				\midrule 
				\multicolumn{8}{l}{\hspace{-1em} \textbf{Panel A. Average redistribution}}  \\  \addlinespace 
				Constant&&0.276&\sym{***}&0.234&\sym{***}&0.042&\sym{**}\\
&&(0.010&)&(0.013&)&(0.016&)\\
 
				\midrule 
				\multicolumn{8}{l}{\hspace{-1em} \textbf{Panel B. Elasticity of redistribution}}  \\  \addlinespace 
				Prob. best performer lost (1$-$$\pi$)&&0.037&\sym{***}&0.020&\sym{***}&0.017&\sym{**}\\
&&(0.005&)&(0.004&)&(0.006&)\\
Constant&&0.331&\sym{***}&0.263&\sym{***}&0.068&\sym{***}\\
&&(0.013&)&(0.014&)&(0.019&)\\
  \midrule
				\multicolumn{8}{l}{\hspace{-1em} \textbf{Panel C. Average redistribution across $\pi$ bins}}  \\  \addlinespace 
				1$-$$\pi=0.00$&&0.131&\sym{***}&0.179&\sym{***}&$-$0.048&\\
&&(0.018&)&(0.018&)&(0.025&)\\
1$-$$\pi\in(0.00,0.05]$&&0.175&\sym{***}&0.198&\sym{***}&$-$0.024&\\
&&(0.018&)&(0.016&)&(0.024&)\\
1$-$$\pi\in(0.05,0.10]$&&0.202&\sym{***}&0.202&\sym{***}&$-$0.000&\\
&&(0.016&)&(0.016&)&(0.022&)\\
1$-$$\pi\in(0.10,0.15]$&&0.240&\sym{***}&0.226&\sym{***}&0.014&\\
&&(0.017&)&(0.017&)&(0.023&)\\
1$-$$\pi\in(0.15,0.20]$&&0.270&\sym{***}&0.208&\sym{***}&0.062&\sym{**}\\
&&(0.017&)&(0.015&)&(0.023&)\\
1$-$$\pi\in(0.20,0.25]$&&0.316&\sym{***}&0.226&\sym{***}&0.090&\sym{***}\\
&&(0.016&)&(0.015&)&(0.022&)\\
1$-$$\pi\in(0.25,0.30]$&&0.345&\sym{***}&0.228&\sym{***}&0.117&\sym{***}\\
&&(0.017&)&(0.016&)&(0.023&)\\
1$-$$\pi\in(0.30,0.35]$&&0.322&\sym{***}&0.240&\sym{***}&0.082&\sym{***}\\
&&(0.016&)&(0.016&)&(0.023&)\\
1$-$$\pi\in(0.35,0.40]$&&0.315&\sym{***}&0.249&\sym{***}&0.066&\sym{**}\\
&&(0.016&)&(0.016&)&(0.023&)\\
1$-$$\pi\in(0.40,0.45]$&&0.336&\sym{***}&0.260&\sym{***}&0.076&\sym{**}\\
&&(0.018&)&(0.017&)&(0.025&)\\
1$-$$\pi\in(0.45,0.50)$&&0.327&\sym{***}&0.291&\sym{***}&0.037&\\
&&(0.019&)&(0.018&)&(0.026&)\\
1$-$$\pi=0.50$&&0.336&\sym{***}&0.298&\sym{***}&0.038&\\
&&(0.020&)&(0.019&)&(0.028&)\\
  \midrule
                \(N\) (Redistributive decisions)&&2,364&&2,328&&4,692&\\

			\end{tabular}
		\end{center}
		\begin{singlespace}  \vspace{-.5cm}
			\noindent \justify \footnotesize \textbf{Notes:} This table shows estimates of redistribution under lucky outcomes (column~1), lucky opportunities (column~2), and the difference (column~3). Panel A shows the mean share of earnings redistributed. Panel B  shows the elasticity of the fraction of earnings redistributed with respect to the likelihood that the winning worker performed better than the losing worker. Panel C shows the relationship between redistribution and the likelihood that the winning worker performed better split into 12 bins. Heteroskedasticity-robust standard errors clustered at the spectator level in parentheses. $^{***}$, $^{**}$ and $^*$ denote significance at the 0.1\%, 1\%, and 5\% level, respectively.
		\end{singlespace} 	
	}
\end{table}

\subsection{Margins of Redistribution and Heterogeneity}

Differences in redistribution across luck environments can be decomposed into differences in the extensive margin (whether anything is redistributed) and the intensive margin (how much is redistributed).\footnote{This analysis of the intensive and extension margins of redistribution was not included in our preregistration and should therefore be considered exploratory in nature.} In Appendix \ref{sec:extensive_intensive}, we investigate the role of the extensive and intensive margins in accounting for the redistribution gap across luck environments. On the extensive margin, we find that a higher fraction of spectators do not redistribute any earnings under lucky opportunities than under lucky outcomes ($p$ < 0.01). On the intensive margin, we still find that redistribution is less responsive to the likelihood that the best performer lost under lucky opportunities than lucky outcomes ($p$ < 0.01). Thus, both the extensive and the intensive margin contribute to the redistribution gap.

We also find that redistribution behavior varies by respondents' gender, income, and political views. Appendix Table \ref{tab:redist-pi-het} presents the results from a heterogeneity analysis. Females tend to favor redistribution more than males across both luck environments. Respondents in households with annual incomes below \$100,000 redistribute more on average across both luck environments. This finding is notable in light of recent cross-country evidence by \cite{marechal2025whose} suggesting that actual redistribution policies are more strongly correlated with the preferences of lower socioeconomic groups.  Finally, Republicans tend to support redistribution less than others. These patterns are consistent with previous evidence and extend to the domain of unequal opportunities \citep{alesina2011preferences, ashok2015support, alesina2018intergenerational, almaas2020cutthroat,cohn2023wealthy}.

\subsection{The Impact of Observable Features on Redistribution}\label{sec:linear}

To understand why spectators are less responsive to changes in the probability of the best performer losing under lucky opportunities compared to lucky outcomes, we conduct a two-step analysis. First, we examine how spectators incorporate observable features of the luck environment into their redistribution decisions (Section \ref{sec:unequal}). Second, we analyze the relationship between these features and the probability that the best performer lost (Section \ref{sec:signals}).

\subsubsection{Spectator Responses to Unequal Opportunities.} \label{sec:unequal} We begin by analyzing how spectators incorporate unequal opportunities into their redistribution decisions. We create two intuitive measures of unequal opportunities: (1) the arithmetic difference between the productivity multipliers of the winner and loser ($m_1 - m_2$), and (2) the ratio of these multipliers ($m_1/m_2$). We then evaluate which measure better predicts spectators' redistribution choices by regressing the fraction of earnings redistributed on each measure (Table \ref{tab:redist-determ}). We find that when either the multiplier difference or ratio is included in the regression by itself, each has a significant impact on redistribution (columns 1 and 2). However, when both measures are included simultaneously in the regression model, only the multiplier difference continues to have a significant effect (column 3).\footnote{The coefficient on the multiplier difference in column 3 is not statistically different from the corresponding estimate in column 1 ($p$ = 0.42). However, we can reject that the coefficient on the multiplier ratio is the same in columns 2 and 3 ($p$ < 0.01).}

These results suggest that spectators primarily rely on the multiplier difference (and not the ratio) when making redistribution decisions in the lucky opportunities environment. For example, spectators treat multipliers $(m_1,m_2)=(2,1)$ the same as $(m_1,m_2)=(3,2)$ when making redistribution decisions, when in fact it is more likely that outcomes were luck-based in the former; specifically, $1-\pi(2/1)=0.46$ compared to $1-\pi(3/2)=0.38$. A worker competing against another with a multiplier of one would only require a multiplier of 1.5 to have an equivalent advantage to a worker with a multiplier of three competing against one with a multiplier of two. 

 \begin{table}[H]\caption{The impact of unequal opportunities on redistribution decisions} \label{tab:redist-determ}
	\vspace{-15pt}
	 	{\footnotesize
		
		 		\begin{center}
			 			\newcommand\w{2}
			 			\begin{tabular}{l@{}lR{\w cm}@{}L{0.43cm}R{\w cm}@{}L{0.43cm}R{\w cm}@{}L{0.43cm}R{\w cm}@{}L{0.43cm}R{\w cm}@{}L{0.43cm}R{\w cm}@{}L{0.43cm}R{\w cm}@{}L{0.43cm}}
				 				\midrule
				 				&& \multicolumn{6}{c}{Outcome: Fraction of earnings redistributed}  \\
				 				\cmidrule{3-8} 
				 				&& (1) && (2) && (3)  \\ \midrule  
				
				 				\addlinespace
				 				Multiplier difference&&0.041&\sym{***}&&&0.051&\sym{***}\\
&&(0.007&)&&&(0.015&)\\
Multiplier ratio&&&&0.042&\sym{***}&$-$0.012&\\
&&&&(0.008&)&(0.015&)\\
Constant&&0.202&\sym{***}&0.171&\sym{***}&0.213&\sym{***}\\
&&(0.014&)&(0.017&)&(0.019&)\\
\(N\) (Redistributive decisions)&&2,328&&2,328&&2,328&\\
  \midrule
				 			\end{tabular}
			 		\end{center}
		 		\begin{singlespace}  \vspace{-.5cm}
			
			 			\noindent \justify \textbf{Notes:} This table shows the average redistribution (from the winner's earnings to the loser) as a function of the difference between the multiplier of the winner and the loser ($m_1 - m_2$) and the ratio between the multiplier of the winner and the multiplier of the loser ($m_1/m_2$) in the lucky opportunities environment. Heteroskedasticity-robust standard errors clustered at the spectator level in parentheses. $^{***}$, $^{**}$ and $^*$ denote significance at the 0.1\%, 1\%, and 5\% level, respectively. 
			 		\end{singlespace} 	
		 	}
	 \end{table}

\begin{figure}[H]\caption{Relationship between the redistribution and the features of the luck environment observed by spectators} \label{fig:red_signals}
	{\scriptsize
		\begin{centering}	
		\protect
			\begin{minipage}{.48\textwidth}
				\captionof*{figure}{Panel A. Average redistribution vs. Coin-flip probability}				\includegraphics[width=\linewidth]{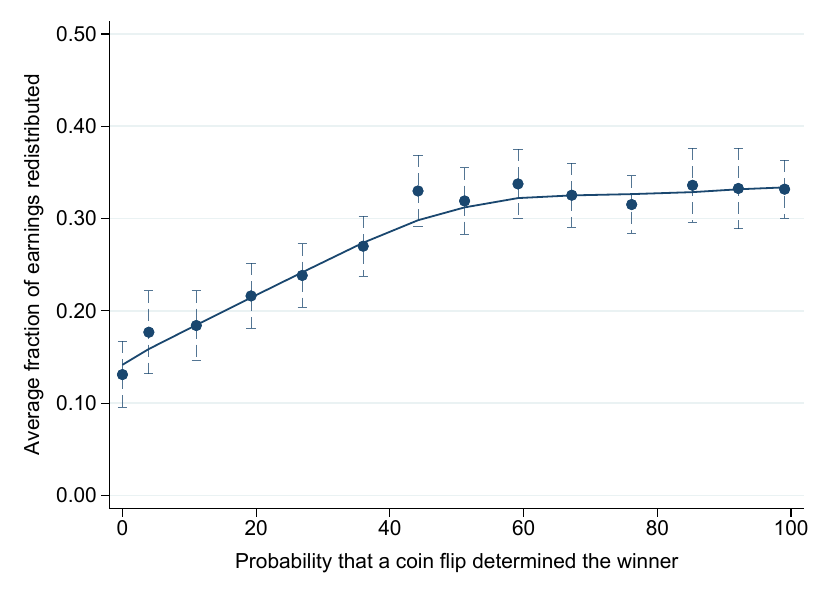}
			\end{minipage}\hspace{1em}
			\begin{minipage}{.48\textwidth}
				\captionof*{figure}{Panel B. Average redistribution vs. multiplier differences}				\includegraphics[width=\linewidth]{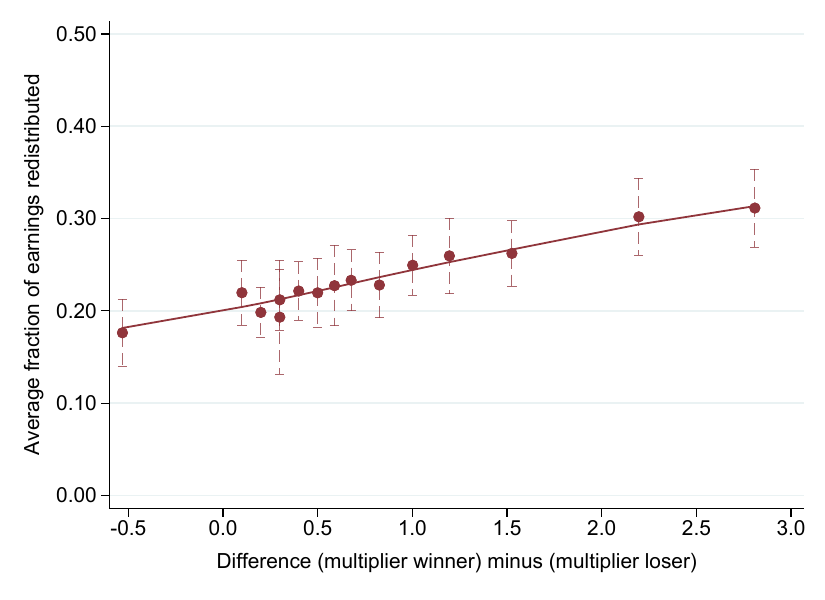}
			\end{minipage}
			\par\end{centering}
		\singlespacing \justify \footnotesize
		\textbf{Notes}: This figure shows the average earnings redistributed from the winner to the loser ($y$-axis) as a function of the probability that a coin flip determined the winner (Panel A) or the difference between the productivity multiplier of the winner and the loser (Panel B), both divided into 15 equal-sized bins. Solid lines plot the predicted values from local polynomial estimates. Dashed vertical lines show 95\% confidence intervals, 

	}
\end{figure}

Figure \ref{fig:red_signals} provides visual evidence of how spectators use the features from the environment in each luck condition to inform their redistribution decisions. The figure presents a series of binned scatterplots plotting average redistribution ($y$-axis) against the information observed by spectators ($x$-axis). Under lucky outcomes (Panel A), the relationship between redistribution and the coin-flip probability is piecewise linear. For low coin-flip probabilities, there is a positive and linear relationship, but for sufficiently large coin-flip probabilities, redistribution is unresponsive to further changes in the coin-flip probability. Under lucky opportunities (Panel B), the relationship between redistribution and the multiplier difference is linear across the entire range of multiplier differences.\footnote{In Appendix Table \ref{tab:redist-poly}, we regress the fraction of earnings redistributed on higher-order polynomials of the multiplier difference and find that none of these terms have a significant effect on redistribution. This provides further evidence suggesting that spectators use the \textit{linear} multiplier difference to inform their redistribution decisions.} On average, a one-unit increase in the difference between the high and low performers' multipliers leads to a four percentage point increase in the share of earnings redistributed.

\subsubsection{Features of the Environment and Source of Inequality.} \label{sec:signals}

\begin{figure}[H]\caption{Relationship between the probability that the best performer lost and features of the luck environment observed by spectators} \label{fig:signals}
	{\scriptsize
		\begin{centering}	
		\protect
			\begin{minipage}{.48\textwidth}
				\captionof*{figure}{Panel A. Likelihood that the best performer lost vs. coin-flip probability}
				\includegraphics[width=\linewidth]{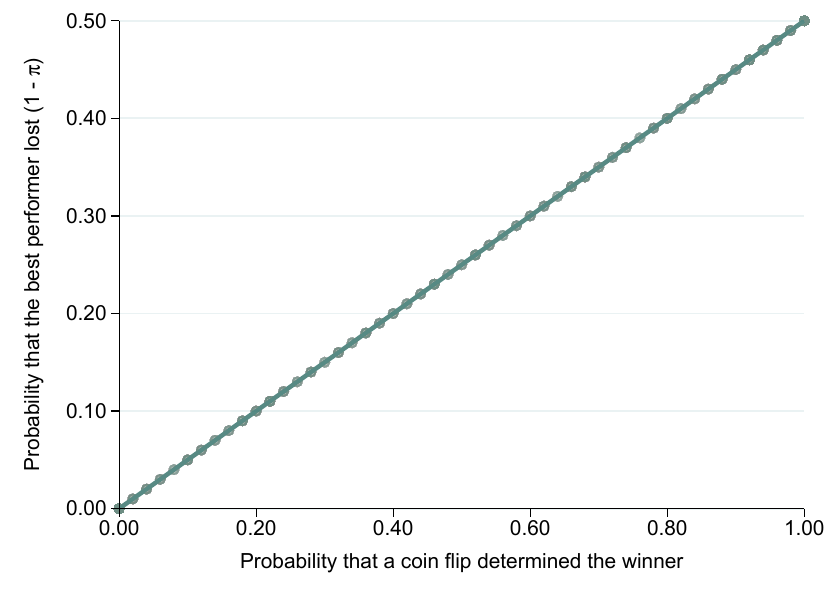}
			\end{minipage}\hspace{1em}
			\begin{minipage}{.48\textwidth}
				\captionof*{figure}{Panel B. Likelihood that the best performer lost vs. multiplier difference}
                \includegraphics[width=\linewidth]{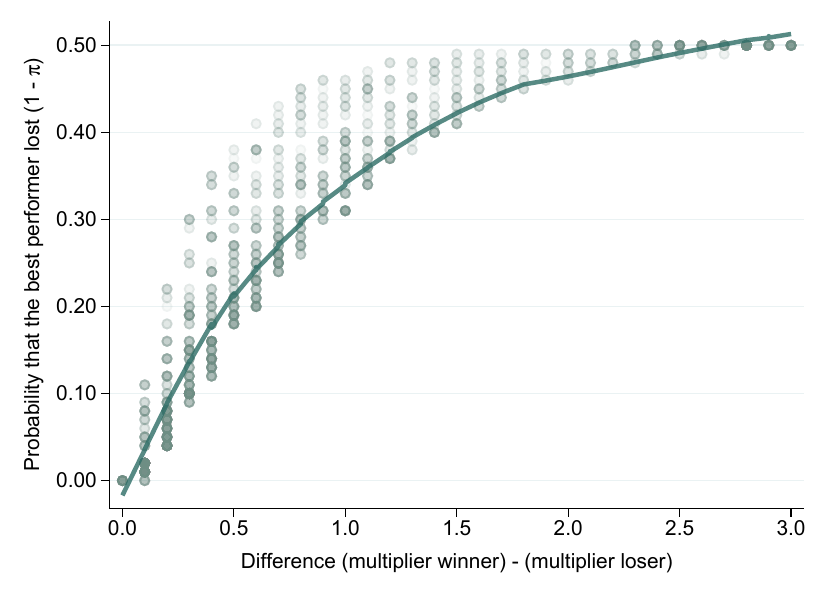}
			\end{minipage}
			\par\end{centering}
		\singlespacing \justify \footnotesize
		\textbf{Notes}:  This figure shows the probability that the best performer lost ($y$-axis) as a function of the probability that a coin flip determined the winner (Panel A) or the difference between the productivity multiplier of the winner and the loser (Panel B). The dots denote the empirical frequencies in our experimental data and the intensity of the color indicates the number of observations at a given point. The solid lines depict the best fit using either linear specification (panel A) or lowess regression (panel B).}
\end{figure}

Next, we study the empirical relationship between the observable features of each luck environment and the probability that the inequality between workers is the result of luck. Under lucky outcomes, there is a linear relationship between the coin-flip probability and the likelihood that the best performer lost (Figure \ref{fig:signals}, Panel A). This relationship starts at the origin, as the best performer never loses when the coin-flip probability is zero, and increases linearly with the coin-flip probability. 

In contrast, under lucky opportunities, there is a concave relationship between the multiplier difference and the likelihood that the best performer lost (Figure \ref{fig:signals}, Panel B). When multipliers are equal, the best performer never loses. However, even a small increase in the multiplier difference leads to a steep rise in the probability of the best performer losing. For example, increasing the multiplier difference from 0 to 0.5 increases the likelihood of the best performer losing from 0 to about 20 percent. As the multiplier difference becomes sufficiently large, the best performer lost whenever they did not receive the high multiplier.

Taken together, the combination of how redistribution responds to the environmental features and how these features relate to the probability that the best performer lost explains why spectators underreact under unequal opportunities. Under lucky outcomes, the coin-flip probability and the likelihood that the best performer lost are linearly related (Figure \ref{fig:signals}, Panel A). As a result, the relationship between redistribution and the coin-flip probability (Figure \ref{fig:red_signals}, Panel A) exactly mirrors the relationship between redistribution and the probability that the best performer lost (Figure \ref{fig:redist-pi}, blue line).

In the lucky opportunities condition, spectators' redistribution decisions respond linearly to the multiplier difference across its entire range (Figure \ref{fig:red_signals}, Panel B). However, this linear response fails to account for the concave relationship between the multiplier difference and the probability that the best performer lost (Figure \ref{fig:signals}, Panel B). As a result, spectators appear to ``underreact'' (redistribute less than expected based on the actual likelihood that the best performer lost) for relatively small values of $(1-\pi)$ and ``overreact'' for larger values (Figure \ref{fig:redist-pi}, green line).\footnote{In Online Appendix Section \ref{sec:headstarts}, we show that the same relationship prevails even if lucky opportunities are not multiplicative but additive. More precisely, we consider the possibility that workers receive unequal, additive boosters, capturing the idea that a relative advantage can resemble a head start sometimes. In such an environment, the probability that the inequality is merit-based is still convex in the relative advantage (the difference in boosters), implying that ($1-\pi$) is concave in the booster difference -- as is the case for our design using productivity multipliers.} In Section \ref{sec:mechanisms}, we investigate whether this linear response to multiplier differences persists when we also provide direct information about $\pi$.

\section{The Role of Information and Effort Responses} \label{sec:mechanisms}

We explore two broad categories of mechanisms that may drive the patterns of redistribution that we observe across luck environments. First, we investigate the role of inaccurate beliefs about the probability the best performer lost through an information intervention. Second, we examine the role of actual or perceived differences in worker performance across luck environments in explaining our results. 

\subsection{Information Intervention}\label{sec:biased}

A key question is why spectators are less responsive to changes in the likelihood that the winner was luck-based in lucky opportunities versus lucky outcomes, which leads to substantial gaps in average redistribution across these environments. To examine whether this is due to spectators facing inferential challenges when assessing the impact of unequal opportunities on earnings inequality, we implement an information treatment. In this treatment, in addition to the information spectators receive in the baseline condition, we provide spectators with the information they would otherwise need to infer. Specifically, for each of their 12 redistribution decisions, we tell spectators the likelihood that the best performer won, $\pi$. 

We begin by examining how the information intervention affects the relationship between redistribution and the likelihood that the outcome is luck-based. If inferential challenges are the primary driver of the linear redistribution pattern, then we expect the treatment to significantly alter this relationship, as spectators no longer need to rely on heuristics or potentially inaccurate inference. In contrast, if the linear pattern persists, this would suggest that other factors shape redistributive behavior. Then, we analyze how the information intervention affects spectators' reliance on observable features of the luck environment.

\begin{figure}[h!]\caption{The effect of providing information about $\pi$} \label{fig:redist-pi-byinfo}
 \vspace{-7pt}
	\centering
	\includegraphics[width=.75\linewidth]{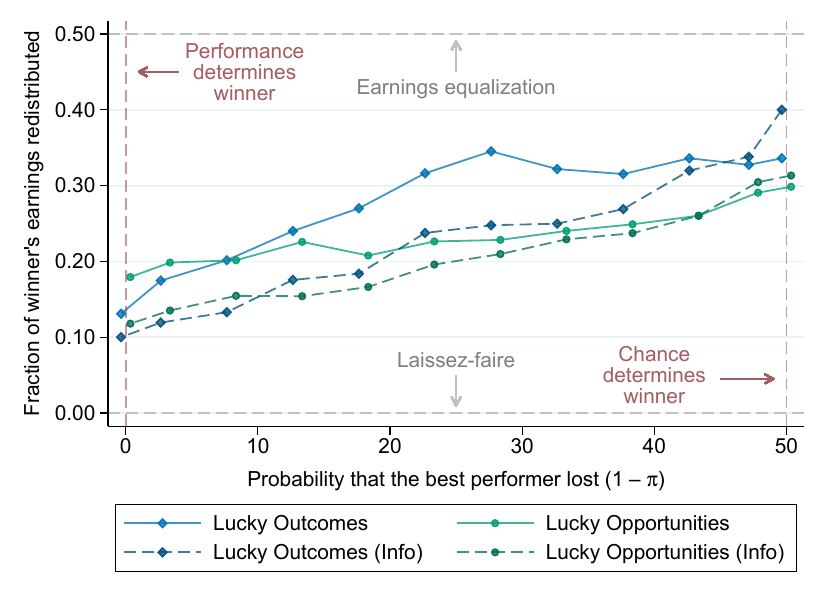}
	\footnotesize
	\singlespacing \justify \footnotesize 
 \vspace{-15pt}
	\textbf{Notes:} This figure shows the average share of earnings redistributed between workers (from the higher-earning winner to the lower-earning loser) as a function of the likelihood that the best performer lost for our two luck environments with and without information about $\pi$. The light blue and green solid lines depict our baseline lucky outcomes and lucky opportunities environments. The dark blue and green dashed lines depict redistribution for the lucky outcomes and lucky opportunities environments with information about $\pi$.  
 \end{figure} 

Providing readily accessible information about $\pi$ made redistribution more responsive to the likelihood that the best performer lost in both luck environments. Figure \ref{fig:redist-pi-byinfo} plots mean redistribution in the baseline and information interventions. In the baseline lucky outcomes condition, a ten percentage point increase in the likelihood that the best performer lost raised redistribution by 3.7 percentage points. In the information-treated group, this effect increases by $1.5$ percentage points ($p<0.05$, Appendix Table \ref{tab:redist-info}, Panel B) to 5.2 percentage points in total. Similarly, in the baseline lucky opportunities condition, a ten percentage point increase in the likelihood that the best performer lost raised redistribution by 2.0 percentage points. In the information-treated group, this effect increases by 1.7 percentage points ($p<0.01$) to 3.7 percentage points in total, implying a 85 percent relative increase in the elasticity of redistribution. In other words, when information about the probability that earnings inequality was due to luck becomes readily accessible, spectators become more sensitive to this probability and adjust their redistribution decisions accordingly.\footnote{The effects of providing information about $\pi$ are robust to again excluding the two extreme situations $1-\pi_{ip}=0$ and $1-\pi_{ip}=0.50$ where spectators face less inferential challenges. Excluding these observations from the analysis of the information treatments, we find that a ten percentage point increase in the likelihood that the best performer lost raises redistribution by 4.8 percentage points under lucky outcomes (versus 3.5 percentage points in the baseline condition) and by 3.6 percentage points under lucky opportunities (versus 1.8 percentage points in the baseline condition).} 

\begin{figure}[H]
\caption{Redistribution gap by $1-\pi$ in the baseline and information treatments} \label{fig:redist-pi-info}
	\centering
\vspace{-7pt}
 \includegraphics[width=.75\linewidth]{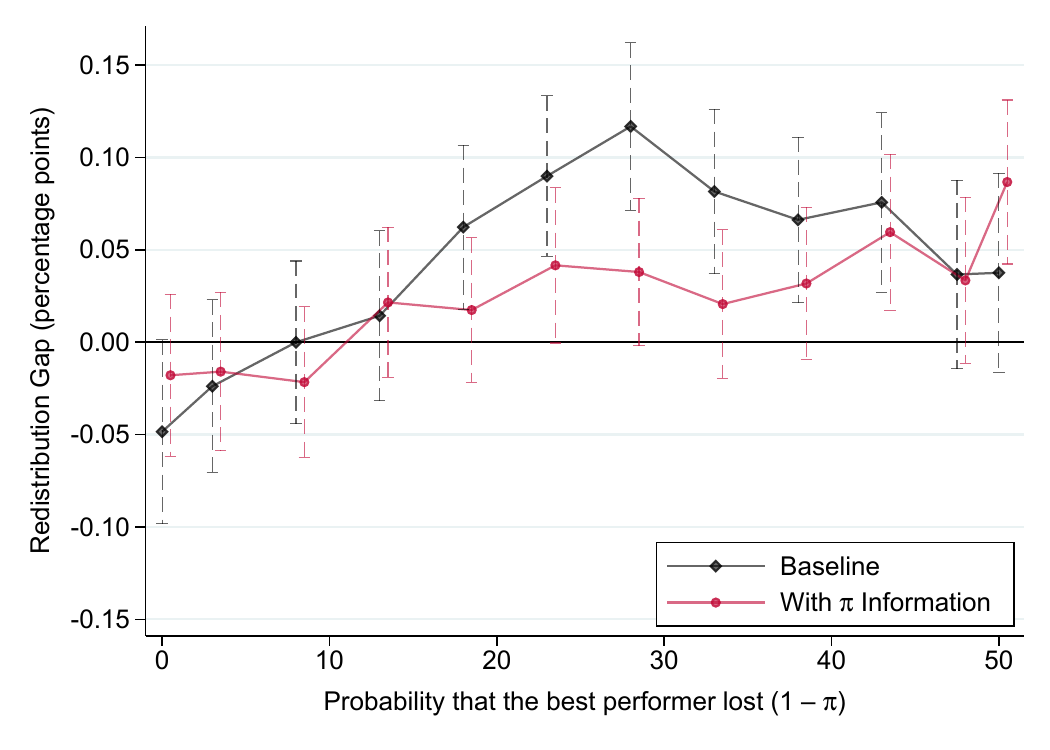}
	\footnotesize
	\singlespacing \justify \footnotesize 
 \vspace{-15pt}
	\textbf{Notes:} This figure shows the gap in the average share of earnings redistributed between workers (from the higher-earning winner to the lower-earning loser) between the lucky outcomes and lucky opportunities environments as a function of the likelihood that the best performer lost. The solid black line depicts the baseline treatments in which spectators do not observe $\pi$. The dashed red line depicts the information treatment in which spectators have direct access to the exact value of $\pi$ when making their redistribution decisions. 
\end{figure}

As a consequence of this increased sensitivity, the information treatment reduced the redistribution gap across luck environments. Figure \ref{fig:redist-pi-info} plots the redistribution gap between the lucky outcomes and lucky opportunities environment in the baseline treatments with no information about $\pi$ (black line) and readily accessible information about $\pi$ (red line). The redistribution gap diminishes substantially when information about $\pi$ is provided, with the most pronounced declines observed over the range $1 - \pi \in [20,40)$. A joint F-test of the null hypothesis that the change in the redistribution gap between lucky opportunities and lucky outcomes across all $\pi$ bins equals zero is rejected at the 1\% significance level (F=2.19, p=0.01). Notably, the gap is small and statistically indistinguishable from zero for ten of the twelve  $\pi$ bins, and is statistically significant at $p < 0.05$ for only two $\pi$ bins (Appendix Table \ref{tab:redist-gap-info}, column 2). This finding is consistent with the interpretation that spectators face inferential challenges when making redistribution decisions. 

\begin{figure}[h!]\caption{Relationship between the redistribution and observable luck features (information intervention)} \label{fig:red_signals_info}
	{\scriptsize
		\begin{centering}	
		\protect
			\begin{minipage}{.48\textwidth}
				\captionof*{figure}{Panel A. Redistribution vs. Coin-flip probability}
    \vspace{-7pt}
		\includegraphics[width=\linewidth]{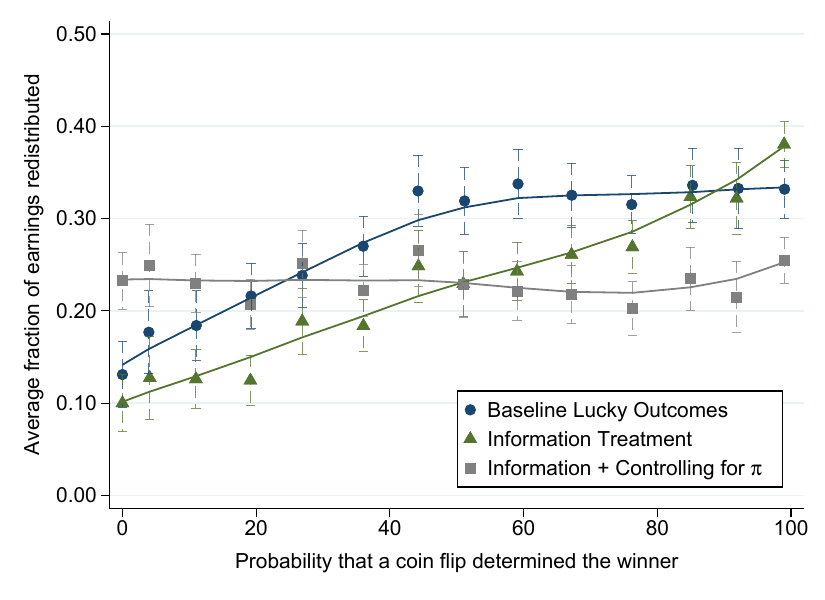}
			\end{minipage}\hspace{1em}
			\begin{minipage}{.48\textwidth}
				\captionof*{figure}{Panel B. Redistribution vs. Multiplier Differences} \vspace{-7pt}
\includegraphics[width=\linewidth]{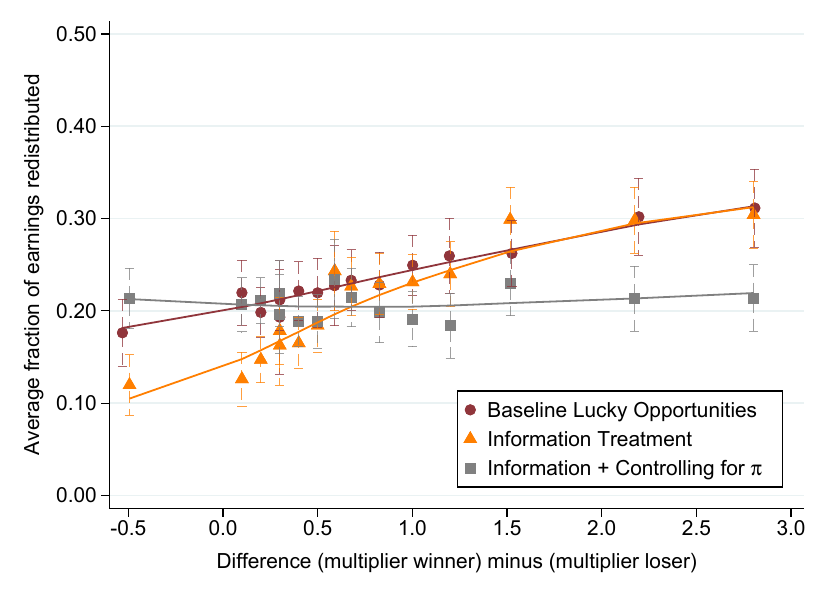}
			\end{minipage}
			\par\end{centering}
   \vspace{-5pt}
		\singlespacing \justify \footnotesize
		\textbf{Notes}: 
		Panel~A shows the average share of winners' earnings redistributed by the probability that a coin flip determined the outcomes. Panel~B shows the average share redistributed by the gap in worker multipliers. Information-treated spectators are provided information about $\pi$ in addition to the observable features of the luck environment.
  
	}
\end{figure}

Next, we investigate how the information intervention affected the spectators' reliance on the other observable features of the two luck environments.\footnote{We did not pre-register this analysis; however, our linearity results from Section \ref{sec:signals} invite further exploration.} Figure \ref{fig:red_signals_info} shows a series of binned scatterplots analogous to those in Figure \ref{fig:red_signals}. Circle markers reproduce the relationship between mean redistribution and the observable features in the baseline luck environments. Triangle markers show this same relationship but under information treatments. Finally, square markers show the relationship between mean redistribution and the observable information, after removing the influence of $\pi$ on redistribution.\footnote{More precisely, we plot the residual of an individual's redistribution after subtracting from it the component explained by the likelihood that the best performer lost. To construct these residuals, we first regress the share of earnings redistributed on the likelihood that the best performer lost (as in equation \eqref{eq:pi-lin}) and estimate the residuals from this regression, adding back the unconditional sample mean to facilitate the interpretation of units.} Solid lines plot the predicted values from local polynomial estimates.

The information treatment altered how spectators incorporated observable features of the luck environment into their redistribution decisions. Figure \ref{fig:red_signals_info} illustrates how the relationship between redistribution and coin-flip probabilities/multiplier differences changed once information on the likelihood that inequality is merit-based became readily available to spectators. In the lucky outcomes environment, the information treatment had two effects: it decreased mean redistribution for most coin-flip probabilities, and it increased the responsiveness of redistribution for relatively high coin-flip probabilities. In the lucky opportunities environment, the information treatment reduced mean redistribution for relatively low multiplier differences. Moreover, it made redistribution more responsive to changes in the multiplier difference for relatively small multipliers, particularly in the range where a small increase in relative multipliers has a substantial impact on worker outcomes.
 
The relationship between mean redistribution and the observable features of the luck environment appears to be mediated by the likelihood that the outcome was merit-based. Controlling for the likelihood that the outcome is merit-based attenuates the relationship between redistribution and the features of the environment observed by spectators (Figure \ref{fig:red_signals_info}, gray lines). Once this likelihood is controlled for, these features have no independent impact on redistribution.\footnote{Appendix Table \ref{tab:redist-determ-pi} provides the corresponding regression estimates and confirms that the multiplier difference has no significant effect on redistribution when controlling for $\pi$ in the information intervention.} This finding further supports the hypothesis that spectators primarily use observable features of the environment as proxies to infer the likelihood of merit-based outcomes.

\subsection{Timing of Luck and Worker Performance}\label{sec:effort}

An important difference between our luck environments is that lucky outcomes occur after completing the task, while unequal opportunities are known before. This aspect of the lucky opportunities condition reflects how luck typically arises in many real-life situations. This difference in the timing of luck could drive the differences in the redistribution decisions that we observe if spectators have different expectations about how workers respond to getting a high or low multiplier.\footnote{What matters for redistribution behavior are spectators' beliefs about worker performance, which we control for using our additional treatments. Empirically, we do not find any evidence that worker performance responds to receiving a high or low multiplier (see Table \ref{tab:worker-eff-m} in the Appendix). We also observe no differences in the overall performance distribution across luck environments (see Appendix Figure \ref{fig:tasks-comp-cf-iop}). A Kolmogorov--Smirnov test for equality of distribution cannot reject the hypothesis that the distribution of worker performance in the lucky outcomes and lucky opportunities environments are equal ($p = 0.909$). 
Notably, other work has found performance responses to differences in opportunities \citep{andre2024shallow}. This difference likely arises because our environment is a winner-takes-all tournament with a fixed working period, while \cite{andre2024shallow} considers differential piece-rate wages and allows workers to choose how long they work. Indeed, \cite{dellavigna2022estimating} find that higher incentives lead to higher output when workers can choose how long they work for but have no effect when there is a fixed working period.}

To examine whether the timing of lucky opportunities affects redistribution, we implemented an additional between-subjects treatment that aligned the timing of when workers learn about their luck across environments. In the \textit{ex-post lucky opportunities} condition, workers learn their multipliers only {after} they complete the task. This is in contrast to our baseline lucky opportunities condition, in which we inform workers of their multipliers before they begin working on the encryption task. However, this timing is in line with when workers learn about the probability that a coin flip determines their earnings in the lucky outcomes environment, namely after completing the task. To ensure this intervention is known to spectators, we provide them a visual timeline of the worker task sequence and require that they pass a comprehension check about whether multipliers are revealed to workers before or after the task.\footnote{The equivalent visual timelines and comprehension checks are provided to all spectator conditions, and not only those spectators in ex-post lucky opportunities. See \href{https://joyzwu.github.io/papers/prsw_supplementary.pdf}{Supplementary Materials}, Figures 5 and 18.}

\begin{figure}[h!]
\caption{Redistribution by $1-\pi$ in the baseline and ex-post lucky opportunities conditions} \label{fig:redist-pi-ea-ep}
	\centering
\vspace{-7pt}
 \includegraphics[width=.75\linewidth]{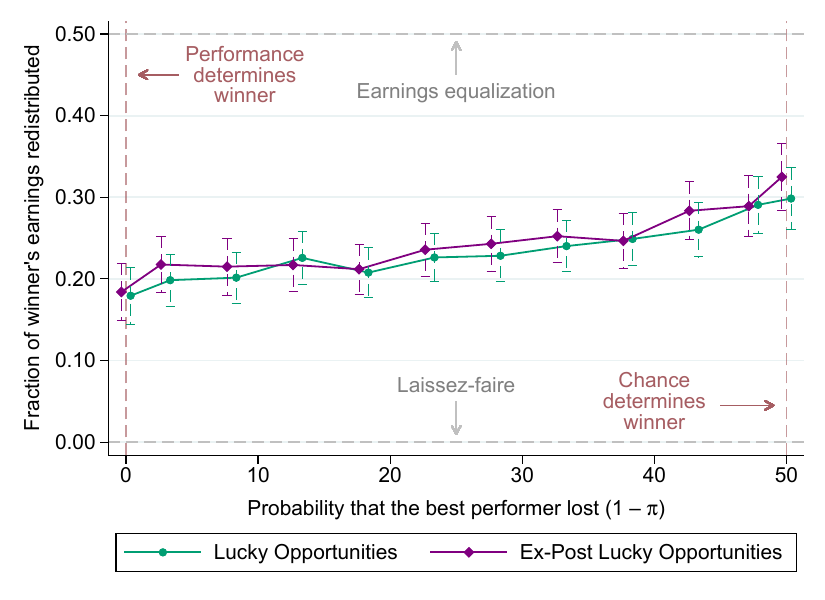}
	\footnotesize
	\singlespacing \justify \footnotesize 
 \vspace{-15pt}
	\textbf{Notes:} This figure shows the average share of earnings redistributed between workers (from the higher-earning winner to the lower-earning loser) as a function of the likelihood that the best performer lost. It depicts two variations of the lucky opportunities condition: Workers in the baseline condition are aware of their multiplier prior to beginning the encryption task, and workers in the ex-post condition learn their multiplier after completing the encryption task.
\end{figure}

We find that whether opportunities are encountered before or after working has no economically or statistically significant impact spectators' redistribution decisions. In Appendix Table \ref{tab:redist-ea-ep}, we re-estimate the primary specifications in Table \ref{tab:redist-pi} but compare redistribution between the baseline and ex-post lucky opportunities. We find no significant differences in the average redistribution: The average amount of income redistributed was 23.4 percent in baseline lucky opportunities versus 24.4 percent in ex-post lucky opportunities ($p = 0.57$). We also find no significant differences in the elasticity of redistribution to changes in luck ($p = 0.89$). Figure \ref{fig:redist-pi-ea-ep} plots our estimates of the average redistribution for both lucky opportunities conditions across each $\pi$ bin. Across the entire range of $\pi$ bins, we find no differences in the level of redistribution.\footnote{Accordingly, there is still a redistribution gap as well as a difference in the elasticity of redistribution between the ex-post lucky opportunities condition and the lucky outcomes condition. For example, if the likelihood that inequality is due to luck increases by ten percentage points, redistribution increases by 1.6 percentage points ($p<0.01$) less under  ex-post lucky opportunities (versus 1.7 percentage points in our main comparison in Section \ref{sub:results}).}

To further assess whether spectators expected differential effort from workers who received a low versus high multiplier, we elicited their stated beliefs about worker performance across the multiplier distribution. To do this we randomly selected a multiplier for each spectator in the baseline lucky opportunities condition and elicited their beliefs about the average number of encryptions completed by workers who received that multiplier. This information enables us to recover the perceived elasticity of worker performance with respect to the productivity multiplier, which we estimate by regressing spectator beliefs about worker performance on the randomly selected multiplier (Appendix Table \ref{tab:beliefs-eff-m}). We find no evidence that spectators expect a significant worker performance response from receiving a high or low multiplier.\footnote{Spectators could also expect that the difference in performance between the winner and loser is not identical across our luck environments. In Appendix \ref{sec:effort_pi}, we show that the effort gap between winners and losers is very similar in the two luck conditions across the entire range of $\pi$ bins. Moreover, we provide evidence that potential differences in the winner-loser effort gap cannot explain the redistribution gap between the two luck environments.}

Finally, it is possible that spectators believed that merely knowing there would be coin flips or multipliers causes workers to put in differential effort across our luck environments, even if the exact coin flip probability or productivity multiplier is revealed to workers only after completing the task. To fully eliminate any scope for differential beliefs about workers' performance responses, we implemented additional subtreatments in which workers did not receive any information about how the winner would be determined \textit{prior} to working on the task. Crucially, spectators in this ``rules-after'' scenario in both the lucky outcomes and ex-post lucky opportunities conditions knew that workers faced identical instructions before completing the task. We analyze the data from these subtreatments in Online Appendix Section \ref{sec:anticipatory-responses}. In Figure \ref{fig:redist-rules}, we show that whether workers learn about the possibility of a coin flip or multipliers affecting the outcome before or after working on the task has no impact on spectators' support for redistribution. In addition, Figure \ref{fig:redist-ep-cf-rules-after} and the estimates in Table \ref{tab:redist-pi-cf-ep-norules} show that there is still a redistribution gap as well as a difference in the elasticity of redistribution between lucky outcomes and lucky opportunities in the rules-after treatment, that is, when workers had identical information before completing the task.

\section{Predicting Political and Social Views}

We document that individuals redistribute less when luck arises through unequal opportunities rather than exogenous sources. But which environment is more predictive of the spectator's political and social views? Our data allows us to examine the external relevance of the experimental measure of redistribution attitudes based on our lucky opportunities environment and compare it to measures based on the lucky outcomes environment used in prior work. To assess this, we estimate the correlation between the average earnings redistributed by spectators in each environment and 14 self-reported political and social views.\footnote{We collected participants' social and political views to conduct validation of our experimental measures and heterogeneity analysis but did not pre-register specific hypotheses. As a result, we consider the analysis in this section as exploratory.}

Redistribution behavior under lucky opportunities correlates more with real-world social and political attitudes than the lucky outcomes environment. We find that mean redistribution in the lucky opportunities condition is more predictive than in lucky outcomes for 13 out of the 14 social and political attitudes we measure (Appendix Table \ref{tab:pred-pol}). To create summary measures of political attitudes, we calculate a $z$-score that averages the 14 self-reported individual attitudes and an index that equals the first component of a principal component analysis. The summary indices reveal that behavior in the lucky opportunities condition is 45 to 62 percent more predictive of attitudes than in the lucky outcomes condition. While the overall explanatory power is modest, with R-squared values ranging from 1.4 to 4.4 percent, the lucky opportunities environment consistently explains nearly three times as much variation in political attitudes as the lucky outcomes environment. This finding suggests that redistribution decisions made in the lab better reflect real-world social and political views when luck arises through unequal opportunities rather than unequal outcomes. 

\section{Discussion}\label{sec:discussion}

Meritocratic fairness ideals contend that individuals are willing to tolerate inequalities due to differences in effort but oppose those arising from chance. But when people face unequal opportunities, this distinction is obfuscated by the fact that luck and effort are intertwined. As a result, individuals may find it difficult to assess the source of inequality and may misattribute success to effort rather than luck. This paper shows that impartial third parties are more tolerant of inequality when luck arises through unequal opportunities, and there is uncertainty about the role of luck in determining outcomes. 

We provide evidence that a key mechanism driving these results is the differential ability to understand the impact of luck across environments, as spectators face greater inferential challenges in mapping features of the luck environment to their consequences for outcomes when luck manifests through unequal opportunities.

Our results offer a potential rationale for the apparent disconnect between the previous experimental literature and observed patterns of U.S. inequality. Research that generates inequality through exogenous variation in outcomes has found that most Americans equalize incomes when income differences are due to luck \citep{almaas2020cutthroat}. However, support for redistribution in the U.S. remained stagnant over a period when differences in opportunities became increasingly important \citep{chetty2014united,ashok2015support}. Consistent with these trends, we show that redistribution is less sensitive to changes in luck when luck arises through unequal opportunities. Similarly, the U.S. remains the most unequal country in the OECD while ranking poorly on equality of opportunity \citep{corak2013income, mitnik2020inequality}. Consistent with these cross-country comparisons, we show that Americans tolerate more inequality when it arises due to differential opportunities.

We conclude by discussing two implications of our results for models that seek to understand and predict attitudes toward redistribution. First, we document that in the absence of readily available information about the likelihood that outcomes are merit-based, spectators rely on simple heuristics when factoring the impact of luck into their redistribution decisions. As a result, people fail to appreciate how small differences in initial circumstances can greatly impact outcomes. Providing information about the importance of luck reduces this reliance on heuristics. Models that seek to accommodate cognitive errors hold some promise for predicting and explaining how beliefs shape redistribution attitudes. 

Second, our results show that readily available information about the likelihood that outcomes are merit-based greatly impacts people's redistribution decisions. This suggests that the information individuals frequently encounter might disproportionately impact their views on inequality and redistribution. For example, popular media coverage (e.g., rags-to-riches stories) may lead individuals to have a greater tolerance for inequality of opportunities, making them less willing to correct this source of unfairness through redistribution. Exploring how salient information shapes individuals' tolerance for inequality is a promising avenue for future research. 

Taken together, our results highlight that redistributive behavior is not invariant to how luck is implemented in the lab. The lucky opportunities environment has several important features that affect redistribution, which the simpler lucky outcomes paradigm cannot capture. We provide a portable, tractable, and rich environment to study income redistribution when there are unequal opportunities that can inform the development of inequality models and the design of optimal redistribution policies.

\clearpage
\begin{spacing}{1}
	\bibliographystyle{apa}
	\bibliography{references}

\begin{thebibliography}{}

\bibitem[\protect\astroncite{Alesina and
  Giuliano}{2011}]{alesina2011preferences}
Alesina, A. and Giuliano, P. (2011).
\newblock Preferences for redistribution.
\newblock In Benhabib, J., Bisin, A., and Jackson, M.~O., editors, {\em
  Handbook of Social Economics}, volume~1, chapter~4, pages 93--131.
  North-Holland.

\bibitem[\protect\astroncite{Alesina
  et~al.}{2018}]{alesina2018intergenerational}
Alesina, A., Stantcheva, S., and Teso, E. (2018).
\newblock Intergenerational mobility and preferences for redistribution.
\newblock {\em American Economic Review}, 108(2):521--54.

\bibitem[\protect\astroncite{Alm\r{a}s et~al.}{2020}]{almaas2020cutthroat}
Alm\r{a}s, I., Cappelen, A.~W., and Tungodden, B. (2020).
\newblock Cutthroat capitalism versus cuddly socialism: Are americans more
  meritocratic and efficiency-seeking than scandinavians?
\newblock {\em Journal of Political Economy}, 128(5):1753--88.

\bibitem[\protect\astroncite{Andre}{2024}]{andre2024shallow}
Andre, P. (2024).
\newblock Shallow meritocracy.
\newblock {\em Review of Economic Studies}, 00(0):1--36.

\bibitem[\protect\astroncite{Armantier et~al.}{2017}]{armantier2017overview}
Armantier, O., Topa, G., Van~der Klaauw, W., and Zafar, B. (2017).
\newblock An overview of the {S}urvey of {C}onsumer {E}xpectations.
\newblock {\em Economic Policy Review}, 23(2):51--72.

\bibitem[\protect\astroncite{Ashok et~al.}{2015}]{ashok2015support}
Ashok, V., Kuziemko, I., and Washington, E. (2015).
\newblock Support for redistribution in an age of rising inequality: New
  stylized facts and some tentative explanations.
\newblock {\em Brookings Papers on Economic Activity}, 46(1):367--405.

\bibitem[\protect\astroncite{Bagnoli and Bergstrom}{2006}]{bagnoli2006log}
Bagnoli, M. and Bergstrom, T. (2006).
\newblock Log-concave probability and its applications.
\newblock In Aliprantis, C.~D., Matzkin, R.~L., McFadden, D.~L., Moore, J.~C.,
  and Yannelis, N.~C., editors, {\em Rationality and Equilibrium}, volume~26 of
  {\em Studies in Economic Theory}, pages 217--41. Springer.

\bibitem[\protect\astroncite{Benjamin}{2019}]{benjamin2019errors}
Benjamin, D.~J. (2019).
\newblock Errors in probabilistic reasoning and judgment biases.
\newblock In Bernheim, B.~D., DellaVigna, S., and Laibson, D., editors, {\em
  Handbook of Behavioral Economics: Applications and Foundations 1}, volume~2,
  chapter~2, pages 69--186. North-Holland.

\bibitem[\protect\astroncite{Benndorf et~al.}{2019}]{benndorf2019minimizing}
Benndorf, V., Rau, H.~A., and Sölch, C. (2019).
\newblock Minimizing learning in repeated real-effort tasks.
\newblock {\em Journal of Behavioral and Experimental Finance}, 22:239--48.

\bibitem[\protect\astroncite{Bhattacharya and
  Mollerstrom}{2022}]{bhattacharyaandmollerstrom:2022}
Bhattacharya, P. and Mollerstrom, J. (2022).
\newblock Lucky to work.
\newblock GMU Working Paper in Economics.
\newblock No. 22-46.

\bibitem[\protect\astroncite{Bozio et~al.}{2024}]{Bozio2024}
Bozio, A., Garbinti, B., Goupille-Lebret, J., Guillot, M., and Piketty, T.
  (2024).
\newblock Predistribution versus redistribution: Evidence from france and the
  united states.
\newblock {\em American Economic Journal: Applied Economics}, 16(2):31--65.

\bibitem[\protect\astroncite{Cappelen et~al.}{2007}]{cappelen2007pluralism}
Cappelen, A.~W., Hole, A.~D., S{\o}rensen, E.~{\O}., and Tungodden, B. (2007).
\newblock The pluralism of fairness ideals: An experimental approach.
\newblock {\em American Economic Review}, 97(3):818--27.

\bibitem[\protect\astroncite{Cappelen et~al.}{2013}]{cappelen2013just}
Cappelen, A.~W., Konow, J., S{\o}rensen, E.~{\O}., and Tungodden, B. (2013).
\newblock Just luck: An experimental study of risk-taking and fairness.
\newblock {\em American Economic Review}, 103(4):1398--413.

\bibitem[\protect\astroncite{Cappelen et~al.}{2022a}]{cappelen2020merit}
Cappelen, A.~W., Moene, K.~O., Skjelbred, S.-E., and Tungodden, B. (2022a).
\newblock The merit primacy effect.
\newblock {\em The Economic Journal}, 133(651):951--70.

\bibitem[\protect\astroncite{Cappelen et~al.}{2022b}]{cappelen2022meritocratic}
Cappelen, A.~W., Mollerstrom, J., Reme, B.-A., and Tungodden, B. (2022b).
\newblock A meritocratic origin of egalitarian behaviour.
\newblock {\em The Economic Journal}, 132(646):2101--17.

\bibitem[\protect\astroncite{Cappelen
  et~al.}{2010}]{cappelen2010responsibility}
Cappelen, A.~W., S{\o}rensen, E.~{\O}., and Tungodden, B. (2010).
\newblock Responsibility for what? {F}airness and individual responsibility.
\newblock {\em European Economic Review}, 54(3):429--41.

\bibitem[\protect\astroncite{Charness and
  Rabin}{2002}]{charness_understanding_2002}
Charness, G. and Rabin, M. (2002).
\newblock Understanding {Social} {Preferences} with {Simple} {Tests}.
\newblock {\em The Quarterly Journal of Economics}, 117(3):817--69.

\bibitem[\protect\astroncite{Chen et~al.}{2016}]{chen2016otree}
Chen, D.~L., Schonger, M., and Wickens, C. (2016).
\newblock otree--an open-source platform for laboratory, online, and field
  experiments.
\newblock {\em Journal of Behavioral and Experimental Finance}, 9:88--97.

\bibitem[\protect\astroncite{Chetty et~al.}{2014}]{chetty2014united}
Chetty, R., Hendren, N., Kline, P., Saez, E., and Turner, N. (2014).
\newblock Is the united states still a land of opportunity? recent trends in
  intergenerational mobility.
\newblock {\em American Economic Review}, 104(5):141–47.

\bibitem[\protect\astroncite{Cohn et~al.}{2023}]{cohn2023wealthy}
Cohn, A., Jessen, L.~J., Kla{\v{s}}nja, M., and Smeets, P. (2023).
\newblock Wealthy americans and redistribution: The role of fairness
  preferences.
\newblock {\em Journal of Public Economics}, 225:104977.

\bibitem[\protect\astroncite{Conlon}{2025}]{conlon2025}
Conlon, J.~J. (2025).
\newblock Attention, information, and persuasion.
\newblock Working Paper.

\bibitem[\protect\astroncite{Corak}{2013}]{corak2013income}
Corak, M. (2013).
\newblock Income inequality, equality of opportunity, and intergenerational
  mobility.
\newblock {\em Journal of Economic Perspectives}, 27(3):79–102.

\bibitem[\protect\astroncite{Cruces et~al.}{2013}]{cruces2013biased}
Cruces, G., Perez-Truglia, R., and Tetaz, M. (2013).
\newblock Biased perceptions of income distribution and preferences for
  redistribution: Evidence from a survey experiment.
\newblock {\em Journal of Public Economics}, 98:100--12.

\bibitem[\protect\astroncite{DellaVigna
  et~al.}{2022}]{dellavigna2022estimating}
DellaVigna, S., List, J.~A., Malmendier, U., and Rao, G. (2022).
\newblock Estimating social preferences and gift exchange at work.
\newblock {\em American Economic Review}, 112(3):1038--74.

\bibitem[\protect\astroncite{Dong et~al.}{2024}]{dongetal:2024}
Dong, L., Huang, L., and Lien, J.~W. (2024).
\newblock ‘they never had a chance’: Unequal opportunities and fair
  redistributions.
\newblock {\em The Economic Journal}, 00(0):ueae099.

\bibitem[\protect\astroncite{Durante et~al.}{2014}]{durante_preferences_2014}
Durante, R., Putterman, L., and van~der Weele, J. (2014).
\newblock Preferences for redistribution and perception of fairness: An
  experimental study.
\newblock {\em Journal of the European Economic Association}, 12(4):1059--86.

\bibitem[\protect\astroncite{Enke}{2020}]{enke2020you}
Enke, B. (2020).
\newblock What you see is all there is.
\newblock {\em The Quarterly Journal of Economics}, 135(3):1363--98.

\bibitem[\protect\astroncite{Erkal et~al.}{2011}]{erkal2011relative}
Erkal, N., Gangadharan, L., and Nikiforakis, N. (2011).
\newblock Relative earnings and giving in a real-effort experiment.
\newblock {\em American Economic Review}, 101(7):3330--48.

\bibitem[\protect\astroncite{Fehr et~al.}{2024}]{fehr2024social}
Fehr, D., M{\"u}ller, D., and Preuss, M. (2024).
\newblock Social mobility perceptions and inequality acceptance.
\newblock {\em Journal of Economic Behavior \& Organization}, 221:366--84.

\bibitem[\protect\astroncite{Frank}{2016}]{frank2016success}
Frank, R.~H. (2016).
\newblock {\em Success and Luck: Good Fortune and the Myth of Meritocracy}.
\newblock Princeton University Press.

\bibitem[\protect\astroncite{Graeber}{2023}]{graeber2023inattentive}
Graeber, T. (2023).
\newblock Inattentive inference.
\newblock {\em Journal of the European Economic Association}, 21(2):560--92.

\bibitem[\protect\astroncite{Konow}{2000}]{konow_fair_2000}
Konow, J. (2000).
\newblock Fair {Shares}: {Accountability} and {Cognitive} {Dissonance} in
  {Allocation} {Decisions}.
\newblock {\em American Economic Review}, 90(4):1072--92.

\bibitem[\protect\astroncite{Kuziemko et~al.}{2015}]{kuziemko_how_2015}
Kuziemko, I., Norton, M.~I., Saez, E., and Stantcheva, S. (2015).
\newblock How {Elastic} {Are} {Preferences} for {Redistribution}? {Evidence}
  from {Randomized} {Survey} {Experiments}.
\newblock {\em American Economic Review}, 105(4):1478--508.

\bibitem[\protect\astroncite{Larrick and Soll}{2008}]{larrick2008mpg}
Larrick, R.~P. and Soll, J.~B. (2008).
\newblock The {MPG} illusion.
\newblock {\em Science}, 320(5883):1593--94.

\bibitem[\protect\astroncite{Levy and Tasoff}{2016}]{levy2016exponential}
Levy, M. and Tasoff, J. (2016).
\newblock Exponential-growth bias and lifecycle consumption.
\newblock {\em Journal of the European Economic Association}, 14(3):545--83.

\bibitem[\protect\astroncite{Mar{\'e}chal et~al.}{2025}]{marechal2025whose}
Mar{\'e}chal, M.~A., Cohn, A., Yusof, J., and Fisman, R. (2025).
\newblock Whose preferences matter for redistribution? cross-country evidence.
\newblock {\em Journal of Political Economy Microeconomics}, 3(1):000--000.

\bibitem[\protect\astroncite{Mart{\'\i}nez-Marquina
  et~al.}{2019}]{martinez2019failures}
Mart{\'\i}nez-Marquina, A., Niederle, M., and Vespa, E. (2019).
\newblock Failures in contingent reasoning: The role of uncertainty.
\newblock {\em American Economic Review}, 109(10):3437--74.

\bibitem[\protect\astroncite{Mirrlees}{1971}]{mirrlees_exploration_1971}
Mirrlees, J.~A. (1971).
\newblock An {Exploration} in the {Theory} of {Optimum} {Income} {Taxation}.
\newblock {\em The Review of Economic Studies}, 38(2):175--208.

\bibitem[\protect\astroncite{Mitnik et~al.}{2020}]{mitnik2020inequality}
Mitnik, P., Hels{\o}, A.-L., and Bryant, V.~L. (2020).
\newblock Inequality of opportunity for income in {D}enmark and the {U}nited
  {S}tates: {A} comparison based on administrative data.
\newblock Working Paper No. 27835, National Bureau of Economic Research.

\bibitem[\protect\astroncite{Mollerstrom et~al.}{2015}]{mollerstrom2015luck}
Mollerstrom, J., Reme, B.-A., and S{\o}rensen, E.~{\O}. (2015).
\newblock Luck, choice and responsibility -- an experimental study of fairness
  views.
\newblock {\em Journal of Public Economics}, 131:33--40.

\bibitem[\protect\astroncite{Rees-Jones and
  Taubinsky}{2020}]{rees2020measuring}
Rees-Jones, A. and Taubinsky, D. (2020).
\newblock Measuring `schmeduling'.
\newblock {\em The Review of Economic Studies}, 87(5):2399--2438.

\end{thebibliography}
\end{spacing}

\clearpage 
\appendix 
	\renewcommand \thesection{\Alph{section}}\setcounter{section}{0}
\renewcommand \thesubsection{\thesection.\arabic{subsection}}\setcounter{subsection}{0}
\renewcommand{\thetable}{\thesection\arabic{table}}\setcounter{table}{0}
\renewcommand{\thefigure}{\thesection\arabic{figure}}\setcounter{figure}{0}
\renewcommand{\theequation}{\thesection\arabic{equation}}\setcounter{equation}{0}

\setcounter{page}{1}

\clearpage
\begin{center}
    \LARGE{\textit{Online Appendix}\\ \bigskip
    
    \textbf{Inequality of Opportunity and Income Redistribution}}
    
    \bigskip
    \normalsize
    
     Marcel Preuss, Germán Reyes, Jason Somerville, and Joy Wu
    
    \bigskip
    
\date{\today}
    
\end{center}

\section{Additional Figures and Tables} \label{app:figs-tabs}

\begin{figure}[H]\caption{Spectator Redistribution Screens for Lucky Outcomes and Lucky Opportunities \\ \underline{with} Information about $\pi$}\label{fig:CF_IOp}
	
	\begin{center}
		\footnotesize\sffamily
		
		\begin{tabular}{|c|c|c|}
			\hline
			\textbf{Worker ID:} & \texttt{sao9rqhr} & \texttt{qeha27vh} \\
			\hline
			\textbf{Coin-Flip Chance:} & \multicolumn{2}{c|}{\texttt{46\%}} \\
			\hline
			\textbf{Result:} & \texttt{won} & \texttt{lost} \\
			\hline
			\textbf{Unadjusted Earnings:} & \texttt{\$5.00} & \texttt{\$0.00} \\		
			\hline
		\end{tabular}
		
		\begin{tabular}{p{6in}}
			\\[-1ex]
			{There was a {46\% }chance that the winner and the loser in this pair were determined by a {coin flip} instead of the number of correct encryptions each worker completed.}\\
			\\
			$\rhd$ {This means that there is a 77\% chance that the winner above completed more transcriptions than the loser.}\\
			\\[2ex]
			\hline
			\\[2ex]
		\end{tabular}
		
		\begin{tabular}{|c|c|c|}
			\hline
			\textbf{Worker ID:} & \texttt{ga2c8k8x} & \texttt{nkqqjd0n} \\
			\hline
			\textbf{Multiplier:} & \texttt{2.9} & \texttt{2.4}\\
			\hline
			\textbf{Result:} & \texttt{won} & \texttt{lost} \\
			\hline
			\textbf{Unadjusted Earnings:}  \hspace{3cm}& \texttt{\$5.00} & \texttt{\$0.00} \\
			\hline
		\end{tabular}
		\begin{tabular}{p{6in}}
			\\[-1ex]
			{The winner had a {higher score} than the loser in this pair. Each worker's score is the number of correct {encryptions} they completed \textit{times} their {multiplier}.}\\
			\\
			$\rhd$ {Based on historical data for these multipliers, there is a 77\% chance that the winner above completed more transcriptions than the loser.}\\
			\\[2ex]
		\end{tabular}
	\end{center}
	\vspace{-15pt}
	\singlespacing \justify \footnotesize \normalfont
	\textbf{Notes}: This figure shows the information for redistribution choices displayed to spectators under the lucky outcomes (top) and lucky opportunities (bottom) conditions. Included directly below the outcomes table is additional text to remind spectators how to interpret the form of luck involved in determining the winner and loser of the pair. The information provision converting the influence of luck as the likelihood that the winner performed better than the loser is only included for information condition spectators (see text next to $\rhd$ symbol). 
\end{figure}

\clearpage
\begin{figure}[H]\caption{Distribution of effort and probability of exerting more effort} \label{fig:worker-eff}
	{\scriptsize
		\begin{centering}	
		\protect
			\begin{minipage}{.48\textwidth}
				\captionof*{figure}{Panel A. Distribution of encryptions completed in  worker task}
				\includegraphics[width=\linewidth]{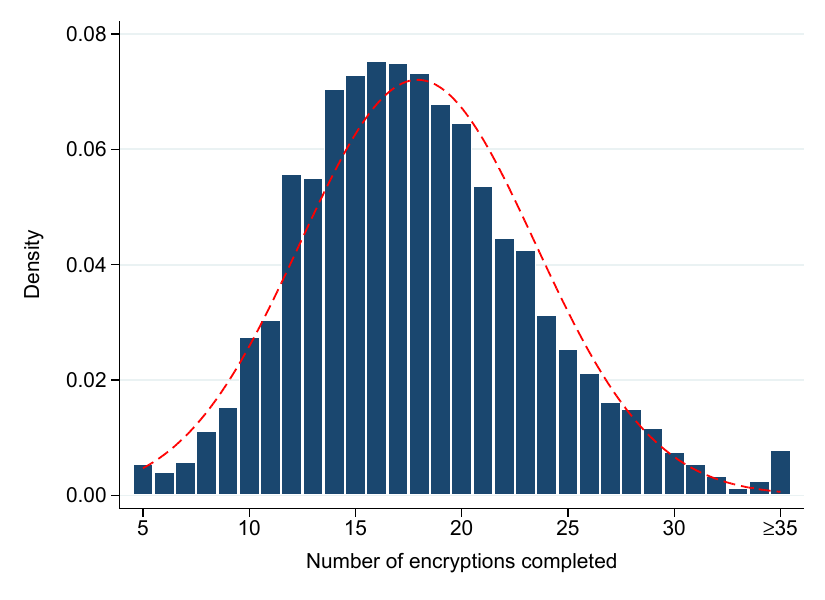}
			\end{minipage}\hspace{1em}
			\begin{minipage}{.48\textwidth}
				\captionof*{figure}{Panel B. Probability that the winner exerted more effort as a function of multiplier ratio}				\includegraphics[width=\linewidth]{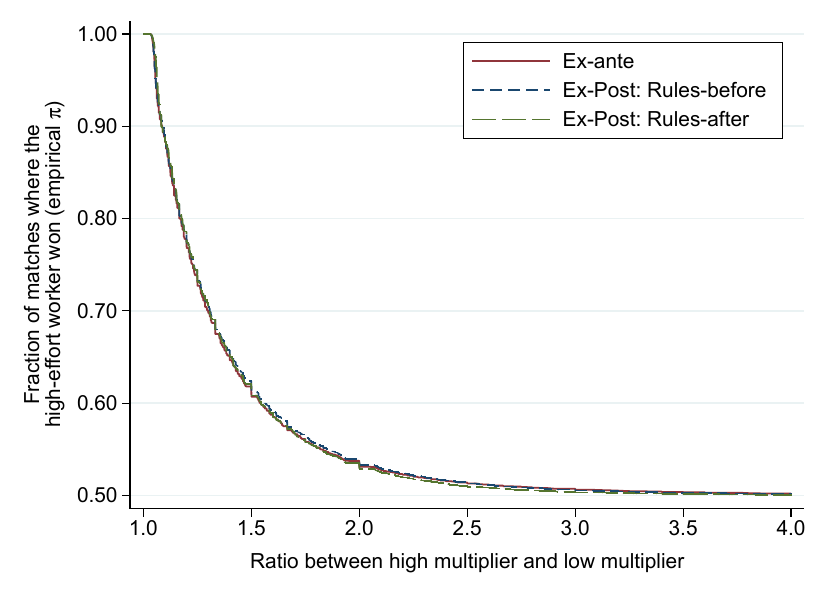}
			\end{minipage}
			\par\end{centering}
		\singlespacing \justify \footnotesize
		\textbf{Notes}: 
		Panel~A shows the distribution of the total number of correct three-word encryptions. The mean number of encryptions completed is 18 and the standard deviation is 5.5. The red dashed line shows the density of a normal random variable that has the same mean and standard deviation as the distribution of tasks completed. 
		In Panel~B, the $y-$axis depicts the fraction of all possible worker pair combinations in which the worker who wins (who has the higher score, i.e., correct encryptions times productivity multiplier) is also the worker who solved more encryptions. The $x-$axis depicts the ratio of the winning worker's to the losing worker's productivity multiplier. Thus, Panel B depicts the empirical relationship between $m$ and $\pi$. At $\pi=0.5$, there is a 50\% chance that the winner of the pair is the worker who solved more encryptions so that the outcome is purely luck-based. The three different lines reflect data from different subtreatments, which we explain in Section \ref{sec:design_timing} and Section \ref{sec:design_rules}, respectively.}
\end{figure}

\clearpage
\begin{figure}[H]\caption{Histogram of tasks completed by condition} \label{fig:tasks-comp-cf-iop} 
	\centering
	\includegraphics[width=.75\linewidth]{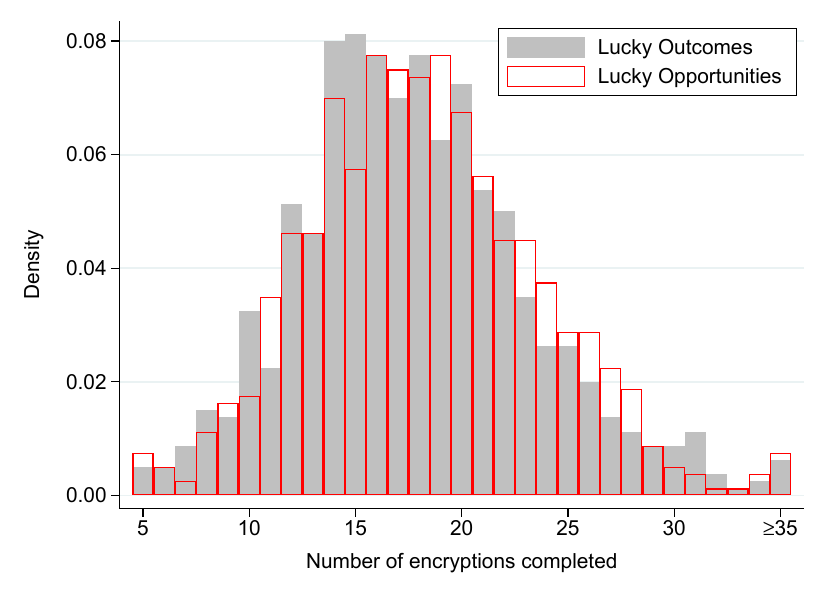}
	\footnotesize
	\singlespacing \justify 
 \vspace{-12pt}
		\textbf{Notes:} This figure shows the distribution of tasks completed by workers in the lucky outcomes and baseline lucky opportunities conditions. A Kolmogorov--Smirnov test for equality of distribution cannot reject the hypothesis that the distributions of worker effort in the two conditions are equal ($p = 0.914$). 
\end{figure}

\clearpage

\begin{landscape}
\vspace*{\fill} 

\begin{table}[ht]
\centering
\caption{Summary of Treatments and Key Features}
\label{tab:treatments}
\resizebox{\linewidth}{!}{%
\begin{tabular}{clcccc}
\hline
\textbf{Treatment} & \textbf{Name} & \textbf{Source of Luck} & \textbf{Timing of Luck} & \textbf{Workers' Knowledge} & \textbf{Spectators' Info} \\
\hline
1 & Lucky Outcomes (rules before) & Coin Flips & After Workers Complete Task & Before Task & Partial \\
2 & Lucky Outcomes (rules before) & Coin Flips & After Workers Complete Task & Before Task & Full \\
3 & Lucky Outcomes (rules after) & Coin Flips & After Workers Complete Task & After Task & Partial \\
4 & Lucky Outcomes (rules after) & Coin Flips & After Workers Complete Task & After Task & Full \\
5 & Baseline Lucky Opportunities (rules before) & Multipliers & Before Workers Start Task & Before Task & Partial \\
6 & Baseline Lucky Opportunities (rules before) & Multipliers & Before Workers Start Task & Before Task & Full \\
7 & Ex-Post Lucky Opportunities (rules before) & Multipliers & After Workers Complete Task & Before Task & Partial \\
8 & Ex-Post Lucky Opportunities (rules before) & Multipliers & After Workers Complete Task & Before Task & Full \\
9 & Ex-Post Lucky Opportunities (rules after) & Multipliers & After Workers Complete Task & After Task & Partial \\
10 & Ex-Post Lucky Opportunities (rules after) & Multipliers & After Workers Complete Task & After Task & Full \\
\hline
\end{tabular}%
}
		\singlespacing\justify\footnotesize \textbf{Notes}: This table summarizes the experimental treatments and their key features. Treatments vary based on the source of luck (coin flips for lucky outcomes versus multipliers for lucky opportunities), the timing of when luck is applied (before or after workers complete the task), workers' knowledge about the role of luck (informed before or after the task), and the level of information provided to spectators (partial or full information about the probability that performance determined the winner). 

\end{table}
\vspace*{\fill} 
\end{landscape}

\clearpage
\begin{table}[H]\caption{Average worker characteristics by treatment condition} \label{tab:summ-wrk}
     \setlength{\tabcolsep}{0.25em} 
	{\small\begin{centering} 
			\protect
			\begin{tabular}{lccccc}
				\hline \addlinespace
				& \multicolumn{2}{C{3.5cm}}{Lucky Outcomes}  & \multicolumn{1}{C{2.5cm}}{Lucky Opportunities} & \multicolumn{2}{C{3.5cm}}{Ex-Post Lucky Opportunities}  \\ \cmidrule{2-3} \cmidrule{5-6}
				Worker characteristics    & Rules-Before & Rules-After   &         & Rules-Before & Rules-After \\\midrule\addlinespace			
				Age &39.08 &37.87 &38.18 &37.95 &38.53 \\
Male &0.56 &0.56 &0.61 &0.57 &0.58 \\
Married &0.62 &0.70 &0.67 &0.58 &0.73 \\
White &0.76 &0.76 &0.79 &0.70 &0.76 \\
Completed college &0.79 &0.85 &0.83 &0.79 &0.84 \\
Income $>$ 75,000 &0.34 &0.31 &0.31 &0.35 &0.31 \\
Has masters certification &0.33 &0.42 &0.34 &0.28 &0.40 \\\midrule
Encryptions attempted &18.11 &18.17 &18.59 &18.43 &17.54 \\
Encryptions completed &17.82 &17.84 &18.27 &18.17 &17.20 \\
Average multiplier &-- &-- &2.58 &2.56 &2.53 \\
Time spent in instructions &121.10 &143.52 &142.99 &153.94 &138.93 \\
Time spent in comprehension screen &110.85 &127.50 &124.49 &125.14 &150.82 \\
Average time spent in each round &17.43 &17.29 &17.03 &17.10 &17.76 \\
Total time in experiment &817.84 &841.46 &875.63 &913.38 &879.16 \\\midrule
\(N\) (Workers) &400 &400 &800 &400 &400 \\
 \addlinespace
				\hline
			\end{tabular}
			\par\end{centering}
		
		\singlespacing\justify\footnotesize
		\textbf{Notes}: \small This table shows summary statistics on our sample of workers. We exclude workers who completed fewer than five encryptions. The time spent in the experiment is measured in seconds.
	}
\end{table}

\clearpage
\begin{table}[H]
\caption{Heterogeneity in redistribution} \label{tab:redist-pi-het}
\vspace{-15pt}
{\footnotesize 
	\begin{center}
		\newcommand\w{1.75}
		\begin{tabular}{l@{}lR{\w cm}@{}L{0.43cm}R{\w cm}@{}L{0.43cm}R{\w cm}@{}L{0.43cm}R{\w cm}@{}L{0.43cm}R{\w cm}@{}L{0.43cm}R{\w cm}@{}L{0.43cm}R{\w cm}@{}L{0.43cm}}
			\midrule
			&  \multicolumn{4}{c}{Lucky Outcomes} && \multicolumn{3}{c}{Lucky Opportunities} \\ \cmidrule{3-5}  \cmidrule{7-10} 
			&& Mean     && Elasticity     && Mean      && Elasticity   \\
			&& Redist.  &&  w.r.t. $1-\pi$  && Redist.   && w.r.t. $1-\pi$ \\
			&& (1)      && (2)            && (3)       && (4)          \\ \midrule
			
			\multicolumn{10}{l}{\hspace{-1em} \textbf{Panel A. Demographic characteristics, education, and income}} \\\addlinespace			
			Female&&0.045&\sym{*}&$-$0.015&&0.054&\sym{*}&$-$0.001&\\
&&(0.020&)&(0.009&)&(0.025&)&(0.007&)\\
 
			Under 50$-$years old&&$-$0.017&&0.007&&$-$0.040&&0.003&\\
&&(0.021&)&(0.009&)&(0.025&)&(0.007&)\\
 
			Married&&$-$0.036&&0.016&&$-$0.028&&0.004&\\
&&(0.021&)&(0.009&)&(0.025&)&(0.008&)\\
 
			White&&0.014&&0.010&&0.027&&$-$0.020&\\
&&(0.025&)&(0.011&)&(0.040&)&(0.011&)\\

			Completed college&&$-$0.015&&0.014&&0.023&&0.011&\\
&&(0.022&)&(0.009&)&(0.027&)&(0.008&)\\
 				
			HH income above 100k&&$-$0.050&\sym{*}&0.013&&$-$0.046&&0.004&\\
&&(0.024&)&(0.009&)&(0.026&)&(0.007&)\\
 \midrule
			
			\multicolumn{10}{l}{\hspace{-1em} \textbf{Panel B. Political and Social preferences}}\\\addlinespace
			Tend to side with republicans&&$-$0.041&&0.005&&$-$0.056&\sym{*}&0.007&\\
&&(0.023&)&(0.010&)&(0.025&)&(0.008&)\\
 
			Oppose gov't interventions&&$-$0.067&\sym{**}&$-$0.008&&$-$0.048&&0.003&\\
&&(0.026&)&(0.011&)&(0.027&)&(0.008&)\\
 				
			Conservative on social issues&&$-$0.019&&0.016&&$-$0.033&&$-$0.005&\\
&&(0.025&)&(0.010&)&(0.026&)&(0.008&)\\
 			
			Influece of hard work is fair&&$-$0.032&&0.047&\sym{***}&$-$0.004&&0.028&\sym{*}\\
&&(0.031&)&(0.011&)&(0.026&)&(0.012&)\\

			Influece of talent is fair&&$-$0.009&&0.045&\sym{***}&$-$0.035&&0.005&\\
&&(0.028&)&(0.010&)&(0.031&)&(0.010&)\\
 
			Influece of luck is fair&&$-$0.018&&0.013&&$-$0.028&&0.005&\\
&&(0.023&)&(0.010&)&(0.026&)&(0.008&)\\
 
			Influece of connections is fair&&$-$0.012&&0.016&&$-$0.053&&$-$0.000&\\
&&(0.027&)&(0.012&)&(0.031&)&(0.009&)\\
 
			Key to success own hands&&$-$0.061&\sym{**}&$-$0.012&&$-$0.090&\sym{***}&0.002&\\
&&(0.020&)&(0.009&)&(0.024&)&(0.008&)\\
  
			Gov't should never redistribute&&$-$0.027&&0.001&&$-$0.056&\sym{*}&$-$0.003&\\
&&(0.021&)&(0.010&)&(0.025&)&(0.008&)\\
 
			Gov't redistribute to correct luck&&0.017&&$-$0.002&&0.056&&0.021&\sym{*}\\
&&(0.031&)&(0.013&)&(0.031&)&(0.011&)\\
 
			Income dist. in the US is fair&&$-$0.006&&$-$0.008&&$-$0.081&\sym{**}&0.007&\\
&&(0.028&)&(0.015&)&(0.026&)&(0.009&)\\
 \midrule

		\end{tabular}
  	\end{center}
		\begin{singlespace}  \vspace{-.5cm}
			\noindent \justify \footnotesize
			\textbf{Notes:} This table shows the difference in mean redistribution and the slope of redistribution across various participant characteristics and stated preferences. Each row shows the result of an independent regression where the coefficient corresponds to the difference between the stated characteristic and the omitted category. All variables in Panel A are indicator variables. All variables in Panel B are indicators equal to one if the participant ``agrees'' or ``strongly agrees'' and zero otherwise. $^{***}$, $^{**}$ and $^*$ denote significance at the 0.1\%, 1\%, and 5\% level, respectively. 
		\end{singlespace} 	
}
\end{table}

\clearpage
\begin{table}[H]
\caption{Fraction redistributed as a function of $1 - \pi$ and information treatment} \label{tab:redist-info}
\vspace{-15pt}
	{\footnotesize 
		\begin{center}
			\newcommand\w{2.5}
			\begin{tabular}{l@{}lR{\w cm}@{}L{0.43cm}R{\w cm}@{}L{0.43cm}R{\w cm}@{}L{0.43cm}R{\w cm}@{}L{0.43cm}R{\w cm}@{}L{0.43cm}R{\w cm}@{}L{0.43cm}}
				\midrule
				&& \multicolumn{6}{c}{Outcome: Fraction of earnings redistributed}  \\
				\cmidrule{3-8} 
					&&  Lucky   &&  Lucky  && \multirow{2}{*}{Difference} \\
					&& Outcomes && Opportunities && \\
				&& (1) && (2) && (3)  \\
				\midrule 
				\multicolumn{8}{l}{\hspace{-1em} \textbf{Panel A. Average Redistribution}}  \\  \addlinespace 
				Info $\pi$&&$-$0.045&\sym{**}&$-$0.027&&$-$0.018&\\
&&(0.014&)&(0.017&)&(0.022&)\\
Constant&&0.276&\sym{***}&0.234&\sym{***}&0.042&\sym{**}\\
&&(0.010&)&(0.013&)&(0.016&)\\
\(N\) (Redistributive decisions)&&4,728&&4,680&&9,408&\\
  \midrule
				\multicolumn{8}{l}{\hspace{-1em} \textbf{Panel B. Linear slope}}  \\  \addlinespace 
				1$-$$\pi$&&0.037&\sym{***}&0.020&\sym{***}&0.017&\sym{**}\\
&&(0.005&)&(0.004&)&(0.006&)\\
Info $\pi$&&$-$0.023&&$-$0.003&&$-$0.021&\\
&&(0.017&)&(0.019&)&(0.026&)\\
1$-$$\pi \times$ Info $\pi$&&0.015&\sym{*}&0.017&\sym{**}&$-$0.002&\\
&&(0.006&)&(0.005&)&(0.008&)\\
Constant&&0.331&\sym{***}&0.263&\sym{***}&0.068&\sym{***}\\
&&(0.013&)&(0.014&)&(0.019&)\\
\(N\) (Redistributive decisions)&&4,728&&4,680&&9,408&\\
  \midrule
			\end{tabular}
		\end{center}
		\begin{singlespace}  \vspace{-.5cm}
			\noindent \justify \footnotesize 
			\textbf{Notes:} Column~1 includes only spectators in the lucky outcomes condition and column~2 includes only spectators in the lucky opportunities condition. Column~3 is the difference in spectator responses between columns~1 and~2. 
		Panel A shows average redistribution. We include a dummy variable indicating whether the spectators were assigned to know $\pi$ (our information intervention). Panel B shows a linear approximation between the fraction of earnings redistributed and the likelihood that the winning worker performed better than the losing worker ($\pi$). We include variables that indicate whether spectators were assigned to know $\pi$ and the interaction of $\pi$ and its provision to spectators. Heteroskedasticity-robust standard errors clustered at the spectator level in parentheses. $^{***}$, $^{**}$ and $^*$ denote significance at the 0.1\%, 1\%, and 5\% level, respectively.
		\end{singlespace} 	
	}
\end{table}

\noindent \textbf{Additional Remarks:} The elasticity of redistribution with respect to changes in the likelihood that the inequality is due to luck as reported in the main text can be calculated using the estimates in Panel B. Specifically, when spectators have information about $\pi$,  the elasticity for each luck environment is given by the sum of the first row ``$1-\pi$'' and the third row ``$(1-\pi)\times \text{Info } \pi$.''

The average difference in redistribution between the lucky outcomes treatment and the lucky opportunities treatment, conditional on receiving information about $\pi$, can be calculated using the estimates in Panel A, Column 3. Subtracting the effect of providing information ($0.018$) from the constant ($0.042$), we arrive at a difference of just 2.4 percentage points.

\begin{table}[H]
	\caption{Gap in mean redistribution between lucky outcomes and lucky opportunities as a function of $1 - \pi$ and information treatment} \label{tab:redist-gap-info}
	\vspace{-15pt}
	{\footnotesize 
		\begin{center}
			\newcommand\w{1.7}
			\begin{tabular}{l@{}lR{\w cm}@{}L{0.43cm}R{\w cm}@{}L{0.43cm}R{\w cm}@{}L{0.43cm}R{\w cm}@{}L{0.43cm}R{\w cm}@{}L{0.43cm}R{\w cm}@{}L{0.43cm}}
				\midrule
				&& \multicolumn{6}{c}{Outcome: Redistributive Gap}  \\
				\cmidrule{3-8} 
				&&  No Info   && Info         &&  {Difference} \\
				&& (1) && (2) && (3)  \\
				\midrule 
				Lucky Opportunities $\times$ 1$-$$\pi=0.00$&&$-$0.048&&$-$0.018&&0.031&\\
&&(0.025&)&(0.022&)&(0.034&)\\
Lucky Opportunities $\times$ 1$-$$\pi\in(0.00,0.05]$&&$-$0.024&&$-$0.016&&0.008&\\
&&(0.024&)&(0.022&)&(0.032&)\\
Lucky Opportunities $\times$ 1$-$$\pi\in(0.05,0.10]$&&$-$0.000&&$-$0.022&&$-$0.022&\\
&&(0.022&)&(0.021&)&(0.031&)\\
Lucky Opportunities $\times$ 1$-$$\pi\in(0.10,0.15]$&&0.014&&0.022&&0.007&\\
&&(0.023&)&(0.021&)&(0.031&)\\
Lucky Opportunities $\times$ 1$-$$\pi\in(0.15,0.20]$&&0.062&\sym{**}&0.017&&$-$0.045&\\
&&(0.023&)&(0.020&)&(0.030&)\\
Lucky Opportunities $\times$ 1$-$$\pi\in(0.20,0.25]$&&0.090&\sym{***}&0.042&&$-$0.048&\\
&&(0.022&)&(0.021&)&(0.031&)\\
Lucky Opportunities $\times$ 1$-$$\pi\in(0.25,0.30]$&&0.117&\sym{***}&0.038&&$-$0.079&\sym{*}\\
&&(0.023&)&(0.020&)&(0.031&)\\
Lucky Opportunities $\times$ 1$-$$\pi\in(0.30,0.35]$&&0.082&\sym{***}&0.021&&$-$0.061&\sym{*}\\
&&(0.023&)&(0.021&)&(0.031&)\\
Lucky Opportunities $\times$ 1$-$$\pi\in(0.35,0.40]$&&0.066&\sym{**}&0.032&&$-$0.034&\\
&&(0.023&)&(0.021&)&(0.031&)\\
Lucky Opportunities $\times$ 1$-$$\pi\in(0.40,0.45]$&&0.076&\sym{**}&0.060&\sym{**}&$-$0.016&\\
&&(0.025&)&(0.022&)&(0.033&)\\
Lucky Opportunities $\times$ 1$-$$\pi\in(0.45,0.50)$&&0.037&&0.033&&$-$0.003&\\
&&(0.026&)&(0.023&)&(0.035&)\\
Lucky Opportunities $\times$ 1$-$$\pi=0.50$&&0.038&&0.087&\sym{***}&0.049&\\
&&(0.028&)&(0.023&)&(0.036&)\\
\(N\) (Redistributive decisions)&&4,692&&4,716&&9,408&\\
  \midrule
			\end{tabular}
		\end{center}
		\begin{singlespace}  \vspace{-.5cm}
			\noindent \justify \footnotesize 
			\textbf{Notes:} This table shows the gap in mean redistribution between the lucky opportunities and the lucky outcomes conditions across different $1-\pi$ bins. Each column shows a different treatment. Column 1 shows the redistributive gap when spectators do not receive information about $\pi$. Column 2 shows the redistributive gap when spectators receive information about $\pi$. Column 3 shows the difference between columns 1 and 2. Heteroskedasticity-robust standard errors clustered at the spectator level in parentheses. $^{***}$, $^{**}$ and $^*$ denote significance at the 0.1\%, 1\%, and 5\% level, respectively. 
		\end{singlespace} 	
	}
\end{table}

\begin{table}[H]
	\caption{Correlation between redistribution behavior and political and social preferences} \label{tab:pred-pol}
	\vspace{-15pt}
	{\footnotesize 
		\begin{center}
			\newcommand\w{2}
			\begin{tabular}{l@{}lR{\w cm}@{}L{0.43cm}R{\w cm}@{}L{0.43cm}R{\w cm}@{}L{0.43cm}R{\w cm}@{}L{0.43cm}R{\w cm}@{}L{0.43cm}R{\w cm}@{}L{0.43cm}}
				\midrule
				&& \multicolumn{6}{c}{ Fraction of earnings redistributed}  \\ \cmidrule{3-8} 
				&& Lucky    && Lucky         && \multirow{2}{*}{Difference} \\
				&& Outcomes && Opportunities &&  \\
	Correlation with... && (1) && (2) && (3)  \\
				\midrule 

			\multicolumn{6}{l}{\hspace{-1em} \textbf{Panel A. Political and social preferences}} \\\addlinespace			
			Tend to side with democrats&&0.081&&0.138&\sym{*}&$-$0.057&\\
&&(0.042&)&(0.055&)&(0.069&)\\
 
			Tend to side with republicans ($-$)&&0.078&&0.154&\sym{**}&$-$0.076&\\
&&(0.045&)&(0.050&)&(0.067&)\\
 
			Oppose gov't interventions ($-$)&&0.124&\sym{**}&0.159&\sym{**}&$-$0.035&\\
&&(0.044&)&(0.058&)&(0.072&)\\
 				
			Conservative on social issues ($-$)&&0.026&&0.111&\sym{*}&$-$0.085&\\
&&(0.041&)&(0.050&)&(0.065&)\\
 			
			Influece of hard work is fair ($-$)&&0.078&\sym{*}&0.116&\sym{**}&$-$0.038&\\
&&(0.039&)&(0.039&)&(0.056&)\\

			Influece of talent is fair ($-$)&&0.039&&0.124&\sym{**}&$-$0.084&\\
&&(0.042&)&(0.044&)&(0.061&)\\
 
			Influece of luck is fair ($-$)&&0.102&\sym{*}&0.059&&0.043&\\
&&(0.043&)&(0.055&)&(0.070&)\\
 
			Influece of connections is fair ($-$)&&0.044&&0.099&&$-$0.055&\\
&&(0.042&)&(0.058&)&(0.072&)\\
 
			Hard work brings a better life ($-$)&&0.115&\sym{**}&0.101&&0.014&\\
&&(0.039&)&(0.058&)&(0.070&)\\
 
			Key to success own hands ($-$)&&0.117&\sym{**}&0.177&\sym{***}&$-$0.060&\\
&&(0.043&)&(0.051&)&(0.067&)\\
  
			Gov't should never redistribute ($-$)&&0.076&&0.139&\sym{*}&$-$0.063&\\
&&(0.043&)&(0.053&)&(0.068&)\\
 
			Gov't redistribute to correct luck&&0.040&&0.156&\sym{**}&$-$0.116&\\
&&(0.043&)&(0.052&)&(0.068&)\\
 
			Gov't eliminate income differences&&0.047&&0.117&\sym{*}&$-$0.070&\\
&&(0.045&)&(0.050&)&(0.067&)\\
 
			Income dist. in the US is fair ($-$)&&0.033&&0.117&\sym{*}&$-$0.084&\\
&&(0.040&)&(0.052&)&(0.065&)\\
\midrule
			
			\multicolumn{6}{l}{\hspace{-1em} \textbf{Panel B. Summary indices}} \\\addlinespace			
			z$-$score ($-$)&&0.064&\sym{**}&0.110&\sym{***}&$-$0.047&\\
&&(0.021&)&(0.027&)&(0.034&)\\
R$-$squared&&0.016&&0.043&&0.030&\\

            \midrule
			PCA first component ($-$)&&0.118&\sym{**}&0.201&\sym{***}&$-$0.083&\\
&&(0.043&)&(0.050&)&(0.066&)\\
R$-$squared&&0.014&&0.040&&0.027&\\
 
			\midrule
			\(N\) (Redistributive decisions)&&2,364&&2,328&&4,692&\\
 \midrule

			\end{tabular}
		\end{center}
		\begin{singlespace}  \vspace{-.5cm}
			\noindent \justify \footnotesize \textbf{Notes:} This table shows the relationship between redistribution behavior in each treatment condition and political and social preferences. Each cell shows the result from a bivariate OLS regression. We normalize variables by their standard deviation so that the coefficients of the regressions can be interpreted as the linear correlation coefficients. ($-$) denotes reverse coded. Heteroskedasticity-robust standard errors clustered at the spectator level in parentheses. $^{***}$, $^{**}$ and $^*$ denote significance at the 0.1\%, 1\%, and 5\% level, respectively.
		\end{singlespace} 	
	}
\end{table}

\clearpage
\begin{table}[H]\caption{Number of encryptions completed and worker multiplier} \label{tab:worker-eff-m}
\vspace{-12pt}
	{\footnotesize 
		\begin{center}		
			\newcommand\w{4}
			\begin{tabular}{l@{}lR{\w cm}@{}L{0.43cm}R{\w cm}@{}L{0.43cm}R{\w cm}@{}L{0.43cm}R{\w cm}@{}L{0.43cm}R{\w cm}@{}L{0.43cm}}
				\midrule
				&& \multicolumn{4}{c}{Outcome: Number of encryptions completed by workers}  \\
				\cmidrule{3-6} 
				&& Linear function && Non-parametric\\					
				&& (1) && (2) \\
				\midrule 
				Multiplier&&0.147&&&\\
&&(0.198&)&&\\
Multiplier $\in[1.0,1.5)$&&&&$-$1.082&\\
&&&&(0.589&)\\
Multiplier $\in[1.5,2.0)$&&&&0.256&\\
&&&&(0.704&)\\
Multiplier $\in[2.0,2.5)$&&&&$-$0.506&\\
&&&&(0.613&)\\
Multiplier $\in[2.5,3.0)$&&&&$-$1.162&\\
&&&&(0.647&)\\
Multiplier $\in[3.0,3.5)$&&&&$-$0.526&\\
&&&&(0.634&)\\
Constant&&17.889&\sym{***}&18.754&\sym{***}\\
&&(0.544&)&(0.386&)\\
\(N\) (Wokers)&&800&&800&\\
  \midrule
			\end{tabular}
		\end{center}
		\begin{singlespace}  \vspace{-.5cm}
			\noindent \justify \footnotesize
			\textbf{Notes:} This table shows the number of encryptions completed by workers in the baseline lucky opportunities condition. Workers are randomly assigned a productivity multiplier $\in[1,4]$ as a rate of return on the number of correct encryptions completed in five minutes. The omitted category in column 2 is multiplier~$\in [3.5, 4.0]$. Heteroskedasticity-robust standard errors clustered at the worker level in parentheses. $^{***}$, $^{**}$ and $^*$ denote significance at the 0.1\%, 1\%, and 5\% level, respectively. 
		\end{singlespace} 	
	}
\end{table}

\clearpage	
	\begin{table}[H]
		\caption{Fraction redistributed as a function of $\pi$ in baseline and ex-post lucky opportunities} \label{tab:redist-ea-ep}
		\vspace{-15pt}
		{\footnotesize 
			\begin{center}
				\newcommand\w{2.5}
				\begin{tabular}{l@{}lR{\w cm}@{}L{0.43cm}R{\w cm}@{}L{0.43cm}R{\w cm}@{}L{0.43cm}R{\w cm}@{}L{0.43cm}R{\w cm}@{}L{0.43cm}R{\w cm}@{}L{0.43cm}}
					\midrule
					&& \multicolumn{6}{c}{Outcome: Fraction of earnings redistributed}  \\
					\cmidrule{3-8} 
					&& Baseline Lucky   && Ex-Post Lucky  && \multirow{2}{*}{Difference} \\
					&& Opportunities && Opportunities && \\
					&& (1) && (2) && (3)  \\
					\midrule 
					\multicolumn{8}{l}{\hspace{-1em} \textbf{Panel A. Average redistribution}}  \\  \addlinespace 
					Constant&&0.234&\sym{***}&0.243&\sym{***}&$-$0.010&\\
&&(0.013&)&(0.014&)&(0.019&)\\
\(N\) (Redistributive decisions)&&2,328&&2,316&&4,644&\\
 
					\midrule 
					\multicolumn{8}{l}{\hspace{-1em} \textbf{Panel B. Linear slope}}  \\  \addlinespace 
					1$-$$\pi$&&0.020&\sym{***}&0.021&\sym{***}&$-$0.001&\\
&&(0.004&)&(0.004&)&(0.005&)\\
Constant&&0.263&\sym{***}&0.274&\sym{***}&$-$0.012&\\
&&(0.014&)&(0.016&)&(0.021&)\\
\(N\) (Redistributive decisions)&&2,328&&2,316&&4,644&\\
  \midrule
					\multicolumn{8}{l}{\hspace{-1em} \textbf{Panel C. Non-parametric estimation}}  \\  \addlinespace 
					1$-$$\pi=0.00$&&0.179&\sym{***}&0.184&\sym{***}&$-$0.005&\\
&&(0.018&)&(0.018&)&(0.025&)\\
1$-$$\pi\in(0.00,0.05]$&&0.198&\sym{***}&0.218&\sym{***}&$-$0.019&\\
&&(0.016&)&(0.018&)&(0.024&)\\
1$-$$\pi\in(0.05,0.10]$&&0.202&\sym{***}&0.215&\sym{***}&$-$0.013&\\
&&(0.016&)&(0.018&)&(0.024&)\\
1$-$$\pi\in(0.10,0.15]$&&0.226&\sym{***}&0.217&\sym{***}&0.009&\\
&&(0.017&)&(0.016&)&(0.023&)\\
1$-$$\pi\in(0.15,0.20]$&&0.208&\sym{***}&0.212&\sym{***}&$-$0.004&\\
&&(0.015&)&(0.016&)&(0.022&)\\
1$-$$\pi\in(0.20,0.25]$&&0.226&\sym{***}&0.236&\sym{***}&$-$0.009&\\
&&(0.015&)&(0.017&)&(0.023&)\\
1$-$$\pi\in(0.25,0.30]$&&0.228&\sym{***}&0.243&\sym{***}&$-$0.015&\\
&&(0.016&)&(0.017&)&(0.024&)\\
1$-$$\pi\in(0.30,0.35]$&&0.240&\sym{***}&0.252&\sym{***}&$-$0.012&\\
&&(0.016&)&(0.017&)&(0.023&)\\
1$-$$\pi\in(0.35,0.40]$&&0.249&\sym{***}&0.247&\sym{***}&0.002&\\
&&(0.016&)&(0.017&)&(0.024&)\\
1$-$$\pi\in(0.40,0.45]$&&0.260&\sym{***}&0.283&\sym{***}&$-$0.023&\\
&&(0.017&)&(0.018&)&(0.025&)\\
1$-$$\pi\in(0.45,0.50)$&&0.291&\sym{***}&0.289&\sym{***}&0.002&\\
&&(0.018&)&(0.019&)&(0.026&)\\
1$-$$\pi=0.50$&&0.298&\sym{***}&0.325&\sym{***}&$-$0.026&\\
&&(0.019&)&(0.021&)&(0.028&)\\
\(N\) (Redistributive decisions)&&2,328&&2,316&&4,644&\\
  \midrule
				\end{tabular}
			\end{center}
			\begin{singlespace}  \vspace{-.5cm}
				\noindent \justify \footnotesize \textbf{Notes:} Column~1 includes only spectators in the baseline lucky opportunities condition and column~2 includes only spectators under the ex-post lucky opportunities condition. Column~3 is the difference in spectator responses between columns~1 and~2. Panel A shows average redistribution. Panel B shows the linear approximation between the fraction of earnings redistributed and the likelihood that the winning worker performed better than the losing worker ($\pi$). Panel C shows the relationship between redistribution and the likelihood that the winning worker performed better ($\pi$) split into 11 bins. The omitted category is $\pi = 0.50$. Heteroskedasticity-robust standard errors clustered at the spectator level in parentheses.  $^{***}$, $^{**}$ and $^*$ denote significance at the 0.1\%, 1\%, and 5\% level, respectively. 
			\end{singlespace} 	
		}
	\end{table}

\clearpage
\begin{table}[H]\caption{Perceived worker effort and worker multiplier} \label{tab:beliefs-eff-m}
	{\footnotesize 
		\begin{center}		
			\newcommand\w{4}
			\begin{tabular}{l@{}lR{\w cm}@{}L{0.43cm}R{\w cm}@{}L{0.43cm}R{\w cm}@{}L{0.43cm}R{\w cm}@{}L{0.43cm}R{\w cm}@{}L{0.43cm}}
				\midrule
				&& \multicolumn{4}{c}{Outcome: Spectator beliefs about encryptions completed}  \\
				\cmidrule{3-6} 
				&& Linear function && Non-parametric\\					
				&& (1) && (2) \\
				\midrule 
				Multiplier&&1.248&&&\\
&&(1.489&)&&\\
Multiplier $\in[1.0,1.5)$&&&&$-$5.599&\\
&&&&(4.532&)\\
Multiplier $\in[1.5,2.0)$&&&&$-$1.195&\\
&&&&(4.559&)\\
Multiplier $\in[2.0,2.5)$&&&&$-$1.740&\\
&&&&(4.266&)\\
Multiplier $\in[2.5,3.0)$&&&&3.162&\\
&&&&(4.450&)\\
Multiplier $\in[3.0,3.5)$&&&&$-$4.195&\\
&&&&(4.252&)\\
Constant&&25.833&\sym{***}&30.478&\sym{***}\\
&&(4.000&)&(2.974&)\\
\(N\) (Spectators)&&390&&390&\\
  \midrule
			\end{tabular}
		\end{center}
		\begin{singlespace}  \vspace{-.5cm}
			\noindent \justify \footnotesize
		\textbf{Notes:} This table shows spectators' perceived effort of workers assigned to each spectator for the luck opportunities condition. Recall that workers are randomly assigned an effort multiplier $\in[1,4]$ as a rate of return on the number of correct encryptions completed in 5 minutes. The omitted category in column (2) is multiplier~$\in [3.5, 4.0]$. Heteroskedasticity-robust standard errors clustered at the spectator level in parentheses. $^{***}$, $^{**}$ and $^*$ denote significance at the 0.1\%, 1\%, and 5\% level, respectively. 
		\end{singlespace} 	
	}
\end{table}

\clearpage
\begin{table}[H]\caption{Fraction redistributed on polynomials of the multiplier difference} \label{tab:redist-poly}
	
	{\footnotesize
		
		\begin{center}
			\newcommand\w{1.2}
			\begin{tabular}{l@{}lR{\w cm}@{}L{0.43cm}R{\w cm}@{}L{0.43cm}R{\w cm}@{}L{0.43cm}R{\w cm}@{}L{0.43cm}R{\w cm}@{}L{0.43cm}R{\w cm}@{}L{0.43cm}R{\w cm}@{}L{0.43cm}}
				\midrule
				&& \multicolumn{10}{c}{Outcome: Fraction of earnings redistributed}  \\
				\cmidrule{3-12} 
				&& (1) && (2) && (3) && (4) && (5) \\ \midrule  
				
				\addlinespace
				(Multiplier difference)$^1$&&0.041&\sym{***}&0.038&\sym{***}&0.042&\sym{***}&0.037&\sym{*}&0.036&\sym{*}\\
&&(0.007&)&(0.010&)&(0.011&)&(0.015&)&(0.015&)\\
(Multiplier difference)$^2$&&&&0.001&&0.009&&0.012&&0.004&\\
&&&&(0.003&)&(0.006&)&(0.009&)&(0.015&)\\
(Multiplier difference)$^3$&&&&&&$-$0.003&&$-$0.002&&0.002&\\
&&&&&&(0.002&)&(0.004&)&(0.009&)\\
(Multiplier difference)$^4$&&&&&&&&$-$0.001&&0.001&\\
&&&&&&&&(0.002&)&(0.002&)\\
(Multiplier difference)$^5$&&&&&&&&&&$-$0.001&\\
&&&&&&&&&&(0.001&)\\
\(N\) (Redistributive decisions)&&2,328&&2,328&&2,328&&2,328&&2,328&\\
  \midrule
			\end{tabular}
		\end{center}
		\begin{singlespace}  \vspace{-.5cm}
			
			\noindent \justify \textbf{Notes:} This figure shows the average redistribution (from the winner's earnings to the loser) as a function of polynomials of the difference between the multiplier of the winner and the loser ($m_1 - m_2$) in the lucky opportunities environment. Heteroskedasticity-robust standard errors clustered at the spectator level in parentheses. $^{***}$, $^{**}$ and $^*$ denote significance at the 0.1\%, 1\%, and 5\% level, respectively. 
		\end{singlespace} 	
	}
\end{table}

\clearpage
\begin{table}[htpb!]\caption{Determinants of redistribution} \label{tab:redist-determ-pi}
	{\footnotesize
		\begin{center}
			\newcommand\w{2}
			\begin{tabular}{l@{}lR{\w cm}@{}L{0.43cm}R{\w cm}@{}L{0.43cm}R{\w cm}@{}L{0.43cm}R{\w cm}@{}L{0.43cm}R{\w cm}@{}L{0.43cm}R{\w cm}@{}L{0.43cm}}
				\midrule
				&& \multicolumn{6}{c}{Outcome: Fraction of earnings redistributed}  \\
				\cmidrule{3-8} 
				&& (1) && (2) && (3)   \\ \midrule  
				
				\multicolumn{6}{l}{\hspace{-1em} \textbf{Panel A. Baseline Lucky Opportunities Condition}}  \\  \addlinespace 		
				Multiplier difference&&0.041&\sym{***}&&&0.028&\sym{**}\\
&&(0.007&)&&&(0.010&)\\
Prob. best performer lost&&&&0.020&\sym{***}&0.007&\\
&&&&(0.004&)&(0.005&)\\
Constant&&0.202&\sym{***}&0.263&\sym{***}&0.223&\sym{***}\\
&&(0.014&)&(0.014&)&(0.020&)\\
\(N\) (Redistributive decisions)&&2,328&&2,328&&2,328&\\
  \midrule

				\multicolumn{6}{l}{\hspace{-1em} \textbf{Panel B. Information Treatment}}  \\  \addlinespace 		
				Multiplier difference&&0.041&\sym{***}&&&0.028&\sym{**}\\
&&(0.007&)&&&(0.010&)\\
Prob. best performer lost&&&&0.020&\sym{***}&0.007&\\
&&&&(0.004&)&(0.005&)\\
Constant&&0.202&\sym{***}&0.263&\sym{***}&0.223&\sym{***}\\
&&(0.014&)&(0.014&)&(0.020&)\\
\(N\) (Redistributive decisions)&&2,328&&2,328&&2,328&\\
  \midrule

			\end{tabular}
		\end{center}
		\begin{singlespace}  \vspace{-.5cm}
			
			\noindent \justify \textbf{Notes:} This figure shows the average redistribution (from the winner's earnings to the loser) as a function of the difference between the multiplier of the winner and the loser ($m_1 - m_2$) and the probability that the best performer lost ($1-\pi$) in the lucky opportunities environment. Heteroskedasticity-robust standard errors clustered at the spectator level in parentheses. $^{***}$, $^{**}$ and $^*$ denote significance at the 0.1\%, 1\%, and 5\% level, respectively. 
		\end{singlespace} 	
	}
\end{table}

 \clearpage
    \section{Additional Results and Analysis}\label{sec:add-analysis}

\setcounter{table}{0}
\setcounter{figure}{0}

\subsection{Luck Perceptions Survey}\label{sec:luck-survey}

We surveyed over 1,000 panelists from the NY Fed's \textit{Survey of Consumer Expectations} in February 2023 about the most important types of luck in determining people's life earnings. We conducted this survey by asking panelists to rank the importance of various luck events that occur in people's lives, including a balanced number of items exemplifying opportunity luck (e.g., access to education, the family or society someone is born into) and outcome luck (e.g., unforeseen events such as financial booms or busts, unexpected windfalls such as winning the lottery).

\begin{figure}[h!]
    \centering\caption{Important Factors for Determining People's Life Earnings}
    \label{fig:lucklikert}
    \includegraphics{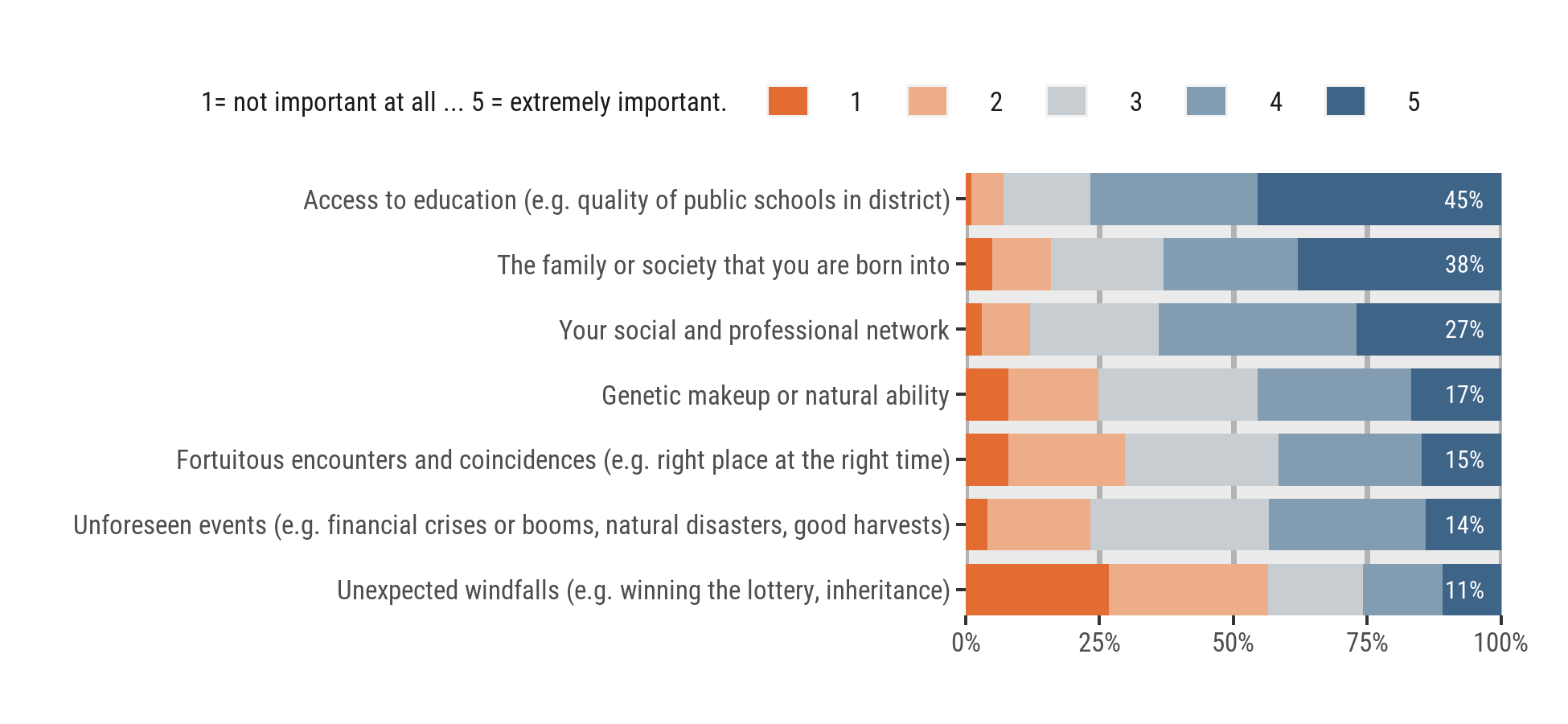}
    \begin{singlespace}  \vspace{-.5cm}
			\noindent \justify \footnotesize \textbf{Notes:} 
   Likert-type responses from the \textit{Survey of Consumer Expecations} ($N=1013$): ``Circumstances beyond a person’s control—or, their luck—can impact their life earnings. Please indicate on a scale from 1 to 5 how important you think the following factors are for determining people’s life earnings \ldots $1 =$ not important at all \ldots $5 =$ extremely important.'' The order of factors displayed to the survey respondent is randomized. This question is revealed \textit{before} the open-ended question eliciting important examples of luck in life.
		\end{singlespace} 	
\end{figure}

Most respondents agree that access to education, the family or society a person is born into, and their social or professional network are important in determining people's lifetime earnings (Figure \ref{fig:lucklikert}). These are all examples of luck as an opportunity, whereby luck confers an individual with a relative advantage. Around 50 percent of people state that unexpected windfalls are unimportant in determining life outcomes---an example of outcome luck, which is independent of effort. Other forms of outcome luck, such as fortuitous encounters and unforeseen events, are rated as less important.

To support these survey responses, we also elicited open-text responses for the most important examples of luck in an individual's life.\footnote{The exact question text was as follows: ``It is often said that luck is important for success in life. In your view, what are the three most important examples of luck in an individual's life?''. We asked this question before presenting panelists with the Likert-scale question to avoid priming them about certain factors.} Overall, 70.4 percent of respondents listed at least one example of lucky opportunities. In contrast, only 26.3 percent of respondents mentioned at least one luck factor that is independent of an individual's effort, such as winning the lottery, being at the right place at the right time, or adverse weather events. A topic analysis of the text largely supports our main survey findings. Panelists indicated examples of opportunity luck in greater frequency. Using a Latent Dirichlet Allocation (LDA) analysis, we find that a majority of topics are related to luck that is non-separable from effort in determining life earnings (e.g., ``family,'' ``education,''  ``birth,''  ``health,'' ``location,'' and ``wealth'').

\subsection{Formal Derivation of the Properties of $\pi$}\label{sec:convexity}

In this section, we derive the properties of how opportunity luck maps to the likelihood that the winner was based on merit. Recall that $\pi = 1$ if the disadvantaged (i.e., lower-multiplier) worker wins. Therefore, we restrict attention here to the more interesting case in which the advantaged worker won. Without loss of generality, suppose the winner is worker $1$ so that $m_1\geq m_2$. As before, we write $m=m_1/m_2$ to denote the relative advantage of worker $1$. 

Let $x(e_1,e_2)$ denote a systematic measure of relative effort comparing the disadvantaged worker to the advantaged worker and denote the distribution of $x$ by $F$. Let $\hat{x}=x(e,e)$ be the equal effort cutoff. Further, let $x^*(m)$ denote the relative effort threshold that $x(e_1,e_2)$ needs to exceed so that worker $2$ wins. In our experiment, $x(e_1,e_2)=e_2/e_1$, $x^*(m)=m$ and $\hat{x}=1$.

Using this notation, we can rewrite the expression for $\pi$ given in the main text as follows: 
\begin{align}
    \pi(m)= \Pr \left(x<\hat{x} \vert x \leq x^*(m)\right)=\frac{F\left(\hat{x}\right)}{F\left(x^*(m)\right)}=\frac{1/2}{F(m)}.
\end{align}

Taking the derivative yields $d \pi(m)/d m=-f(m)/F(m)^2<0$, which shows that $\pi$ decreases in $m$ as stated in the main text. Next, consider the second derivative:
\begin{align}
    \frac{d^2 \pi(m)}{d m^2 }=\frac{-f'(m)F(m)^2-2F(m)f(m)(-f(m))}{F(m)^4}.
\end{align}
Thus, convexity of $\pi$ follows if 
\begin{align}\label{app:equ:convexity}
    -f'(m)F(m)+2f(m)^2\geq 0 \qquad \Leftrightarrow  \qquad 2f(m)^2\geq f'(m)F(m).
\end{align}
The second inequality is implied by log-concavity of $F$. Thus, to prove that $\pi$ is convex in $m$, it is sufficient to show that $F$ is log-concave. Notably, $F$ is log-concave if $x=e_2/e_1$ is log-normally distributed \citep[see Table 3 in][for example]{bagnoli2006log}. Since both $e_1$ and $e_2$ are log-normal, it follows that $\log (e_2/e_1)=\log(e_2)-\log(e_1)$ is the difference of two normally distributed variables ($\log(e_2)$ and $\log(e_1)$), which is itself normally distributed. That is,  $x=e_2/e_1$ is log-normally distributed and as a result, $\pi$ is convex. 

Expression \eqref{app:equ:convexity} permits a straightforward assessment of the properties of $\pi(m)$ for other effort distributions. For example, for a uniform distribution over $[0,1]$, notice that the CDF $F$ for $x=e_2/e_1\geq 1$ is given by $1/2+1/2 (1-1/x)$. Based on that expression, it is easy to verify that (\ref{app:equ:convexity}) holds, implying convexity of $\pi$. In fact, the result readily extends to any uniform distribution over $[0,c]$. To see this, simply multiply any relative advantage $m$ by $c$ so that the effective relative advantage $c\cdot m$ is uniformly distributed over $[0,c]$. To replicate the analysis above, note that $f$ does not change since the $c$ term cancels out when taking the ratio. The only change is that $x^*(m)=c m$ now. However, the constant $c$ does not alter the necessary condition  (\ref{app:equ:convexity}) for $\pi$ to be convex. Therefore, even in situations in which effort is uniformly dispersed in the population, there is a convex mapping from relative opportunities to the likelihood of success. 

Similar conclusions emerge if we instead consider additive lucky headstarts instead of multipliers. Suppose the advantaged worker $1$ receives a relative headstart $b>0$. To analyze this case, we can simply redefine $x(e_1,e_2)=e_2-e_1$, and therefore the equal-effort threshold becomes $\hat{x}=0$. Since the effort threshold required to win is still a linear function of the relative advantage, i.e., $x^*(b)=b$, we can simply substitute $m$ with $b$ in the analysis above. If worker effort is normally distributed, then $x(e_1,e_2)$ is also normally distributed, and therefore log-concavity holds via equation B2.\footnote{While assuming a log-normal distribution of effort may be preferred to assuming normal, the distribution of $e_2-e_1$ is analytically intractable if $e_1$ and $e_2$ are log-normally distributed.} As in the case above, we obtain the same prediction when effort is uniformly distributed over any interval $[0,c]$. To see why, notice that $e_2-e_1$ follows a triangular distribution in this case, whose density is concave and thus log-concave. By \cite{bagnoli2006log}, log-concavity of the density implies log-concavity of the distribution function $F$, which is sufficient for $\pi$ being convex by the above arguments. Hence, convexity is also an inherent feature of lucky headstarts under reasonable assumptions about the effort distribution. While the exact properties of $\pi$ will depend on the exact distribution of worker effort, we confirm in the following section that convexity holds for both additive and multiplicative opportunities for the empirical distribution of effort we observe. 

\subsection{Alternative Experiment with Additive Opportunities or ``Headstarts''}\label{sec:headstarts}

The critical feature of opportunity luck that we emphasize is that it induces a convex relationship between opportunities and how impactful merit is on individual outcomes. Intuitively, this relationship implies that relatively small differences in opportunities can substantially impact outcomes. Our theoretical framework and experimental design consider the case of productivity multipliers that amplify or dampen the returns to exerting effort. Panel B of Figure \ref{fig:worker-eff} shows that this type of opportunity luck indeed generates a convex association between the relative opportunities (the ratio of multipliers, $m$) and the likelihood that the winner was due to effort rather than luck, $\pi$. 

Alternative forms of opportunity luck are similarly convex in terms of their impact on the outcome. At the end of Section \ref{sec:convexity}, we described a hypothetical ``lucky headstarts'' treatment in which opportunity luck instead takes an additive form. Formally, suppose that workers $i$ and $j$ receive additive boosts, $b_1$, $b_2$ $\in [0,1,...B]$. The final score for each worker is simply the sum of their headstart and the number of solved encryptions: $b_j + e_j$ for $j\in\{1,2\}$. Without loss of generality, suppose that worker 1 is the winner of the pair and define relative boost as $b \equiv b_1 - b_2$. 

\begin{figure}[h!]
    \centering
    \caption{Additive vs. Multiplicative Opportunities}    \label{fig:headstarts}
    \includegraphics[scale=0.6]{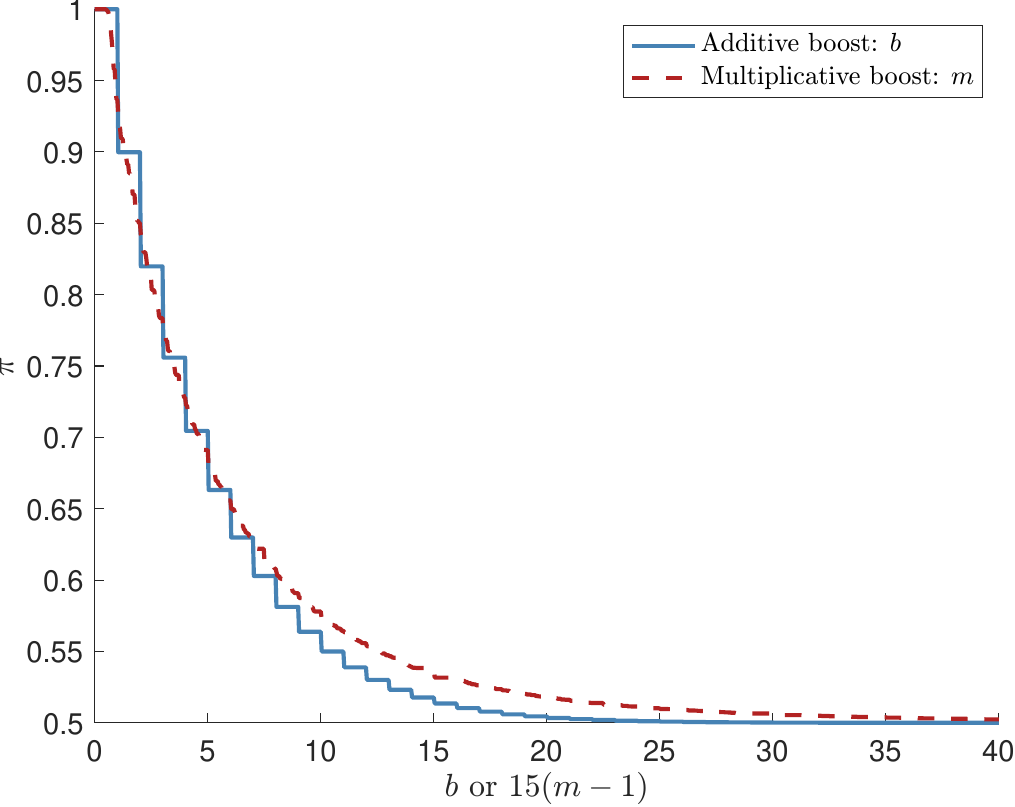}
    	\singlespacing \justify \footnotesize
		\textbf{Notes}: 
		This figure shows the fraction of paired workers in which the worker who won the match completed more encryptions. We determine the winner by comparing the final scores and selecting the worker with the higher score. The solid blue line depicts the case with additive score boosts in which the final score is the number of correct encryptions plus an additive boost. The dashed red line depicts the case with multiplicative score boosts which the final score is the number of correct encryptions times a score multiplier. Values near 0.5 are worker matches in which luck has a greater influence on the final outcome. Values near 1.0 are worker matches in which luck has little influence on the final outcome.
\end{figure}

We can map the relative headstart, $b$, to the likelihood that the winner solved more encryptions, $\pi$, using the same procedure that we used for the relative multiplier ratio $m$ in the main text. Specifically, for a given $b$, we can take the empirical effort distribution from our worker task and consider all possible pairings. For each pairing, we can assign a relative boost of $b$ to either worker and then compute the empirical likelihood that the winner was the worker who solved more encryptions. The solid blue line in Figure \ref{fig:headstarts} plots the empirical $\pi$ as a function of the relative boost $b$. It depicts a highly convex relationship between the relative headstart and its impact on the outcome. A single-point headstart leads to a drop in $\pi$ from 100 to 90 percent. A two-score advantage leads to an additional eight percentage point drop in $\pi$, while a three-score advantage leads to an additional six percentage point drop. In contrast, moving from a ten-score advantage to an 11-score advantage only leads to a one percentage point change in $\pi$. 

The dashed red line in Figure \ref{fig:headstarts} reproduces the relationship between the relative multiplier ratio $m$ and $\pi$ from our lucky opportunities treatment. The additive and multiplicative forms of opportunity luck are similar in their relationship with $\pi$. If anything, the additive headstart case is slightly more convex in $\pi$ than the multiplicative one. More broadly, this highlights that the convex relationship between opportunities and the importance of luck for the outcome is not due to the multiplicative nature of our lucky opportunities environment.

\subsection{Extensive vs. Intensive Margin of Redistribution}\label{sec:extensive_intensive}

To further understand changes in redistributive behavior under unequal opportunities, we distinguish between the intensive and extensive margins of redistribution. The extensive margin refers to whether or not spectators redistribute any earnings. The intensive margin refers to how much spectators redistribute, conditional on redistributing anything. Each of these channels has a mechanical effect on the aggregate level of redistribution and how responsive redistribution is to changes in the likelihood that the best performer lost. For example, a higher share of spectators who never redistribute shifts the average level of redistribution down and makes redistribution less responsive to $1 - \pi$.

We first explore whether the likelihood of redistributing a positive amount differs across luck environments. In Table \ref{tab:libert-det}, we estimate regressions where the outcome is a binary variable equal to one if a spectator never redistributes anything across all 12 decisions. Column (1) shows that 9.6 percent of spectators never redistribute when there are lucky outcomes. In contrast, 15.9 percent of spectators do not redistribute when there are lucky opportunities. The difference of 6.3 percentage points is statistically significant ($p<0.01$) and equates to a 66 percent increase in the share of spectators who never redistribute. Thus, the extensive margin of redistribution is substantially lower when there are unequal opportunities than when chance directly influences outcomes. 

\begin{table}[H]
	\caption{Fraction of spectators who do not redistribute across conditions} \label{tab:libert-det}
	\vspace{-15pt}
	{\footnotesize
		\begin{center}		
			\newcommand\w{1.9}
			\begin{tabular}{l@{}lR{\w cm}@{}L{0.43cm}R{\w cm}@{}L{0.43cm}R{\w cm}@{}L{0.43cm}R{\w cm}@{}L{0.43cm}R{\w cm}@{}L{0.43cm}R{\w cm}@{}L{0.43cm}R{\w cm}@{}L{0.43cm}R{\w cm}@{}L{0.43cm}R{\w cm}@{}L{0.43cm}}
				\midrule
				\addlinespace					
				&& \multicolumn{8}{c}{Outcome: $=1$ if spectator does not redistribute in any round}  \\
				\cmidrule{3-10}
				
				&& (1) && (2) && (3) && (4)   \\
				\midrule 
				Lucky Opportunities&&0.063&\sym{**}&0.063&\sym{**}&0.063&&0.068&\\
&&(0.024&)&(0.024&)&(0.034&)&(0.035&)\\
Info $\pi$&&&&$-$0.011&&$-$0.010&&$-$0.014&\\
&&&&(0.024&)&(0.030&)&(0.031&)\\
Lucky Opportunities $\times$ Info $\pi$&&&&&&$-$0.002&&$-$0.004&\\
&&&&&&(0.048&)&(0.049&)\\
Constant&&0.096&\sym{***}&0.102&\sym{***}&0.102&\sym{***}&0.115&\\
&&(0.015&)&(0.019&)&(0.022&)&(0.109&)\\
\(N\) (Redistributive decisions)&&9,408&&9,408&&9,408&&9,396&\\
  
				\midrule
				Spectator-level controls && No && No && No && Yes \\
				\midrule
			\end{tabular}
		\end{center}
		\begin{singlespace}  \vspace{-.5cm}
			\noindent \justify \textbf{Notes:} The dependent variable is the fraction of spectators who do not redistribute in any round. In column 4, we control for age, gender, marital status, number of children in the household, educational attainment, numerical literacy, race, indicators for working part-time and full-time, homeownership, income, region, the time spectators spent on the experiment, indicators for passing the comprehension and attention checks, an indicator that equals one if the spectator completed the survey on a mobile device, the probability that the winner exerted more effort on each worker-pair, and round number fixed effects (to control for possible fatigue effects). Heteroskedasticity-robust standard errors clustered at the spectator level in parentheses. $^{***}$, $^{**}$ and $^*$ denote significance at the 0.1\%, 1\% and, 5\% level, respectively.
		\end{singlespace} 	
	}
\end{table}

Next, we analyze differences in redistribution decisions among spectators that redistribute some amount in at least one of their 12 decisions. Table \ref{tab:redist-pi-nolibert} reproduces the analysis in Panels A--B of Table \ref{tab:redist-pi} but excludes spectators who do not redistribute anything in all 12 decisions. We continue to find differences in the average level of redistribution across the lucky outcomes and lucky opportunities conditions for this sub-sample: On average, spectators redistribute 30.7 percent when luck emerges through coin flips (column 1) and 28.0 percent when luck arises through productivity multipliers (column 2). This difference is statistically significant at the ten percent level (column 3). 

We also continue to find that spectators are less sensitive to changes in the probability that the best performer lost in the lucky opportunities condition. In Panel B, columns (1) and (2) show that a ten percentage point increase in $1-\pi$ increases redistribution by 4.1 percentage points in the lucky outcomes condition and 2.4 percentage points in the lucky opportunities condition. This difference in slope is statistically significant ($p < 0.01$, column 3). Notably, the magnitude of this difference is similar to the baseline estimates in Table \ref{tab:redist-pi}. Thus, the diminished overall sensitivity to luck that we observe when luck stems from unequal opportunities is not merely due to more spectators deciding to redistribute nothing. Instead, changes in the responsiveness to the importance of luck in determining workers' outcomes among spectators who redistribute drive the result. 

\begin{table}[H]\caption{Fraction redistributed as a function of $\pi$ for spectators who redistribute something}
\vspace{-15pt}
\label{tab:redist-pi-nolibert}
	{\footnotesize 
		\begin{center}
			\newcommand\w{2.5}
			\begin{tabular}{l@{}lR{\w cm}@{}L{0.43cm}R{\w cm}@{}L{0.43cm}R{\w cm}@{}L{0.43cm}R{\w cm}@{}L{0.43cm}R{\w cm}@{}L{0.43cm}R{\w cm}@{}L{0.43cm}}
				\midrule
				&& \multicolumn{6}{c}{Outcome: Fraction of earnings redistributed}  \\
				\cmidrule{3-8} 
				&& Lucky && Lucky && \multirow{2}{*}{Difference} \\
				&& Outcomes && Opportunities  &&  \\
				&& (1) && (2) && (3)  \\
				\midrule 
				\multicolumn{8}{l}{\hspace{-1em} \textbf{Panel A. Average redistribution}}  \\  \addlinespace 
				Constant&&0.307&\sym{***}&0.280&\sym{***}&0.027&\\
&&(0.009&)&(0.012&)&(0.015&)\\
\(N\) (Redistributive decisions)&&2,124&&1,944&&4,068&\\
  \midrule
            	\multicolumn{8}{l}{\hspace{-1em} \textbf{Panel B. Linear slope}}  \\  \addlinespace 
				1$-$$\pi$&&0.041&\sym{***}&0.024&\sym{***}&0.018&\sym{**}\\
&&(0.005&)&(0.004&)&(0.007&)\\
Constant&&0.368&\sym{***}&0.315&\sym{***}&0.053&\sym{**}\\
&&(0.011&)&(0.014&)&(0.018&)\\
\(N\) (Redistributive decisions)&&2,124&&1,944&&4,068&\\
  \midrule
			\end{tabular}
		\end{center}
		\begin{singlespace}  \vspace{-.5cm}
			\noindent \justify \footnotesize \textbf{Notes:} Panel A shows the mean share of earnings redistributed under lucky outcomes (column~1), lucky opportunities (column~2), and the difference (column~3). Panel B shows estimates of redistribution as a linear function of the probability that the winner was the best performer lost ($1 - \pi$) on each treatment. The sample is restricted to spectators who redistributed a strictly positive amount in at least one of their 12 decisions. Heteroskedasticity-robust standard errors clustered at the spectator level in parentheses. $^{***}$, $^{**}$ and $^*$ denote significance at the 0.1\%, 1\%, and 5\% level, respectively.
		\end{singlespace} 	
	}
\end{table}

\subsection{Anticipatory Effort Responses}\label{sec:anticipatory-responses}

In the main text, we show that whether workers learn their multipliers before or after working on the task has no impact on spectators' redistribution decisions. While this removes much of the scope for different beliefs about workers' effort responses, spectators might still expect the distribution of effort levels under lucky opportunities to be meaningfully different from lucky outcomes environments. In particular, spectators may believe that a worker who learns the rules of how effort multipliers determine outcomes may be motivated to exert different levels of effort independent of their knowledge about their assigned multiplier (i.e., under the ex-post opportunities condition) than a worker who learns how a coin flip impacts outcomes. For example, workers in the ex-post lucky opportunities environment might work harder to insure against the possibility of drawing a bad multiplier, which, in turn, could shape redistribution preferences if spectators anticipate such behavior. Alternatively, spectators might suspect that workers believe their efforts matter less and, thus, become less motivated to exert effort when there are lucky outcomes. 

To control for and test this mechanism experimentally, we vary the timing of when workers learn how luck plays a role in determining outcomes in subtreatments. In the ``rules-before'' condition, we inform workers that effort multipliers or a coin flip will influence the outcome \textit{before} they start the task. In the ``rules-after'' condition, we inform workers that multipliers or a coin flip will influence the outcome \textit{after} they complete the task. Crucially, spectators in the rules-after treatments in both the lucky outcomes and ex-post lucky opportunities conditions knew that workers had identical information before beginning the task. Between these two conditions, there is thus no scope for differences in beliefs about the distribution of worker effort. 

\begin{figure}[H]\caption{Redistribution and awareness of rules in the ex-post lucky opportunities and lucky outcomes conditions} \label{fig:redist-rules}
			\begin{centering}	
		\protect
			\begin{minipage}{.48\textwidth}
				\captionof*{figure}{Panel A. Ex-Post Lucky Opportunities}
    \vspace{-7pt}
				\includegraphics[width=\linewidth]{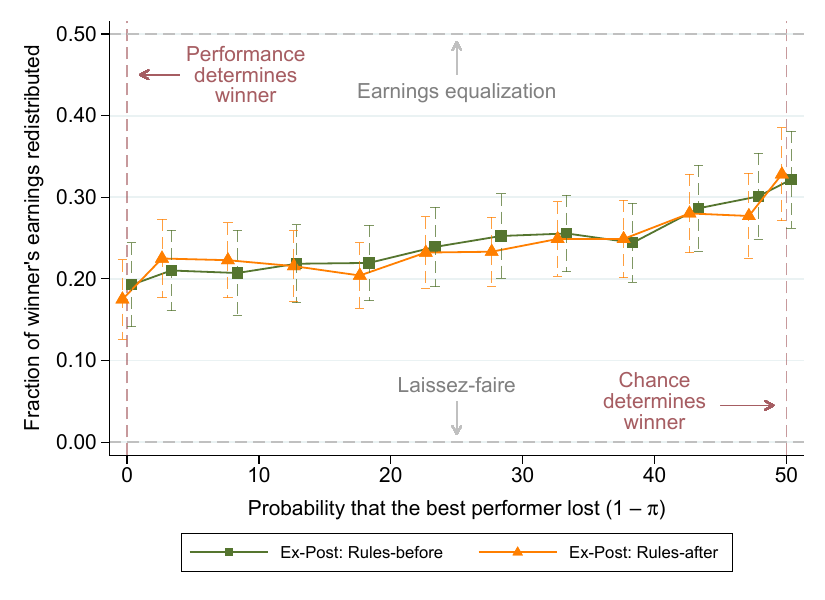}
			\end{minipage}\hspace{1em}
			\begin{minipage}{.48\textwidth}
				\captionof*{figure}{Panel B. Lucky Outcomes}
        \vspace{-7pt}
                \includegraphics[width=\linewidth]{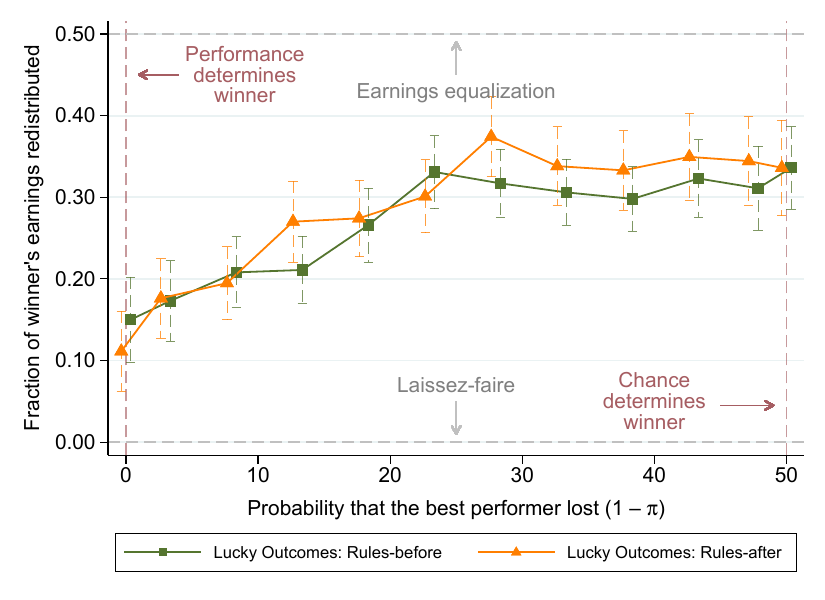}
			\end{minipage}
			\par
	\end{centering}
     \vspace{-7pt}
	\singlespacing \justify \footnotesize
	\textbf{Notes:} This figure shows the average share of earnings redistributed between workers as a function of the likelihood that the better performer lost, split by our rules-before and rules-after subtreatments. In rules-before, workers are aware of their multiplier prior to their encryption task, and in rules-after, workers are aware of their multiplier after completing their encryption task. Panel A depicts data from the ex-post lucky opportunities condition, and Panel B depicts data from the lucky outcomes condition.  
\end{figure}

Figure \ref{fig:redist-rules} plots the average redistribution for each $\pi$ bin separately for our rules-before and rules-after subtreatments. Panel A shows that the redistribution decisions of spectators in the ex-post lucky opportunities condition are very similar and do not depend on workers learning about how luck matters before or after working. Similarly, Panel B shows that whether the rules are revealed before or after working has no impact on the overall pattern of redistribution in the lucky outcomes environment. 

\begin{figure}[H]\caption{Redistribution in the rules-after treatments: lucky outcomes vs ex-post lucky opportunities} \label{fig:redist-ep-cf-rules-after}
\vspace{-7pt}
	\centering
	\includegraphics[width=.75\linewidth]{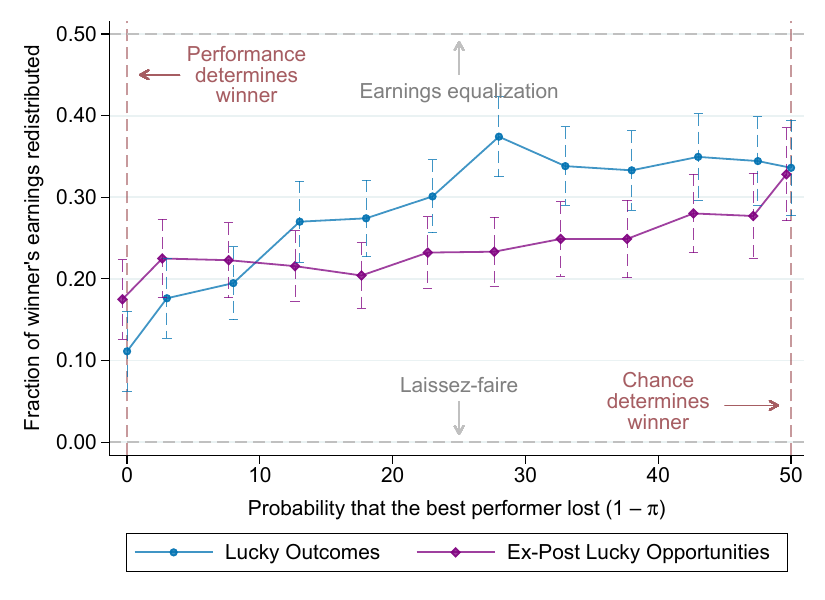}
	\footnotesize
	\singlespacing \justify \footnotesize
 \vspace{-12pt}
	\textbf{Notes:} This figure shows the average share of earnings redistributed between workers as a function of the likelihood that the winner exerted more effort for the rules-after subtreatments for lucky outcomes and ex-post lucky opportunity luck. Note that these conditions are observationally identical to workers until after they perform their tasks. 
\end{figure}

Figure \ref{fig:redist-ep-cf-rules-after} compares average redistribution in the lucky outcomes and ex-post lucky opportunities environments for only the rules-after subtreatments. Even when workers faced identical information prior to exerting effort, spectators redistribute less when luck manifests itself through unequal opportunities than directly via a coin flip. Moreover, spectators continue to be less responsive to changes in the importance of luck. Table \ref{tab:redist-pi-cf-ep-norules} re-estimates our main specifications in Table \ref*{tab:redist-pi} from the main text but only compares lucky outcomes and ex-post lucky opportunities for the rules-after scenario. We continue to find significant differences in the level and slope of redistribution. The estimated coefficients are similar in magnitude to the baseline results. 

Finally, we compare spectators' stated beliefs about average worker effort across the ex-post lucky opportunities and lucky outcomes conditions. We find no differences across these conditions: The median number of tasks spectators believe workers completed is 20 encryptions in both the lucky opportunities and lucky outcomes environment. We also find no differences based on whether workers learned about the tournament rules before or after completing the task: the median number of tasks spectators believe workers completed is 20 in both rules-before variant of lucky outcomes and ex-post lucky opportunities. Overall, we find no evidence that differences in spectators' beliefs about the distribution of effort can explain the differences in redistribution across luck environments. 

\clearpage
\begin{table}[htpb!]
\caption{Fraction redistributed as a function of $\pi$ in ex-post lucky opportunities and lucky outcomes conditions (only rules-after)} \label{tab:redist-pi-cf-ep-norules}
\vspace{-15pt}
	{\footnotesize 
		\begin{center}
			\newcommand\w{2.4}
			\begin{tabular}{l@{}lR{\w cm}@{}L{0.43cm}R{\w cm}@{}L{0.43cm}R{\w cm}@{}L{0.43cm}R{\w cm}@{}L{0.43cm}R{\w cm}@{}L{0.43cm}R{\w cm}@{}L{0.43cm}}
				\midrule
				&& \multicolumn{6}{c}{Outcome: Fraction of earnings redistributed}  \\
				\cmidrule{3-8} 
				&& Lucky    && Ex-Post Lucky  && Difference \\
				&& Outcomes && Opportunities && \\
				&& (1) && (2) && (3)  \\
				\midrule 
				\multicolumn{8}{l}{\hspace{-1em} \textbf{Panel A. Average redistribution}}  \\  \addlinespace 
				Constant&&0.284&\sym{***}&0.241&\sym{***}&0.043&\\
&&(0.016&)&(0.018&)&(0.024&)\\
\(N\) (Redistributive decisions)&&1,164&&1,152&&2,316&\\
 
				\midrule 
				\multicolumn{8}{l}{\hspace{-1em} \textbf{Panel B. Linear slope}}  \\  \addlinespace 
				$\pi$&&0.041&\sym{***}&0.020&\sym{***}&0.020&\sym{*}\\
&&(0.006&)&(0.005&)&(0.008&)\\
Constant&&0.344&\sym{***}&0.271&\sym{***}&0.073&\sym{*}\\
&&(0.020&)&(0.021&)&(0.029&)\\
\(N\) (Redistributive decisions)&&1,164&&1,152&&2,316&\\
  \midrule
				\multicolumn{8}{l}{\hspace{-1em} \textbf{Panel C. Non-parametric estimation}}  \\  \addlinespace 
				1$-$$\pi=0.00$&&0.111&\sym{***}&0.175&\sym{***}&$-$0.064&\\
&&(0.025&)&(0.025&)&(0.035&)\\
1$-$$\pi\in(0.00,0.05]$&&0.176&\sym{***}&0.225&\sym{***}&$-$0.049&\\
&&(0.025&)&(0.025&)&(0.035&)\\
1$-$$\pi\in(0.05,0.10]$&&0.195&\sym{***}&0.223&\sym{***}&$-$0.028&\\
&&(0.023&)&(0.024&)&(0.033&)\\
1$-$$\pi\in(0.10,0.15]$&&0.270&\sym{***}&0.216&\sym{***}&0.054&\\
&&(0.025&)&(0.022&)&(0.034&)\\
1$-$$\pi\in(0.15,0.20]$&&0.274&\sym{***}&0.204&\sym{***}&0.070&\sym{*}\\
&&(0.024&)&(0.021&)&(0.032&)\\
1$-$$\pi\in(0.20,0.25]$&&0.301&\sym{***}&0.232&\sym{***}&0.069&\sym{*}\\
&&(0.023&)&(0.022&)&(0.032&)\\
1$-$$\pi\in(0.25,0.30]$&&0.374&\sym{***}&0.233&\sym{***}&0.141&\sym{***}\\
&&(0.025&)&(0.022&)&(0.033&)\\
1$-$$\pi\in(0.30,0.35]$&&0.338&\sym{***}&0.249&\sym{***}&0.089&\sym{**}\\
&&(0.025&)&(0.023&)&(0.034&)\\
1$-$$\pi\in(0.35,0.40]$&&0.333&\sym{***}&0.249&\sym{***}&0.084&\sym{*}\\
&&(0.025&)&(0.024&)&(0.035&)\\
1$-$$\pi\in(0.40,0.45]$&&0.349&\sym{***}&0.280&\sym{***}&0.069&\\
&&(0.027&)&(0.025&)&(0.037&)\\
1$-$$\pi\in(0.45,0.50)$&&0.344&\sym{***}&0.277&\sym{***}&0.067&\\
&&(0.028&)&(0.027&)&(0.039&)\\
1$-$$\pi=0.50$&&0.336&\sym{***}&0.328&\sym{***}&0.008&\\
&&(0.030&)&(0.029&)&(0.042&)\\
\(N\) (Redistributive decisions)&&1,164&&1,152&&2,316&\\
  \midrule
			\end{tabular}
		\end{center}
		\begin{singlespace}  \vspace{-.5cm}
			\noindent \justify \footnotesize \textbf{Notes:} This table includes only spectators in the rules-after condition. Column 1 includes only spectators in the lucky outcomes condition. Column 2 includes only spectators under the ex-post lucky opportunities condition. Column 3 shows the difference between columns 1 and 2. Panel A shows average redistribution. Panel B shows a linear approximation between the fraction of earnings redistributed and the likelihood that the winning worker performed better than the losing worker ($\pi$). Panel C shows the relationship between redistribution and the likelihood that the winning worker performed better ($\pi$) split into 11 bins. The omitted category is $\pi = 0.50$. $^{***}$, $^{**}$ and $^*$ denote significance at the 0.1\%, 1\%, and 5\% level, respectively. 
		\end{singlespace} 	
	}
\end{table}

\clearpage 

\subsection{Effort Differences Across the $\pi$ Distribution}\label{sec:effort_pi}

The ex-post lucky opportunities treatment shows that spectators' expectations about workers' potential effort responses do not drive our main treatment effect. Nonetheless, the performance levels of winners and losers conditional on $\pi$ could differ across our luck treatments. This difference might arise because only worker pairs with certain performance levels can be selected for a particular $\pi$ in the lucky opportunities condition. In contrast, workers with any performance level can be selected for a particular $\pi$ in the lucky outcomes condition. For example, when a coin flip determines the winner independently of worker effort, a worker could have won with despite an extremely low performance. Conversely, winning always requires some minimum performance in the lucky opportunities environment. Sophisticated spectators may be aware of these possible differences in the winner-loser performance gap, which could explain their greater propensity to redistribute in the lucky outcomes condition relative to the lucky opportunities condition.\footnote{Note that small differences in the winner-loser performance gap for certain values of $\pi$ across luck environments are the result of the inherent features of outcome luck and opportunity luck, regardless of whether the latter is realized as additive boosters or multipliers. This is because the expected effort gap shrinks linearly in $\pi$ under lucky outcomes. Under lucky opportunities, by contrast, the convex relationship between the winning worker's relative advantage and $\pi$ leads to a convex relationship between $\pi$ and the expected winner-loser effort gap.}

\begin{figure}[H]
\caption{Number of encryptions completed by workers across values of $\pi$} \label{fig:effort_pi}
		{\scriptsize
			\begin{centering}	
				\protect
				\begin{minipage}{.48\textwidth}
					\captionof*{figure}{Panel A. Tasks completed by winners and losers and winner-loser effort gap}
					\includegraphics[width=\linewidth]{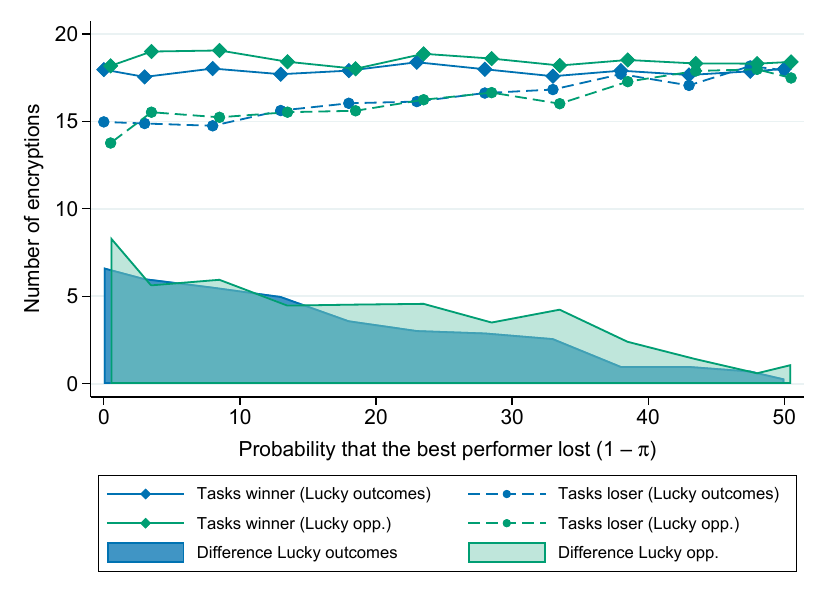}
				\end{minipage}\hspace{1em}
				\begin{minipage}{.48\textwidth}
					\captionof*{figure}{Panel B. Difference in winner-loser effort gap across conditions}				
					\includegraphics[width=\linewidth]{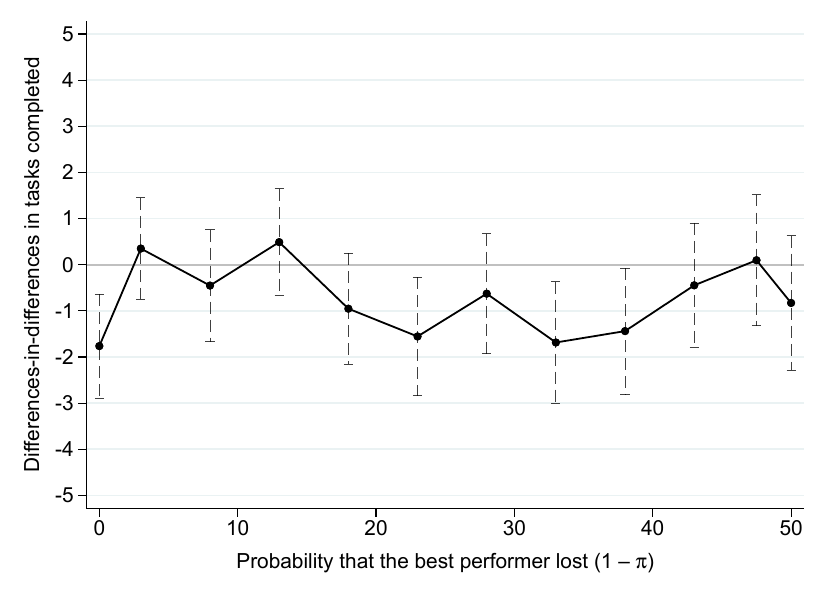}
				\end{minipage} \vspace{.25cm}
				
				\par\end{centering}
			\singlespacing \justify \footnotesize \textbf{Notes:} Panel A shows the average number of encryptions completed across all worker pairs as a function of the luck treatment and the value of $\pi$. Shaded areas denote the average winner-loser effort gap, as measured by the difference between the number of encryptions completed by the winner and the loser. Panel B shows the difference between the winner-loser effort gap in the lucky outcomes condition and the lucky opportunities condition. Dashed lines denote 95 percent confidence intervals.
		}
	\end{figure}

We provide several pieces of evidence against this explanation. First, across all values of $(1-\pi)$, the winner-loser performance gap is quantitatively similar in the two conditions (Figure \ref{fig:effort_pi}). Panel A of Figure \ref{fig:effort_pi} depicts the average number of encryptions completed by winners and losers in each luck treatment as a function of $\pi$, as well as the difference between the number of tasks completed by the winners and losers in each condition or, equivalently, the ``winner-loser effort/ performance gap.'' Panel B of Figure \ref{fig:effort_pi} shows the difference between the winner-loser effort gap in the lucky opportunities condition and the lucky outcomes condition. 

There are no meaningful differences in the number of encryptions completed by winners and losers across luck treatments. On average, winners in the lucky outcomes condition completed a similar number of encryptions as winners in the lucky opportunities condition across the entire range of $\pi$ bins. For example, in the lucky outcomes condition, the average number of encryptions completed by winners ranges from 17.5 to 18.3 (depending on the value of $\pi$), whereas, in the lucky opportunities condition, the corresponding value ranges from 18.0 to 19.1. Similarly, losers in the two conditions completed a similar number of encryptions. As $\pi$ increases, the winner-loser effort gap increases in both luck environments. Crucially, these effort differences are similar between our luck environments. For all values of $\pi$, the difference in the winner-loser effort gap between the two conditions is smaller than two encryptions (Figure \ref{fig:effort_pi}, Panel B).

\begin{figure}[H]\caption{Redistribution and differences in the winner-loser effort gap} \label{fig:red_effort}
		{\scriptsize
			\begin{centering}	
				\protect
				\begin{minipage}{.48\textwidth}
					\captionof*{figure}{Panel A. Earnings redistributed and differences in encryptions completed by winners and losers}
					\includegraphics[width=\linewidth]{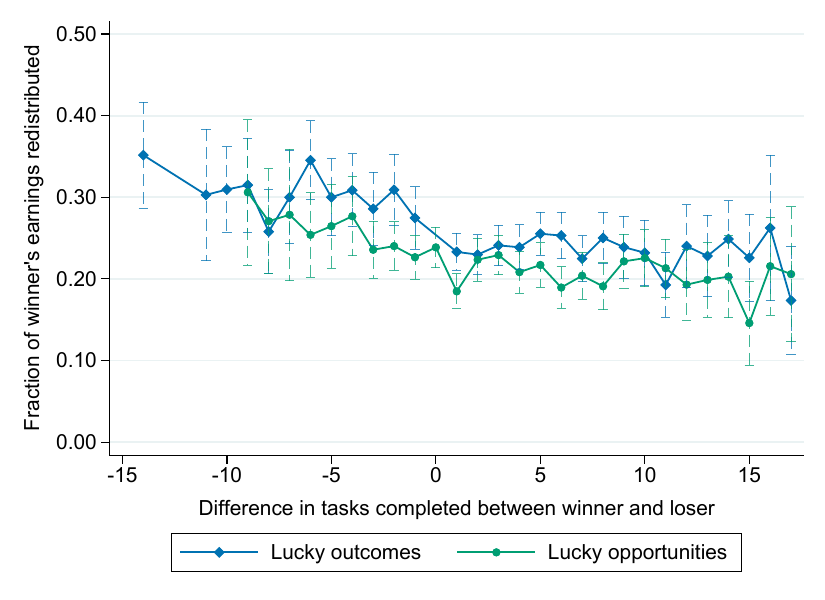}
				\end{minipage}\hspace{1em}
				\begin{minipage}{.48\textwidth}
					\captionof*{figure}{Panel B. Redistributive gap predicted by differences in winner-loser effort gap}				
					\includegraphics[width=\linewidth]{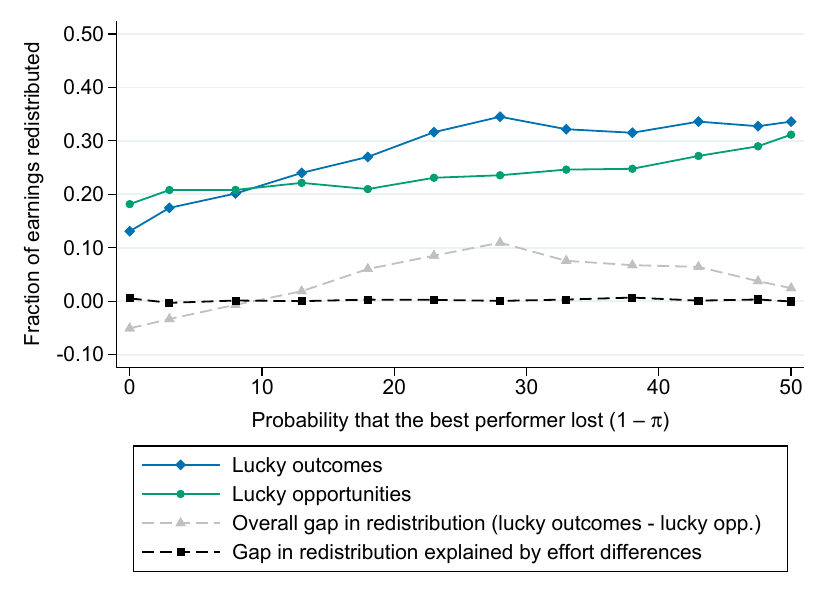}
				\end{minipage} 
				\par\end{centering}
			\singlespacing \justify \footnotesize \textbf{Notes:} Panel A shows the average share of earnings redistributed between workers (from the higher-earning winner to the lower-earning loser) relative to the difference in encryptions completed between the winner and the loser. Panel B shows the average share of earnings redistributed between workers relative to the likelihood that the winner exerted more effort across the two luck conditions, as well as the difference between the fraction of earnings redistributed in the two conditions. This panel also shows estimates of how much of the difference in redistribution can be attributed to differences in the winner-loser effort gap across conditions.
		}
	\end{figure}

Next, we empirically show that the slight differences in the winner-loser effort gap across conditions cannot quantitatively account for the stark differences in redistribution observed across conditions (even if spectators correctly foresee those differences). First, we measure the elasticity of redistribution with respect to the winner-loser effort gap. Panel A of Figure \ref{fig:red_effort} depicts the relationship between the fraction of earnings redistributed ($y$-axis) and the difference in tasks completed by the winner and loser ($x$-axis), separately by luck treatment. To measure the elasticity of redistribution with respect to the winner-loser effort gap, we estimate equation \eqref{eq:pi-lin} replacing $1-\pi_{ip}$ with the difference in tasks completed between the winner and the loser. We estimate that a one-encryption increase in the winner-loser effort gap decreases the share of earnings redistributed by 0.3 percentage points ($p<0.01$). This means that increasing the winner-loser effort gap by three encryptions decreases redistribution by approximately one percent.

To assess how much of the observed differences in redistribution across conditions can be attributed to winner-loser gap differences, we multiply this elasticity by the actual winner-loser effort gaps we find in our data (Figure \ref{fig:effort_pi}, Panel B). We find that the winner-loser performance gap can account for only 0.2 percentage points of the difference in redistribution between the conditions on average across all values of $\pi$. This contribution is negligible compared to the average overall redistribution difference, which equals 4.2 percentage points. The small influence of the winner-loser effort gap primarily stems from the fact that empirically this gap is too small to impact redistributive behavior substantially, given the spectators' limited sensitivity to this gap. Panel B of Figure \ref{fig:red_effort} visually compares these estimates along with the observed difference in redistribution. We conclude that differences in the winner-loser effort gap across our luck environments cannot explain our results.

 \clearpage

\end{document}